\documentclass[twocolumn]{aastex63}

\newcommand{\cc}{C$^{18}$O}
\newcommand{\ct}{$^{13}$CO}
\newcommand{\cseven}{C$^{17}$O}
\newcommand{\kms}{ km s$^{-1}$}
\newcommand{\nco}{N$_{\rm CO}$~}
\usepackage[version=3]{mhchem}

\shorttitle{MAPS V: CO gas distribution}
\shortauthors{Ke Zhang et al.}
\graphicspath{{./}{figures/}}

\begin{document}

\title{Molecules with ALMA at Planet-forming Scales (MAPS) V: CO gas distributions}

\correspondingauthor{Ke Zhang}
\email{ke.zhang@wisc.edu}

\author[0000-0002-0661-7517]{Ke Zhang}
\altaffiliation{NASA Hubble Fellow}

\affiliation{Department of Astronomy, University of Wisconsin-Madison, 
475 N Charter St, Madison, WI 53706}
\affiliation{Department of Astronomy, University of Michigan, 
323 West Hall, 1085 S. University Avenue, 
Ann Arbor, MI 48109, USA}

\author[0000-0003-2014-2121]{Alice S. Booth}
\affiliation{Leiden Observatory, Leiden University, 2300 RA Leiden, the Netherlands}
\affiliation{School of Physics and Astronomy, University of Leeds, Leeds, LS2 9JT, UK}

\author[0000-0003-1413-1776]{Charles J. Law}
\affiliation{Center for Astrophysics \textbar\, Harvard \& Smithsonian, 60 Garden St., Cambridge, MA 02138, USA}

\author[0000-0003-4001-3589]{Arthur D. Bosman}
\affiliation{Department of Astronomy, University of Michigan,
323 West Hall, 1085 S. University Avenue,
Ann Arbor, MI 48109, USA}

\author[0000-0002-6429-9457]{Kamber R. Schwarz}
\altaffiliation{NASA Hubble Fellowship Program Sagan Fellow}
\affiliation{Lunar and Planetary Laboratory, University of Arizona, 1629 E. University Blvd, Tucson, AZ 85721, USA}

\author[0000-0003-4179-6394]{Edwin A. Bergin}
\affiliation{Department of Astronomy, University of Michigan,
323 West Hall, 1085 S. University Avenue,
Ann Arbor, MI 48109, USA}

\author[0000-0001-8798-1347]{Karin I. \"Oberg} 
\affiliation{Center for Astrophysics \textbar\, Harvard \& Smithsonian, 60 Garden St., Cambridge, MA 02138, USA}

\author[0000-0003-2253-2270]{Sean M. Andrews} 
\affiliation{Center for Astrophysics \textbar\, Harvard \& Smithsonian, 60 Garden St., Cambridge, MA 02138, USA}

\author[0000-0003-4784-3040]{Viviana V. Guzm\'{a}n}
\affiliation{Instituto de Astrof\'isica, Pontificia Universidad Cat\'olica de Chile, Av. Vicu\~na Mackenna 4860, 7820436 Macul, Santiago, Chile}

\author[0000-0001-6078-786X]{Catherine Walsh}
\affiliation{School of Physics and Astronomy, University of Leeds, Leeds, LS2 9JT, UK}

\author[0000-0001-8642-1786]{Chunhua Qi} 
\affiliation{Center for Astrophysics \textbar\, Harvard \& Smithsonian, 60 Garden St., Cambridge, MA 02138, USA}

\author[0000-0002-2555-9869]{Merel L. R. van 't Hoff}
\affiliation{Department of Astronomy, University of Michigan,
323 West Hall, 1085 S. University Avenue,
Ann Arbor, MI 48109, USA}

\author[0000-0002-7607-719X]{Feng Long}
\affiliation{Center for Astrophysics \textbar\, Harvard \& Smithsonian, 60 Garden St., Cambridge, MA 02138, USA}

\author[0000-0003-1526-7587]{David J. Wilner}
\affiliation{Center for Astrophysics \textbar\, Harvard \& Smithsonian, 60 Garden St., Cambridge, MA 02138, USA}

\author[0000-0001-6947-6072]{Jane Huang}
\altaffiliation{NASA Hubble Fellowship Program Sagan Fellow}
\affiliation{Department of Astronomy, University of Michigan,
323 West Hall, 1085 S. University Avenue,
Ann Arbor, MI 48109, USA}
\affiliation{Center for Astrophysics \textbar\, Harvard \& Smithsonian, 60 Garden St., Cambridge, MA 02138, USA}

\author[0000-0002-1483-8811]{Ian Czekala}
\altaffiliation{NASA Hubble Fellowship Program Sagan Fellow}
\affiliation{Department of Astronomy and Astrophysics, 525 Davey Laboratory, The Pennsylvania State University, University Park, PA 16802, USA}
\affiliation{Center for Exoplanets and Habitable Worlds, 525 Davey Laboratory, The Pennsylvania State University, University Park, PA 16802, USA}
\affiliation{Center for Astrostatistics, 525 Davey Laboratory, The Pennsylvania State University, University Park, PA 16802, USA}
\affiliation{Institute for Computational \& Data Sciences, The Pennsylvania State University, University Park, PA 16802, USA}
\affiliation{Department of Astronomy, 501 Campbell Hall, University of California, Berkeley, CA 94720-3411, USA}

\author[0000-0003-1008-1142]{John~D.~Ilee} 
\affiliation{School of Physics and Astronomy, University of Leeds, Leeds, UK, LS2 9JT}

\author[0000-0002-2700-9676]{Gianni Cataldi}
\affiliation{National Astronomical Observatory of Japan, 2-21-1 Osawa, Mitaka, Tokyo 181-8588, Japan}
\affil{Department of Astronomy, Graduate School of Science, The University of Tokyo, 7-3-1 Hongo, Bunkyo-ku, Tokyo 113-0033, Japan}

\author[0000-0002-8716-0482]{Jennifer B. Bergner} 
\altaffiliation{NASA Hubble Fellowship Program Sagan Fellow}
\affiliation{University of Chicago Department of the Geophysical Sciences, Chicago, IL 60637, USA}

\author[0000-0003-3283-6884]{Yuri Aikawa}
\affil{Department of Astronomy, Graduate School of Science, The University of Tokyo, Tokyo 113-0033, Japan}

\author[0000-0003-1534-5186]{Richard Teague}
\affiliation{Center for Astrophysics \textbar\, Harvard \& Smithsonian, 60 Garden St., Cambridge, MA 02138, USA}

\author[0000-0001-7258-770X]{Jaehan Bae}
\altaffiliation{NASA Hubble Fellowship Program Sagan Fellow}
\affil{Department of Astronomy, University of Florida, Gainesville, FL 32611, USA}
\affil{Earth and Planets Laboratory, Carnegie Institution for Science, 5241 Broad Branch Road NW, Washington, DC 20015, USA}

\author[0000-0002-8932-1219]{Ryan A. Loomis}
\affiliation{National Radio Astronomy Observatory, 520 Edgemont Rd., Charlottesville, VA 22903, USA}

\author[0000-0002-0150-0125]{Jenny K. Calahan} 
\affiliation{Department of Astronomy, University of Michigan, 
323 West Hall, 1085 S. University Avenue, 
Ann Arbor, MI 48109, USA}

\author[0000-0002-2692-7862]{Felipe Alarc\'on }
\affiliation{Department of Astronomy, University of Michigan,
323 West Hall, 1085 S. University Avenue,
Ann Arbor, MI 48109, USA}

\author[0000-0002-1637-7393]{Fran\c cois M\'enard}
\affiliation{Univ. Grenoble Alpes, CNRS, IPAG, F-38000 Grenoble, France}

\author[0000-0003-1837-3772]{Romane Le Gal}
\affiliation{Center for Astrophysics \textbar\, Harvard \& Smithsonian, 60 Garden St., Cambridge, MA 02138, USA}
\affil{IRAP, Universit\'{e} de Toulouse, CNRS, CNES, UT3, 31400 Toulouse, France}
\affil{IPAG, Universit\'{e} Grenoble Alpes, CNRS, IPAG, 38000 Grenoble, France}
\affil{IRAM, 300 rue de la piscine, F-38406 Saint-Martin d'H\`{e}res, France}

\author[0000-0002-5991-8073]{Anibal Sierra} \affiliation{Departamento de Astronom\'ia, Universidad de Chile, Camino El Observatorio 1515, Las Condes, Santiago, Chile}

\author[0000-0003-4099-6941]{Yoshihide Yamato} 
\affiliation{Department of Astronomy, Graduate School of Science, The University of Tokyo, 7-3-1 Hongo, Bunkyo-ku, Tokyo 113-0033, Japan}

\author[0000-0002-7058-7682]{Hideko Nomura}
\affiliation{National Astronomical Observatory of Japan, 2-21-1 Osawa, Mitaka, Tokyo 181-8588, Japan}

\author[0000-0002-6034-2892]{Takashi Tsukagoshi} \affiliation{National Astronomical Observatory of Japan, 2-21-1 Osawa, Mitaka, Tokyo 181-8588, Japan}

\author[0000-0002-1199-9564]{Laura M. P\'erez} \affiliation{Departamento de Astronom\'ia, Universidad de Chile, Camino El Observatorio 1515, Las Condes, Santiago, Chile}

\author[0000-0002-8623-9703]{Leon Trapman}
\affiliation{Department of Astronomy, University of Wisconsin-Madison, 
475 N Charter St, Madison, WI 53706}

\author[0000-0002-7616-666X]{Yao Liu}
\affiliation{Purple Mountain Observatory \& Key Laboratory for Radio Astronomy, Chinese Academy of Sciences, Nanjing 210023, China}

\author[0000-0002-2026-8157]{Kenji Furuya} 
\affiliation{National Astronomical Observatory of Japan, 2-21-1 Osawa, Mitaka, Tokyo 181-8588, Japan}



\begin{abstract}

Here we present high resolution (15-24 au) observations of CO isotopologue lines from the Molecules  with  ALMA on Planet-forming Scales (MAPS) ALMA Large Program. Our analysis employs $^{13}$CO and C$^{18}$O ($J$=2-1), (1-0), and C$^{17}$O (1-0) line observations of five protoplanetary disks. We retrieve CO gas density distributions, using three independent methods: (1) a thermo-chemical modeling framework based on the CO data, the broadband spectral energy distribution, and the mm-continuum emission; (2) an empirical temperature distribution based on optically thick CO lines; and (3) a direct fit to the C$^{17}$O hyperfine lines. Results from these methods generally show excellent agreement. The CO gas column density profiles of the five disks show significant variations in the absolute value and the radial shape. Assuming a gas-to-dust mass ratio of 100, all five disks have a global CO-to-H$_2$ abundance of 10-100 times lower than the ISM ratio. The CO gas distributions between 150-400 au match well with models of viscous disks, supporting the long-standing theory. CO gas gaps appear to be correlated with continuum gap locations, but some deep continuum gaps do not have corresponding CO gaps.
The relative depths of CO and dust gaps are generally consistent with predictions of planet-disk interactions, but some CO gaps are 5-10 times shallower than predictions based on dust gaps. 
This paper is part of the MAPS special issue of the Astrophysical Journal Supplement.

\end{abstract}


\section{Introduction} \label{sec:intro}

In protoplanetary disks, H$_2$ is the primary mass carrier, but it does not have strong transitions in the temperature range that most of the disk mass resides. Instead, CO is the most widely used gas tracer, as it is abundant and has rotational transitions sensitive to cold, 10-50\,K, gas \citep[e.g.,][]{bergin17}. CO is also chemically stable \citep{vanDishoeck88,visser09}, and stays in the gas phase at temperature warmer than $\sim$20\,K  \citep{Molyarova17}. Besides being a tracer of gas mass, CO is one of the primary carbon and oxygen carriers in protoplanetary disks, a stepping-stone to a series of chemical reactions that eventually lead to the formation of complex organics \citep[e.g.,][]{Walsh14}. 

Observations of (sub)mm CO ($J\leq3$) lines have been widely used to study fundamental physical properties of protoplanetary disks, including total disk gas mass \citep{williams14,ansdell16,miotello17,long17}, gas depletion in cavities and gaps \citep{vanderMarel15,Facchini18, Favre19}, temperature structures \citep{Qi06, Qi11, pietu07, schwarz16, kama16, Pinte18_imlup}, kinematics in disks \citep{Teague18a,Pinte18_hd163, Teague19Nat}, and turbulence levels \citep{teague16,flaherty15,flaherty17,Flaherty20}. Therefore, accurately measuring the CO gas distribution in disks is essential for our understanding of the physical and chemical evolution of protoplanetary disks as well as the planet formation processes occurring within them. 

Over the past few years, one of the most extensive debates is whether CO is a robust tracer of H$_2$ gas mass in disks. The gas mass derived from CO observations hinges on the crucial assumption that the CO-to-H$_2$ ratio is the canonical ISM ratio of 10$^{-4}$ in the warm molecular layer throughout the entire disk life-time \citep{bergin17}. However, observations are in tension with this assumption. Surveys of protoplanetary disks in nearby star-forming regions show that gas masses derived from CO lead to a gas-to-dust ratio $\le$10 in the majority of disks \citep{ansdell16,miotello17,long17}. In contrast, the high gas accretion rates onto the central star suggest many disks are still gas-rich \citep{manara16}. The most direct evidence of CO depletion comes from independent gas mass measurements with the HD $J$=1-0 line in three disks \citep{bergin13, mcclure16,Trapman17}. Their CO-to-H$_2$ ratios are one to two orders of magnitude lower than the canonical ISM ratio \citep{Favre13,schwarz16,zhang17,Zhang19,Calahan21_twhya}.

To explain low CO abundances seen in protoplanetary disks, two broad types of processes have been proposed. The first type is chemical processes --- CO is chemically processed into other molecules \citep[e.g.,][]{aikawa98,bergin14,Schwarz18,Bosman18}. The second type is physical processes --- CO gas is gradually removed from the disk atmosphere, as icy dust grains grow and settle towards the mid-plane \citep{xu17,krijt16,Krijt18, Krijt20}. Recent comparisons between the CO gas mass in Class I and II disks suggest that the CO-to-\ce{H_2} abundance ratio decreases rapidly on a timescale of $\sim$1\,Myr \citep{Zhang20_evolution,Bergner_20evolution}. In short, current observations and theoretical studies suggest that the CO gas abundance evolves along with the disk evolution and planet formation, rather than a simple heritage of the molecular cloud.

Another important question is whether the CO gas distribution has substructures at dust gap locations. Substructures (e.g., rings, gaps, and spirals) have been commonly seen in (sub)mm continuum emission of protoplanetary disks \citep[e.g.,][]{Zhang16, Isella16, Andrews18b, long18, Cieza21_ODISEAIII}. 
Various mechanisms have been proposed as possible causes, including planet-disk interaction \citep{dong15, Rosotti16,Zhang_S18}, dust property changes at snowlines \citep{zhang15b,okuzumi16, Pinilla17}, and magnetic instability \citep{Flock15, Suriano18, Riols19}. As gas and mm-sized dust grains have different dynamic behaviors, gas substructures can provide an important complementary insight into possible causes. In particular, planet-disk interaction models predict that giant planets can open deep gas gaps that manifest as gaps in CO line emissions \citep[e.g.,][]{Facchini18, vanderMarel18, Alarcon20}.

Most of previous studies of the CO gas mass distributions in Class II disks have relatively low spatial resolutions, between 25-100\,au \citep[e.g.,][]{schwarz16, nomura16, Isella16, Fedele17, Zhang19, Favre19, Rosotti21}. Although local coincidences in CO and dust gap locations have been reported in a few disks, previous observations often did not have sufficient spatial resolution to fully characterize CO substructures. Also, studies often focused on local perturbations and thus lacked a holistic view of the CO gas distribution across the entire disk.  

Here we study the CO gas mass distribution in five protoplanetary disks observed as part of the Molecules with ALMA on Planet-forming Scales (MAPS) ALMA Large Program. We use five CO isotopologue lines to constrain CO gas distribution at a scale of $\sim$15-24\,au. We aim to robustly measure CO gas distributions in these five disks, focusing on three important questions: (1) How does the CO column density vary with radius? (2) How much does the CO column density change across continuum substructures? (3) What are the lower limits of gas masses in these disks? 

This paper is structured as follows. In Section~\ref{sec:obs}, we briefly present the observations of CO lines in the MAPS program and the data reduction process. We describe the three methods used to constrain the CO column density distributions and present the results in Section~\ref{sec:methods}. In Section~\ref{sec:discussion}, we discuss how these results answer the three questions above. Then we summarize our findings in Section~\ref{sec:summary}.

\section{Observations} \label{sec:obs}

\begin{figure*}[htbp]
    \centering
    \includegraphics[width=0.95\textwidth]{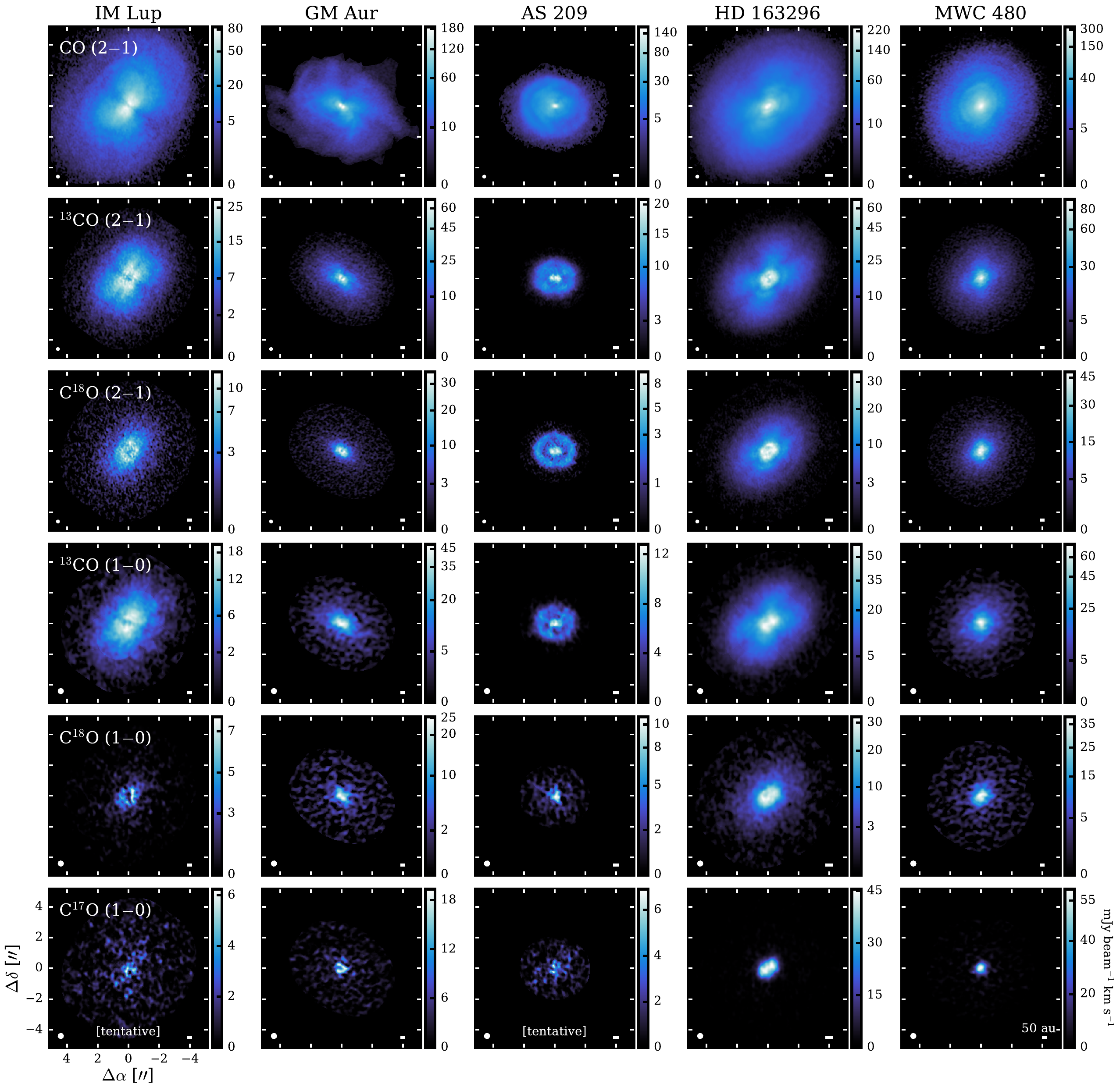}
    \caption{Integrated intensity maps of CO,\ct, \cc, and \cseven~lines for the MAPS disks. Axes are angular offsets from the disk center with each white tick mark representing a spacing of 2$^{\prime \prime}$. The synthesized beam is plotted in the lower left corner of each panel. The white scale bar at the lower right corner of each panel represents 50\,au scale.} 
        \label{fig:mom0_maps}
\end{figure*}

\begin{figure*}[htbp]
\centering
\includegraphics[width=0.95\textwidth]{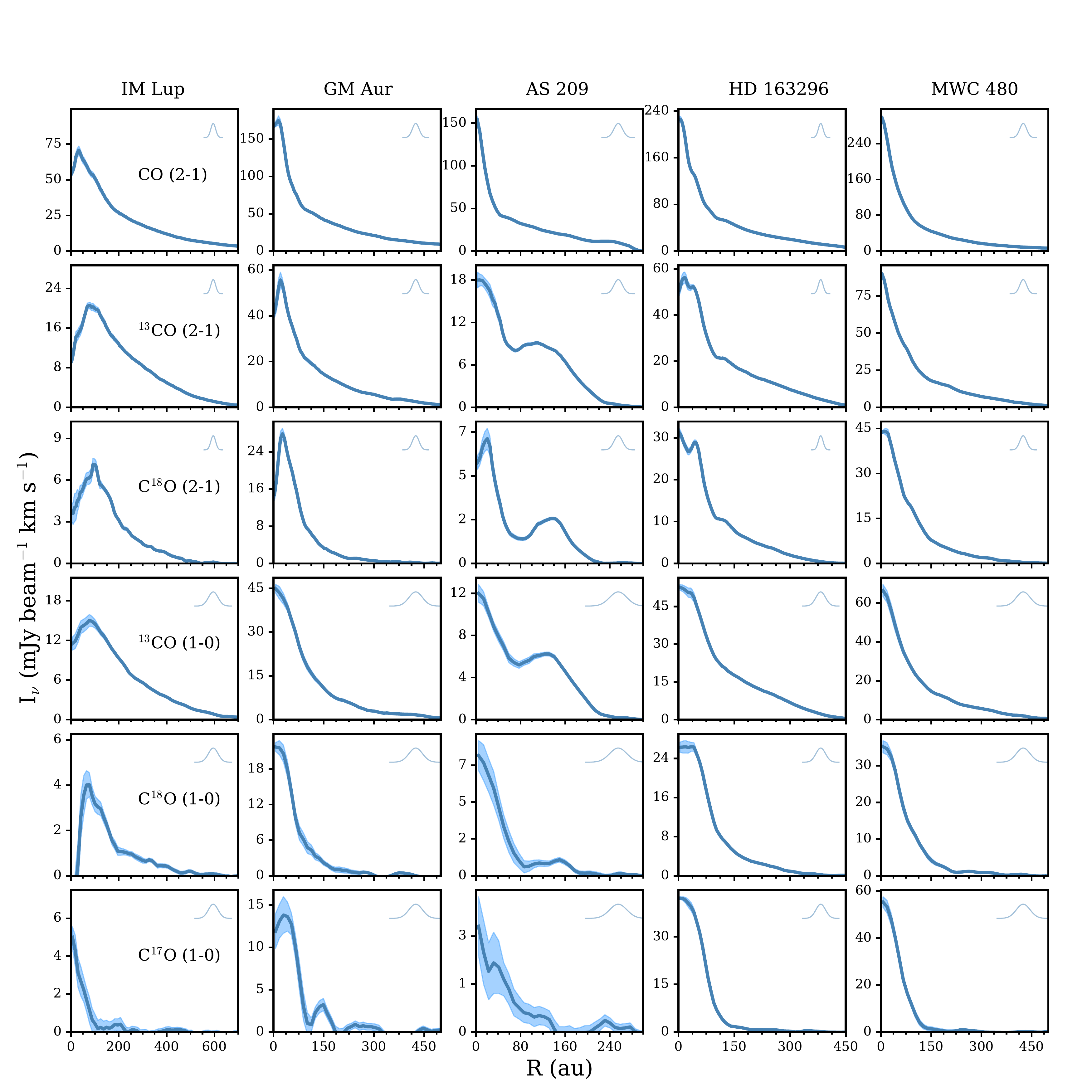}
\caption{Deprojected radial intensity profiles of CO lines, for the MAPS sample, ordered by increasing stellar mass from left to right. The radial profiles are azimuthally-averaged, see details in \citet{law_maps_radial}. The only exception is the AS 209 CO (2-1) line, which is derived from a 55$^{\circ}$ wedge to avoid cloud contamination. Blue shaded regions show the standard error on the mean in the annulus or arc over which the emission was averaged.} The individual beam size is plotted at the upper right corner of each panel. The beam sizes are 0\farcs15$\times$0\farcs15 for all (2-1) lines, and 0\farcs3$\times$0\farcs3 for (1-0) lines, respectively. A plot of the intensity profiles on a logarithmic scale is shown in the Appendix Figure~\ref{fig:co_obs_log}. \label{fig:co_obs}
\end{figure*}

This study uses CO isotopologue line data covered in the MAPS ALMA Large Program (2018.1.01055.L) for five disk sources: IM Lup,  GM~Aur, AS~209, HD~163296, and MWC~480. The ALMA Band 6 setting targeted the \ce{CO}, \ce{^{13}CO}, \ce{C^{18}O}
$J$=2-1 transitions and the Band 3 setting targeted the \ce{^{13}CO}, \ce{C^{18}O}, \ce{C^{17}O} $J$=1-0 transitions \citep{oberg20_maps}. The analysis presented here is based on the MAPS fiducial images, which have 0\farcs15 and 0\farcs30 circularized beams for lines in Bands 6 and 3, respectively. The full details of the data calibration and imaging are described in \citet{Czekala20_maps}. The observational logs, line frequencies, and image statistics are listed in \citet{oberg20_maps}.
We use the integrated intensity maps and radial emission profiles from \citet{law_maps_radial}. Figures~\ref{fig:mom0_maps} and~\ref{fig:co_obs} show integrated line intensity maps and radial profiles used in this study. 

We also present the \ce{C^{17}O} (1-0) data which are not covered by \citet{law_maps_radial}.  
The \ce{C^{17}O} (1-0) line has three hyperfine components which are blended in the integrated intensity map. 
As shown in Figure~\ref{fig:mom0_maps} and \ref{fig:co_obs}, the \ce{C^{17}O} (1-0) lines were robustly detected in the HD~163296 and MWC~480 disks, and tentatively in the IM Lup, GM Aur, and AS 209 disks.
For the tentative detections, we used the matched filter method developed by \citet{Loomis2018} to estimate the signal-to-noise ratio of their detections. The \ce{C^{17}O} data were searched with a Keplerian mask extending to 100\,au for all five disks. This mask size is chosen because most of the \ce{C^{17}O} emission is expected to be compact ($\leq$~100~au) originating from within the CO snowline \citep{zhang17, Booth19}. The matched filter responses for the GM~Aur, IM~Lup, and AS~209 disks are presented in the Appendix in Figure~\ref{c17o_filter}.
The \ce{C^{17}O} (1-0) line is robustly detected (20\,$\sigma$) in the GM~Aur disk, but only tentatively (3-4\,$\sigma$) in the AS~209 and IM~Lup disks. 

\section{Methods and Results} \label{sec:methods}

In this work, our goal is to measure how CO gas column density varies with radial distance from the central star in the five MAPS disks (N$_{\rm CO}$(R)), using spatially resolved CO isotopologue line images. 

\begin{figure*}[htbp]
\centering
\vspace{-0.2cm}
\includegraphics[width=0.9\textwidth]{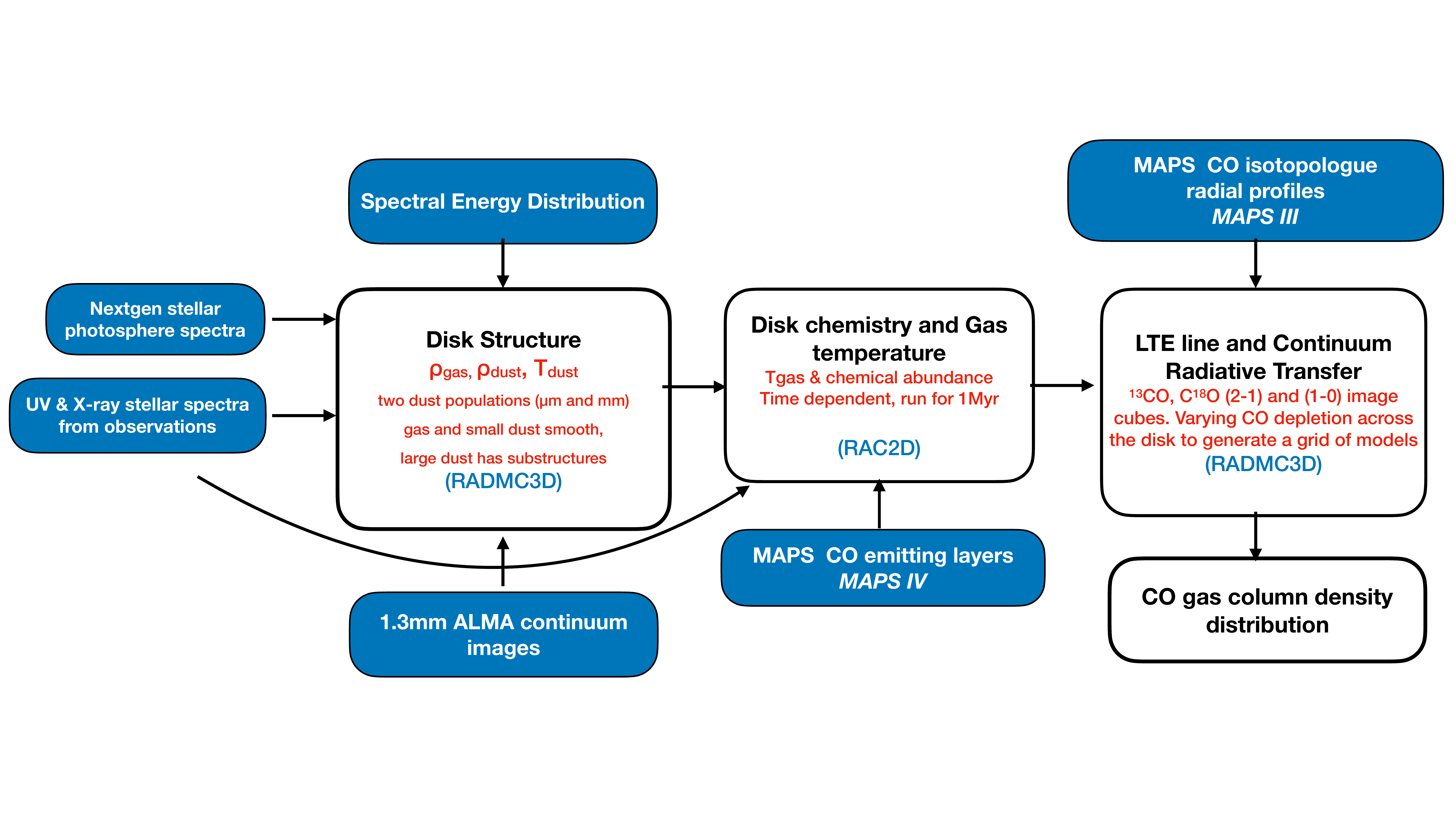}
\vspace{-1.5cm}
\caption{An outline of the modeling processes. Codes used are in light blue color, input data are highlighted in solid blue boxes, and the output of each step is highlighted in red. \label{fig:flowchart}}
\end{figure*}

To test the robustness of the results, we employ three independent methods to constrain N$_{\rm CO}$(R) profiles:

Method 1: we build thermo-chemical models for individual disks to constrain the temperature structures of gas and dust in these disks. These models are based on fitting broad band spectral energy distribution (SED), (sub)mm continuum images, and vertical locations of CO emission surfaces. We then use the thermal structure to retrieve the \nco at each radius by matching the observed radial intensity profiles of lines. The workflow of the modeling processes is shown in Figure~\ref{fig:flowchart}, and the details are described in Section~\ref{sec:thermo-chemical models}.

Method 2: we replace the gas temperature in thermo-chemical models with empirically constructed temperature structures based on optically thick CO lines, and then redo the \nco retrieval.
This process is described in Section~\ref{sec:empirical_temperature}.

Method 3: we use line ratios and absolute intensities of \cseven~(1-0) hyperfine components to constrain optical depth and excitation temperature. This analysis is described in Section~\ref{sec:c17o_col}. 

In the following subsections, we describe the detailed setup of each method. We note that all of CO data and thermochemical models used in this paper can be downloaded from the official website of the MAPS program: www.alma-maps.info.

\begin{deluxetable*}{lcccccccccc}
\tablecaption{Stellar and Disk Properties \label{tab:sources}}
\tablehead{
\colhead{Source} & \colhead{SpT} &\colhead{d} & \colhead{incl} & \colhead{PA} & \colhead{T$_{\rm eff}$} &\colhead{$L_{\rm *}$} &\colhead{$M_*$}  &\colhead{log$_{10}$($\dot{M}$)} &\colhead{$v_{\rm sys}$}  &\colhead{References}\\
\colhead{}  & \colhead{} &\colhead{(pc)}&\colhead{(deg)}&\colhead{(deg)}&\colhead{(K)} &\colhead{($L_{\rm \odot}$)} &\colhead{($M_{\rm \odot}$)} &\colhead{($M_{\rm \odot}$ yr$^{-1}$)} & \colhead{(km~s$^{-1}$)}
}
\startdata
IM Lup  &K5  &158    &47.5   &144.5  &4266   &2.57   &1.1    &$-$7.9   &4.5    &1,2,3,4,5\\
GM Aur  &K6 &159 &53.2  &57.2  &4350   &1.2    &1.1    &$-$8.1   &5.6  &1,6,7,8,9\\
AS 209  &K5   &121    &35.0  &85.8  &4266   &1.41   &1.2    &$-$7.3   &4.6    &1,2,5,10,11\\
HD 163296   &A1  & 101 &46.7   &133.3  &9332   &17.0 &2.0  &$-$7.4   &5.8   &1,2,5,12,13\\
MWC 480 &A5 &162   &37.0 &148.0    &8250   &21.9   &2.1   &$-$6.9   &5.1    &1,14,15,16,17,18\\
\enddata
\tablecomments{Reproduced from \citet{oberg20_maps}, where further details about the source characteristics are available.\\\\References are (1) \citet{gaia2}; (2) \citet{Huang18b}; (3) \citet{Alcala17}; (4) \citet{Pinte18_imlup}; (5) \citet{Andrews18b}; (6) \citet{Huang20}; (7) \citet{Macias18}; (8) \citet{Ingleby15}; (9) \citet{Espaillat10}; (10) \citet{salyk13}; (11) \citet{Huang17}; (12) \citet{Fairlamb15}; (13) \citet{Teague19Nat}; (14) \citet{Liu19}; (15) \citet{Montesinos09}; (16) \citet{Simon19}; (17) \citet{pietu07}; (18) \citet{Mendigutia13}}
\end{deluxetable*}

\subsection{Themo-chemical models}\label{sec:thermo-chemical models}

\subsubsection{Density structure}

Our disk model is axisymmetric and consists of three mass components -- gas, a small dust population, and a large dust population. We assume that gas and the small dust population are spatially coupled and thus their densities differ only by a constant factor across the disk. The mass surface distribution of gas and small dust is set as a self-similarly viscous disk \citep[e.g.,][]{lynden-bell74,andrews11}.

\begin{equation}\label{eq:sigma_profile}
\Sigma (R) = \Sigma_c \Big(\frac{R}{R_c}\Big)^{-\gamma} {\rm exp} \Big[ -\Big(\frac{R}{R_c}\Big)^{2-\gamma}\Big] 
\end{equation}
\noindent where R$_c$ is the characteristic scaling radius and $\gamma$ is the gas surface density exponent.

For the large dust population (pebbles), its surface density is set by matching mm continuum images. The details of the process are described in Section~\ref{sec:cont_images}.

The vertical density structure is assumed to be a Gaussian function characterized by a scale height $H(R)$ that is a power-law function of radius. 
\begin{equation}\label{eq:rho}
\rho_i(R, Z) = f_i \frac{\Sigma (R)}{\sqrt{2\pi} H_i(R)}  {\rm exp} \Big[ -\frac{1}{2}\Big(\frac{Z}{H_i(R)}\Big)^2\Big], 
\end{equation}
\begin{equation}\label{eq:hr}
H_i (R) = \chi_i H_{100} (R/100~{\rm au} )^\psi 
\end{equation}
where $f_i$ is the mass fraction of each mass component, $H_{100}$ is the scale height at 100\,au, $\psi$ is a parameter that characterizes the radial dependence of the scale height. The large grain population is expected to be more settled compared to the gas and small grains \citep{Nakagawa86,dullemond04_settling}. This settling effect is characterized with the parameter $\chi_i$. Here we fix $\chi$=1 for the gas and the small grain population, and $\chi$=0.2 for the large grain population \citep{andrews11}.

\begin{deluxetable*}{lccccccccccc}
\tablecaption{Best-fit disk Model Parameters\label{tab:disk_para}}
\tablehead{
\colhead{Source} & \colhead{M$_g$} &\colhead{$\Sigma_{c}^g$} &\colhead{M$_{mm}$} & \colhead{M$_{\mu m}$} & \colhead{$\Sigma_{c}^{\mu m}$} &
\colhead{R$_{c}^g$} & \colhead{$\gamma_g$} & \colhead{H$_{100}$} & \colhead{$\psi$} & \colhead{r$_{\rm in}$} & \colhead{r$_{\rm out}$} \\
\colhead{} & \colhead{(M$_\odot$)} & \colhead{(g cm$^{-2}$)} &\colhead{(M$_{\odot}$)} &\colhead{(M$_{\odot}$)} & \colhead{(g cm$^{-2}$)} &\colhead{(au)} & & \colhead{(au)} & & \colhead{(au)} & \colhead{(au)} 
}
\colnumbers
\startdata
IM Lup      &         0.2  &  28.4 &     1.97e-03  & 2.02e-05  &   2.9e-03 &     100  &  1.0  &10.0  &    1.17  &0.20  &1200 \\
GM Aur      &         0.2  &   9.4 &     5.94e-04  & 1.03e-04  &   6.0e-03 &     176  &  1.0  &7.5  &    1.35  &1.00  &650 \\
AS 209      &      0.0045  &   1.0 &     4.50e-04  & 5.23e-05  &   1.2e-02 &      80  &  1.0  &6.0  &    1.25  &0.50  &300 \\
HD 163296   &        0.14  &   8.8 &     2.31e-03  & 2.00e-04  &   1.3e-02 &     165  &  0.8  &8.4  &    1.08  &0.45  &600 \\
MWC 480     &        0.16  &   5.8 &     1.41e-03  & 1.69e-04  &   6.1e-03 &     200  &  1.0  &10.0  &    1.08  &0.45  &750 \\
\enddata
\tablecomments{(1) Source name; (2) total gas mass of the disk; (3) surface density of gas at R$_{c}^g$; (4) total mass of the large grain population; (5) total mass of the small grain population; (6) surface density of the small grain population at R$_{c}^g$;  (7) characteristic radius of the gas and the small grain population; (8) surface density exponent in eq.~(\ref{eq:sigma_profile}); (9) scale height at 100\,au; (10) power index of the radial dependence of scale height, eq.~(\ref{eq:hr}); (11) the inner edge of the disk model; (12) the outer edge of the disk model. }
\end{deluxetable*}

\subsubsection{Dust Opacity}
We employ the dust size distribution of $n(a)\propto a^{-3.5}$ from the Mathis, Rumpl, \& Nordsieck model \citep{Mathis77}, for both small and large dust populations. The minimum grain size  a$_{\rm min}$ is fixed at 0.005\,$\mu$m. The maximum size a$_{\rm max}$ of the small and large dust populations is set to be 1\,$\mu$m and 1\,mm, respectively. The a$_{\rm max}$ of the large dust population is consistent with multi-wavelength analysis of these five disks. Based on the 1.3\,mm and 3\,mm continuum images, \citet{Sierra20} showed that all five MAPS disks have an a$_{\rm max}$ between 1\,mm and 3\,mm beyond 25\,au.

For the composition of the large dust population, we adopt the recommendation of the DSHARP collaboration \citep{Birnstiel18}, which is a mixture of water ice, astronomical silicates, troilite, and refractory organics. For the small dust population, we adopt a 50\%-50\% mixture of silicates and refractory organics because current observations and dust evolution models suggest that water ice is destroyed by photodesorption at higher disk atmosphere region \citep{hogerheijde11_water,krijt16,du17}. The \texttt{dsharp\_opac} package from \citet{Birnstiel18} is used to compute the wavelength dependence of dust opacity. We show the dust opacity profiles in the Appendix Figure~\ref{fig:stellar_spec}. 

\subsubsection{Stellar Spectrum}
Young stars accrete disk materials actively and produce strong X-ray and UV radiation \citep{Hartmann16_ARAA}. This high energy radiation is particularly important for chemistry in disks \citep{Bergin07}. Here we make a composite spectrum for each source by combining a model stellar photosphere spectrum with UV and X-ray spectra from observations.

The stellar photospheric spectra are computed by interpolating the standard Nextgen spectral model library in the (log$_{10}$\,T$_{\rm eff}$, log$_{10}$\,$g$) space, assuming a solar metallicity \citep{Nextgen_giant}. The UV and X-ray spectra of individual sources are taken from the DIANA database \citep{Dionatos19}. As UV spectra are highly variable with time, we use averaged spectra. The observed UV spectra may suffer from extinction of circumstellar materials, and therefore do not represent the UV spectra received by the disk. We apply the extinction at UV wavelength based on A$_{\rm V}$ of individual sources and the wavelength dependent dust opacity of our small dust grain population. Some of the sources do not have short wavelength coverage, and we replace these missing parts by scaling the UV spectrum of TW Hya \citep{Herczeg02,Herczeg04}. The UV spectra are then scaled to fit the stellar spectra. The final composite spectra are shown in the Appendix in Figure~\ref{fig:stellar_spec}.

\subsubsection{Thermo-chemical modeling}

We use the thermo-chemical code RAC2D to compute the CO abundance structure and gas temperature. The detailed description of RAC2D can be found in \citet{du14} and we only briefly describe the code here. 

Taking a static density structure of gas and dust as inputs, RAC2D first uses a Monte Carlo approach to simulate absorption and scattering events of photons in the disk. The first step produces a dust thermal structure and radiation field (from X-ray to centimeter wavelengths). The cosmic-ray ionization inside the disk is simulated with an attenuation length of 96\,g cm$^{-2}$ \citep{Umebayashi81, bergin17}. With the computed radiation field, the code then simultaneously solves the chemical evolution and gas thermal structure in the disk. The chemical network has 484 species and 5830 reactions, including the full gas-phase network from the UMIST 2006 database\footnote{http://udfa.ajmarkwick.net}, dissociation of H$_2$O and OH by Ly$\alpha$ photons, adsorption and desorption of species on the dust grain surface either thermally or induced by cosmic rays and UV photons, and two-body reactions on the dust grain surface. The self-shielding effects of H$_2$O and CO are included in the models. 

The initial chemical abundances are listed in Table~1 of \citet{du14}. We adopt a relatively low cosmic-ray rate of 1.36$\times10^{-18}$\,s$^{-1}$ at the disk surface, because previous studies suggest the rate may be low in protoplanetary disks \citep{cleeves15}.

We let the chemistry evolve for 1 Myr for all models. Isotopologue fractionation is not considered in the chemical network, as fractionation is expected to be insignificant in massive disks \citep{miotello14}. We scale the output CO abundance by the local ISM ratios of CO/\ct = 69, CO/\cc = 557, CO/\cseven = 2005 to generate CO isotopologue abundances \citep{wilson99}. We will  discuss how this assumption may affect our results in Section~\ref{sec:photodissociation}.

\subsubsection{Radiative Transfer modeling}

For a given dust density structure and stellar spectrum, we calculate the dust temperature structure using the Monte Carlo radiative transfer code RADMC3D \citep{radmc3d}. These structures are then used to calculate synthetic SEDs and mm continuum images, using the ray-tracing module of RADMC3D. The disk inclination angle ($i$) and the major axis position angle (PA) are adopted from Table~\ref{tab:sources}. Dust scattering is not included in our models of SED and continuum images, as it does not affect our results except for the inner 15\,au region. The analysis of dust scattering effects of MAPS continuum images is explored in \citet{Sierra20}.

 RADMC3D is also used to generate line image cubes. The CO line transitions are assumed to be in local thermal equilibrium (LTE). We generate model image cubes at 2-3\,au resolution and 0.04\,\kms~spectral resolution. The model channels are rebinned to the spectral resolution of MAPS observations and then are convolved with a Gaussian that has the same size and position angle of the beam of CO line observations. Then azimuthally-averaged radial intensity profiles of the model images are generated using the \texttt{radial\_profile} function in the Python package \texttt{GoFish} \citep{GoFish}, following the same procedure as those generated from the observations \citep{law_maps_radial}.

\subsubsection{Constraining Disk Parameters} \label{sec:cont_images}

\begin{figure*}[htbp]
\centering
\includegraphics[width=0.95\textwidth]{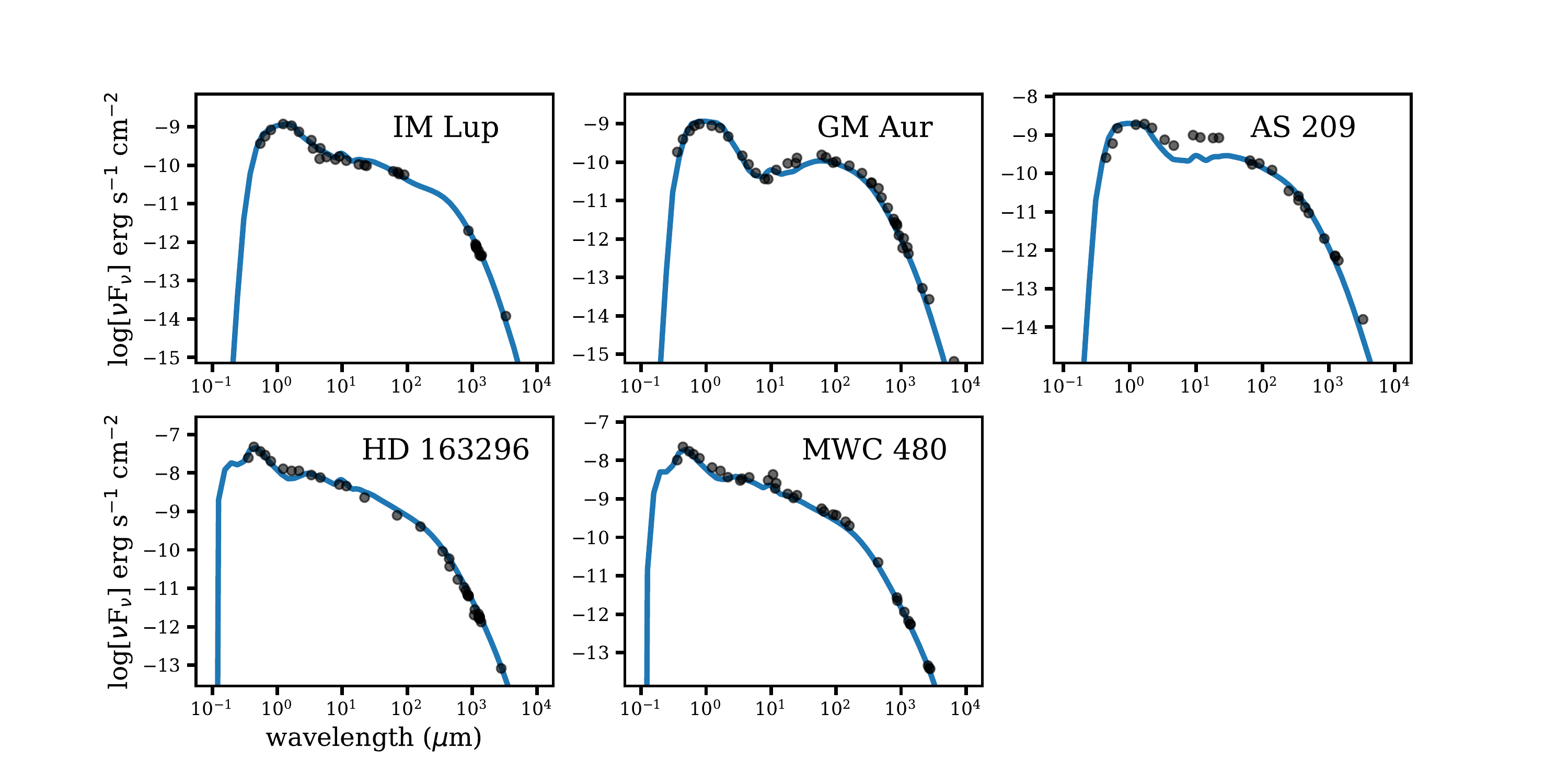}
\vspace{-0.5cm}
\caption{Best-fit SED models of MAPS disks. The models are in blue and photometric data are in black. The photometric data of IM Lup, AS 209, and HD 163296 are adopted from \citet{Andrews18b}, GM Aur from \citet{Macias18}, and MWC 480 from \citet{anderson13}. \label{fig:sed}}
\end{figure*}

\begin{figure*}[htbp]
\centering
\vspace{-0.2cm}
\includegraphics[width=0.95\textwidth]{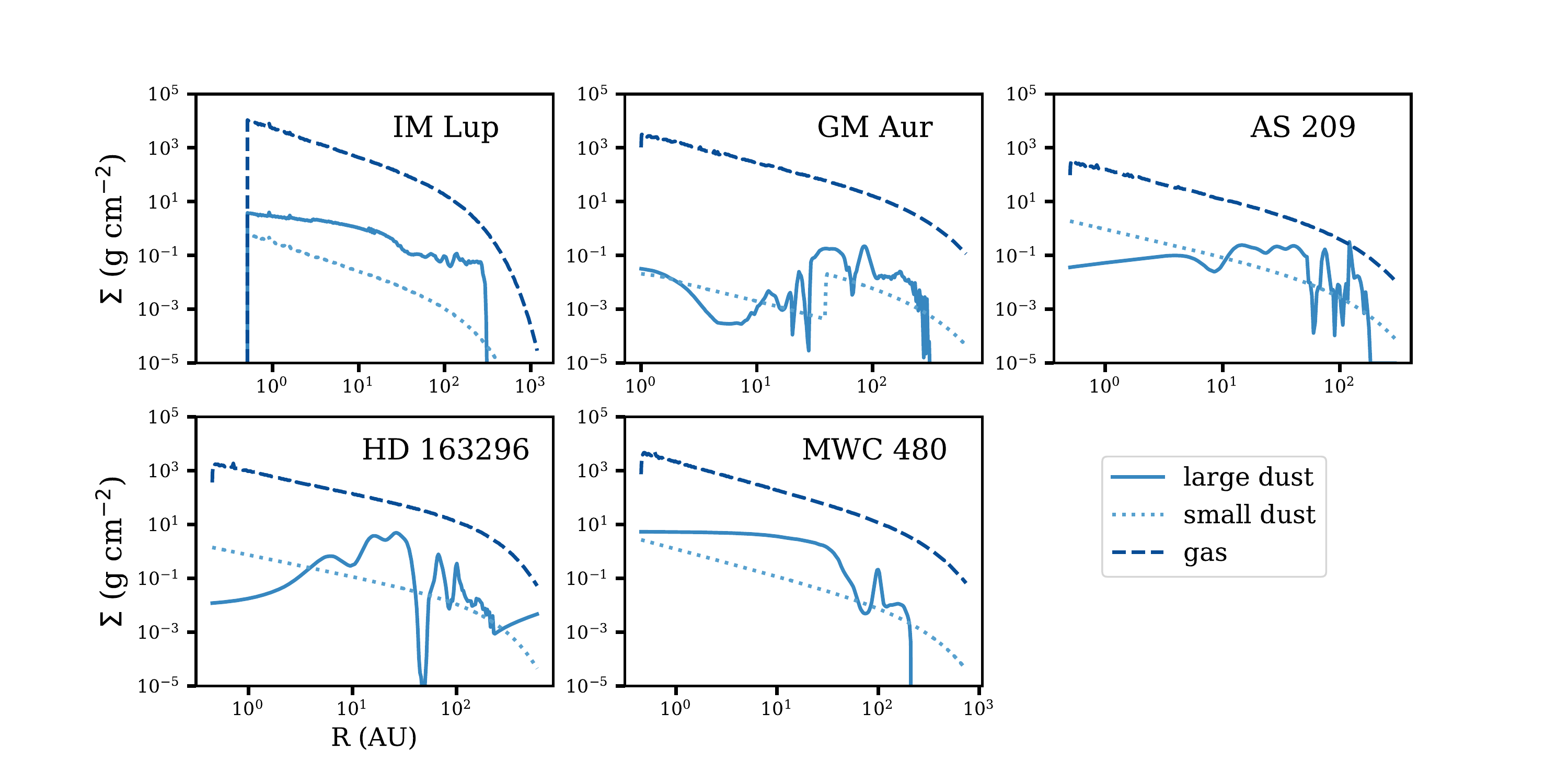}
\vspace{-0.4cm}
\caption{Surface density profiles of the gas, small, and large dust grain populations in our best-fit models of the five MAPS disks. The distributions of the gas and small dust grains are characterized by eq.~(\ref{eq:sigma_profile}), while the distributions of the large dust population are customized to match high-resolution continuum observations (see more details in Section~\ref{sec:cont_images}).  \label{fig:sigma}}
\end{figure*}

\begin{figure*}[htbp]
\centering
\includegraphics[width=\textwidth]{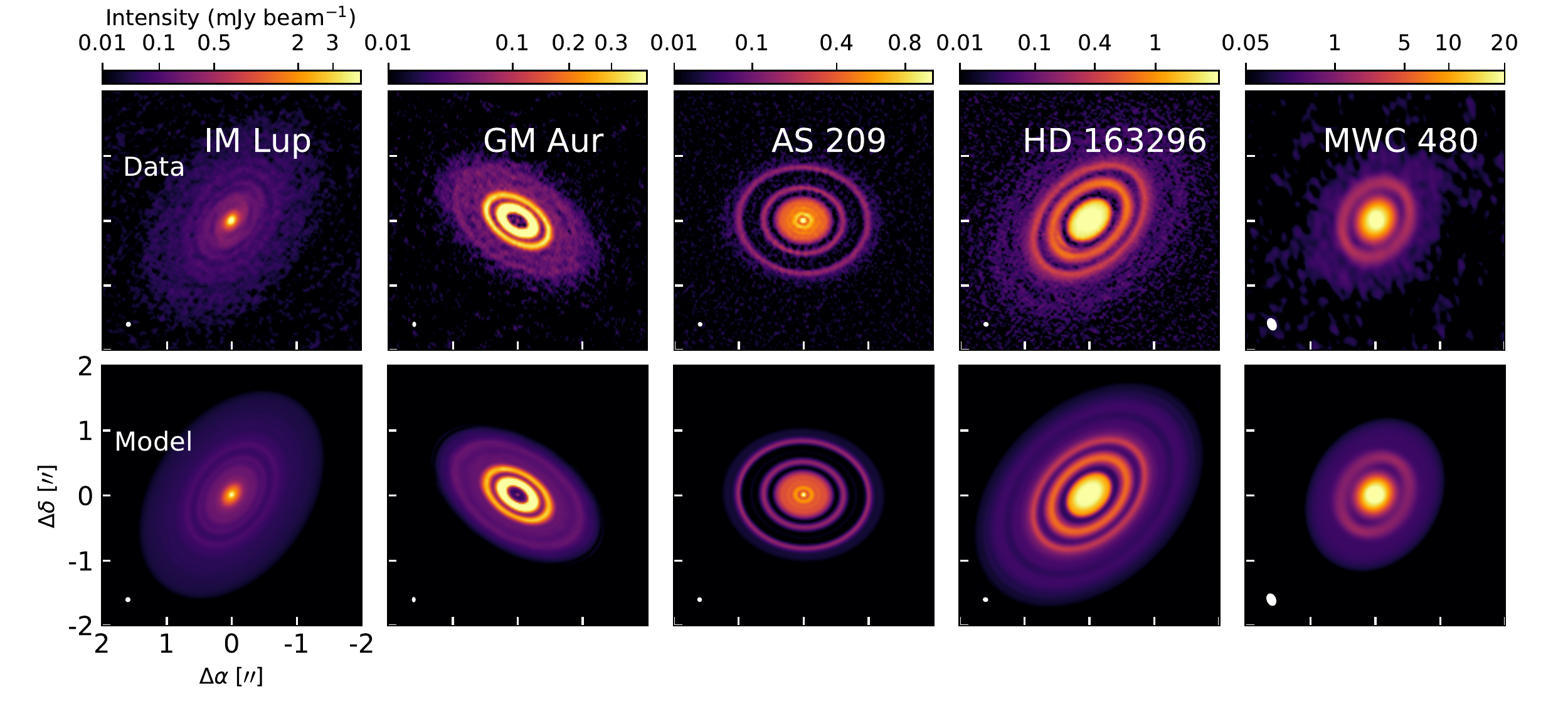}
\caption{Comparison between the observed 1.3\,mm continuum images (top row) and simulated images from our thermo-chemical models (bottom row). The continuum images of the IM Lup, AS 209, and HD 163296 disks are from the DSHARP program \citep{Andrews18b,Huang18b}; the MWC 480 disk image from \citet{long18,Liu19}, and the GM Aur disk image from \citet{Huang20}. For each disk, the data and model panels use the same color scale. The model fitting processes are described in Section~\ref{sec:cont_images}. \label{fig:cont_model}}          
\end{figure*}

The disk structure model has 6 free parameters: 4 parameters to characterize the surface density profiles of gas and small dust, $\big\{\Sigma_c^g, \Sigma_c^{\mu m}, R_c, \gamma\big\}$, and 2 parameters to describe the scale height distribution $\big\{H_{\rm 100},\psi \big\}$. Moreover, the surface density profile of the large grain population is determined by an iterative process to fit the mm continuum images (more details below).

The initial values of parameters are taken from literature for each disk \citep{cleeves16,Macias18,Fedele18,Zhang19,Liu19}. We then update the parameters by comparing model outputs with three types of observations: SEDs, mm-continuum images, and the CO emission surfaces (see detailed explanation below). The general modeling processes are outlined in Figure~\ref{fig:flowchart}.

The mid and far-IR profile of the SED is mostly determined by the $\psi$ and H$_{100}$ parameters. Therefore, we run a small grid of $\psi$ and H$_{100}$ values to search for the best fitting value. The scale height of the large grain population is fixed to be 20\% of the gas scale height \citep{dubrulle95,andrews11}. The SEDs from our best-fitting models are shown in Figure~\ref{fig:sed}. 

The second type of observational constraints is the spatially resolved mm-continuum images of the five disks. For the large dust grain population, we iterate an initial surface density profile until the azimuthally averaged surface brightness distribution matches with the high-resolution ALMA observations. The continuum images of IM Lup, AS 209, and HD 163296 are from the DSHARP program \citep{Andrews18b,Huang18b}, MWC 480 image from \citet{long18} and GM Aur image from \citet{Huang20}. 

Because our model is axisymmetric, we only compare azimuthally averaged radial profiles with observations. The ratio between the model image and ALMA observations is then used to update the input surface density profile and generate the next model. After 3-4 iterations, the model image generally converges to a good match to observations. In some cases, the inner~10\,au becomes optically thick, and thus increasing surface density alone cannot significantly increase the surface brightness. It suggests that our model underpredicts the mid-plane temperature inside 10\,au, and the real dust scale height may deviate from the general power-law function we used to match SEDs. Given our line observations have a spatial resolution of 15-24\,au, we do not perturb the dust scale height to match the brightness in the inner 10\,au. The surface density profiles and continuum images of best-fitting models are shown in Figure~\ref{fig:sigma} and Figure~\ref{fig:cont_model}, respectively. 

\begin{figure*}[htbp]
\centering
\includegraphics[width=1\textwidth]{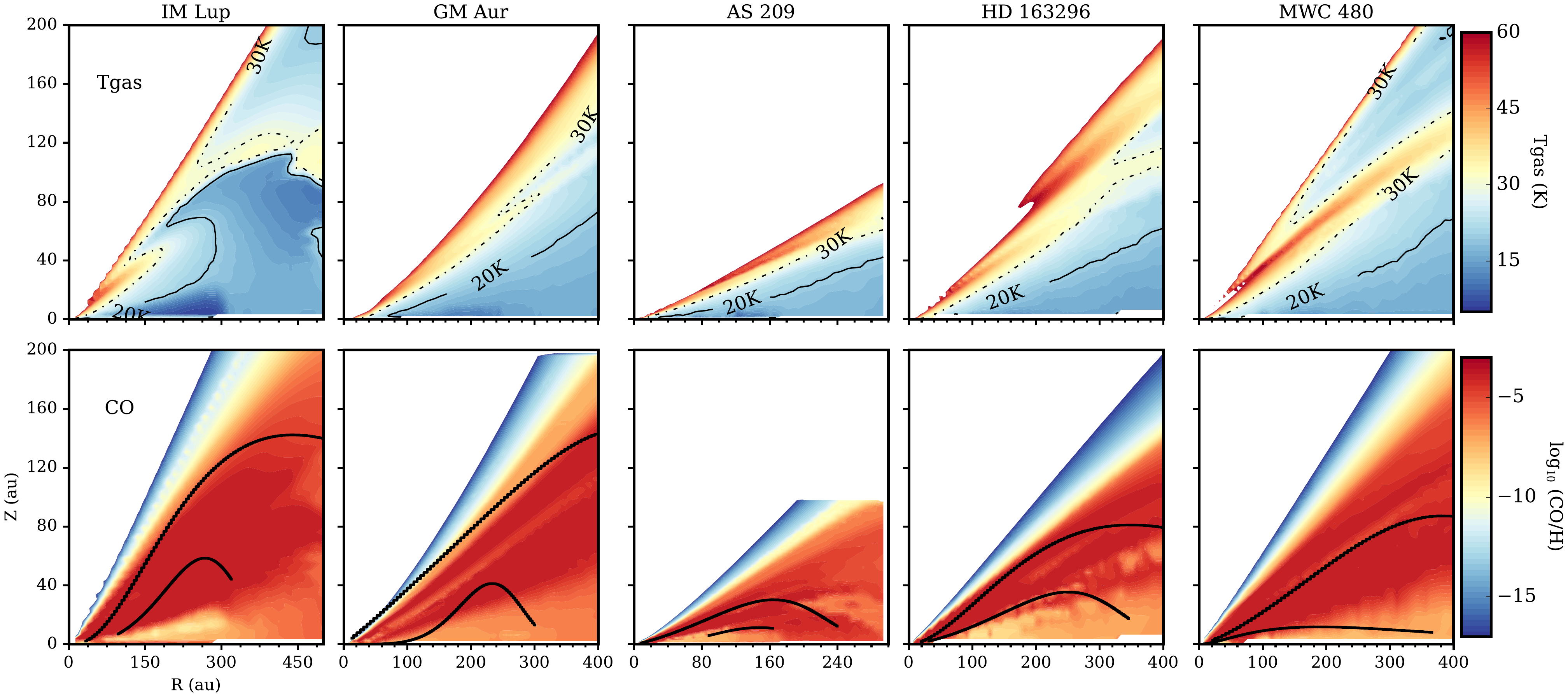}
\caption{Gas temperature (top row) and CO abundance structure (bottom row) of the best-fit thermo-chemical models. The solid black lines are emission surfaces of $^{12}$CO and \ct~(2-1) lines measured from \citet{law20_maps_surface}. CO abundance structures agree with the observed emission surfaces relatively well, except for the MWC 480 disk.   \label{fig:disk_structure}}
\end{figure*}

The third type of observational constraint  is the emission surfaces of optically thick CO and \ct~(2-1) lines. These emission surfaces are the vertical locations where a line has most of its emission at each radius. See Figure~\ref{fig:disk_structure} for examples of emission surfaces. For optically thick lines, emission surfaces are roughly their $\tau\sim$1 layer. With the high spatial resolution of MAPS data, the vertical locations of these surfaces can be directly measured using the asymmetry of the emission relative to the major axis of the disk. Please see \citet{law20_maps_surface} for the exact procedure used for MAPS measurements. 

Here, we use CO and \ct~(2-1) emission surfaces to constrain the vertical locations of the warm molecular layer of CO gas in our thermo-chemical models. Because the CO (2-1) line is highly optically thick, its emission surface is expected to be only slightly below the CO photodissociation layer \citep[e.g.,][]{Pinte18_imlup}. The \ct~(2-1) emission surface is expected to be slightly higher than the CO freeze-out surface. Therefore, these two emission surfaces provide rough upper and lower boundaries of the warm molecular layer of CO gas. Comparing these emission surfaces to the CO abundance structures in thermo-chemical models provides a zeroth-order check on whether the models reasonably match with the observations. For example, the CO freeze-out surface in models should be always below the observed \ct~emission surface.

For a given stellar luminosity, the warm molecular layer of CO is mainly regulated by the scale height distribution and the amount of small grains in the disk.  The amount of small grains affects how deep the stellar light penetrates and thus the T$_{\rm{gas}}\simeq$~20~K layer where CO freezes out. 
We start with  H$_{100}$ and $\psi$ parameters that match the SEDs. The large dust mass distribution is then constrained by mm-continuum images. For the third step, we run five models for each disk in which the mass fraction of the small grains takes 1\%, 5\%, 10\%, 15\%, and 20\% of the total dust mass. We then compare the CO abundance structures of the five models with the observed CO emission surfaces and choose the one that best matches the observed surfaces by eye.

For the IM Lup, GM Aur, and HD 163296 disks, we find solutions that reasonably match both the SED and CO emission surfaces. For the AS 209 disk, we find that the thermo-chemical models with $\psi$ and H$_{100}$ reproducing the observed SED have a CO-freezeout layer much higher than the observed \ct~(2-1) emission surface. Changing the mass of small grains alone cannot solve the problem. We therefore choose to use scale height parameter values that match in CO emission surfaces instead of the SED, because our analysis is mainly based on the CO images. For the MWC 480 disk, our model matches the \ct~(2-1) emission surface but over predict the CO (2-1) emission surface. The best-fitting model parameters are listed in Table~\ref{tab:disk_para}.  Figure~\ref{fig:disk_structure} shows the gas temperature and CO abundance structure in our best-fitting models.

\subsubsection{Deriving CO column density}
\begin{figure}[htbp]
\centering

\includegraphics[width=0.485\textwidth]{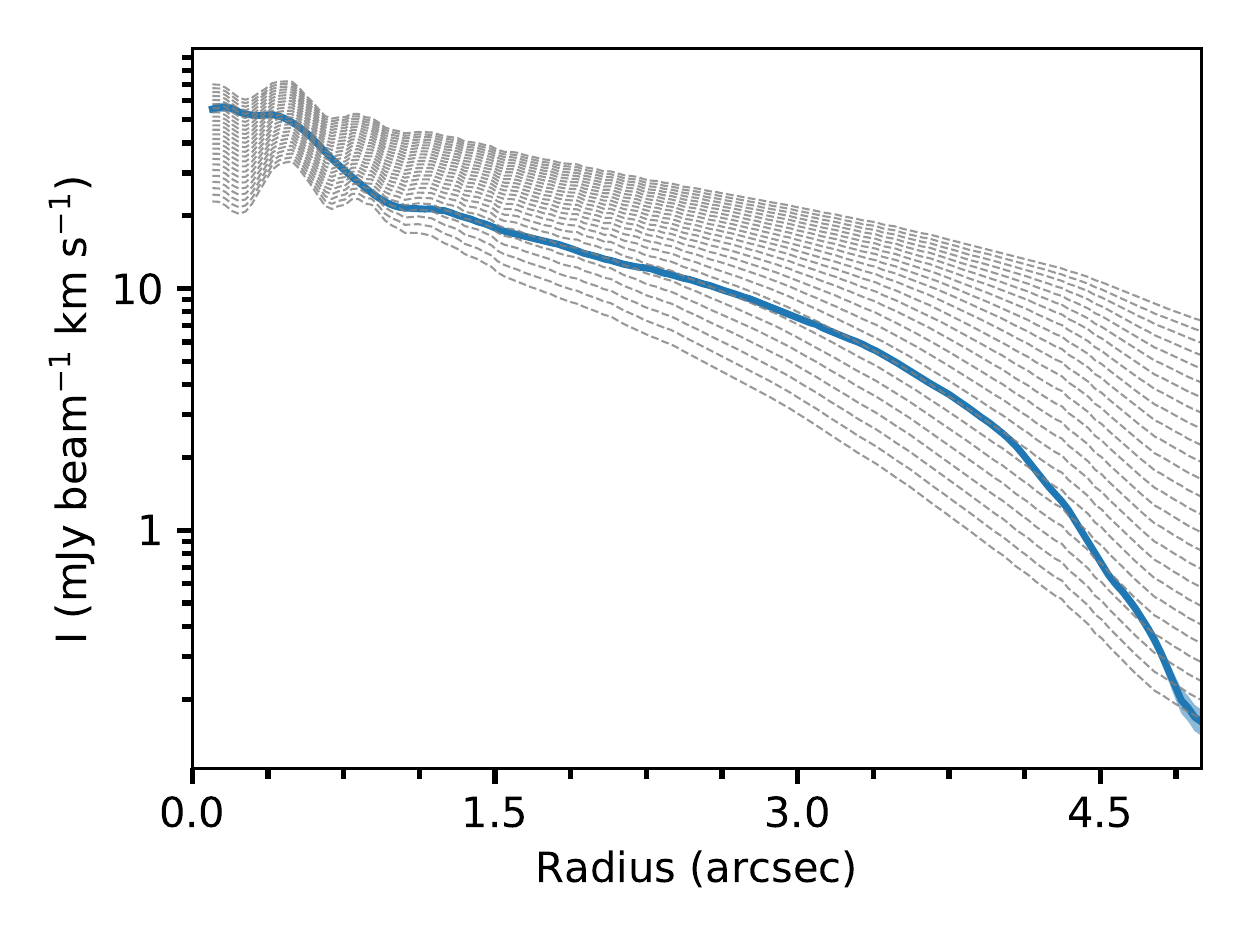}
\vspace{-0.7cm}
\caption{An example of a model grid used to derive the CO depletion profile. The example is for the \ct~(2-1) line of the HD 163296 disk. The grey dashed lines are models with different CO depletion factors, and the blue line is the radial profile from MAPS observations. At each radius,  we adopt the local scaling factor of the  best-fit  model  and  then  create  a  composite 1D depletion profile of the CO gas. \label{fig:co_grid}} 
\end{figure}

For a given disk, we derive the CO column density profiles (\nco) from each CO line. 
Following the approach used in \citet{Zhang19}, we generate a grid of synthetic line images, by scaling the CO gas abundance in each model with a constant scaling factor throughout the disk. The model grid spans a wide range of scaling factor between 0.001 to 50, and the grid increases with a step-size of 1.1 in a logarithmic scale. For a given CO transition and a disk, we compare radial profiles from the model grid with the radial profile from observations. An example of the model grid is shown in Figure~\ref{fig:co_grid}. At each radius, we adopt the local scaling factor from the best-fitting model and then create a composite 1D depletion profile of the CO gas. The depletion profile is then applied to the CO gas column density from our thermo-chemical model. A robustness test of this method is presented in the Appendix~\ref{sec:app:co_retrieval_test}. For a consistency check, we also apply the best-fitting 1D depletion profiles to CO abundance structures in thermo-chemical models and then generate simulated CO line radial profiles. Figure~\ref{fig:co_profile_empi} in the Appendix shows that our best-fitting models match well with the observed profiles.

\subsubsection{CO column density from thermo-chemical models} \label{sec:co_column_thermo}
\begin{figure*}[htbp]
\centering
\includegraphics[width=1\textwidth]{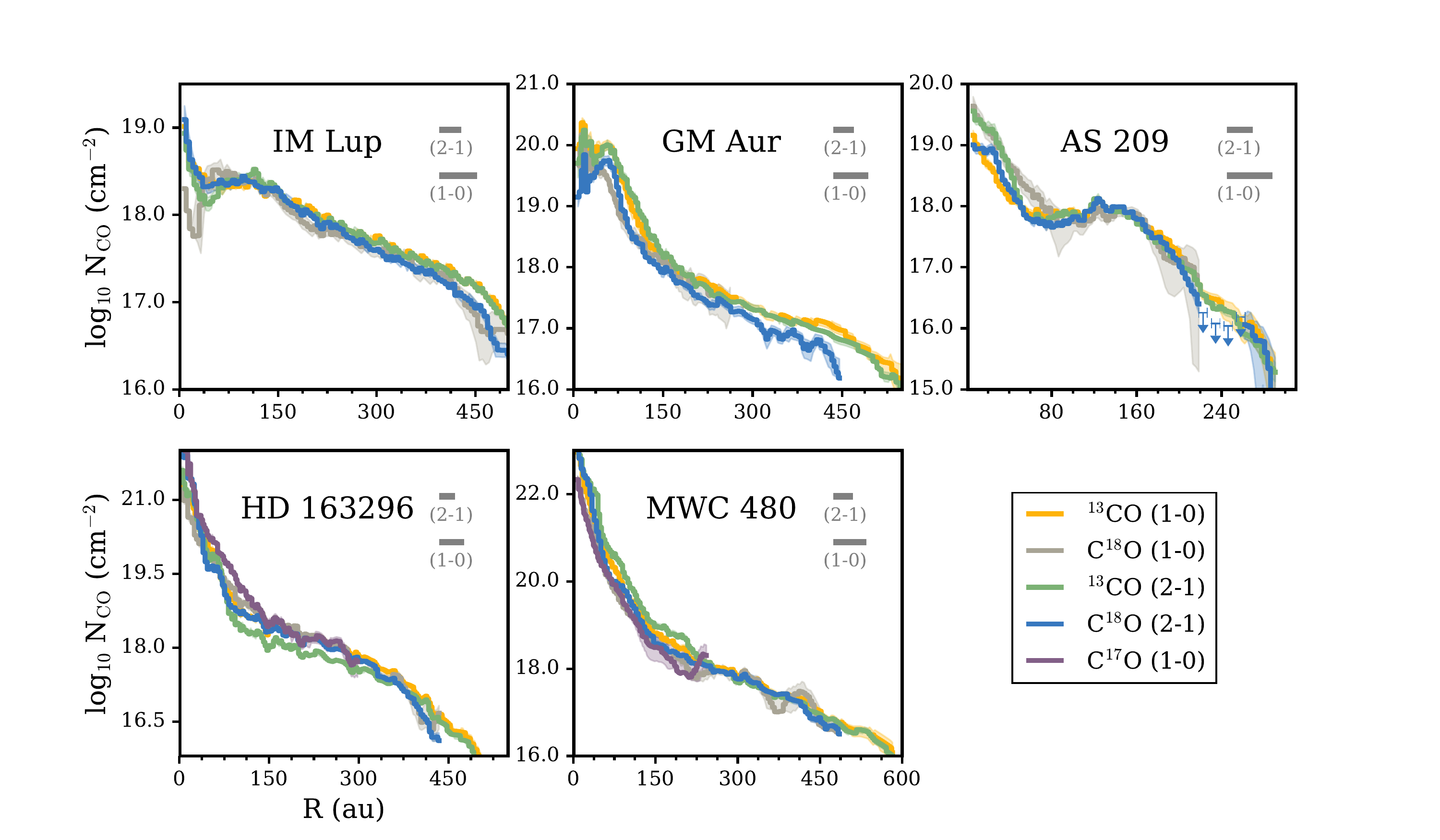}
\vspace{-0.8cm}
\caption{CO column density profiles of five MAPS disks, based on \ct~, \cc~(2-1), (1-0) and \cseven~(1-0) lines. The \nco are derived from thermo-chemical models described in section~\ref{sec:co_column_thermo}. More substructures can be seen in the \nco derived from (2-1) lines than that of (1-0) lines, because (2-1) line images have a factor of two higher resolutions ( 0\farcs15 beams for (2-1) lines vs. 0\farcs3 for (1-0) lines). The beam sizes are plotted at the upper right corner of each panel. \label{fig:Nco_profiles}}
\end{figure*}

In Figure~\ref{fig:Nco_profiles}, we show profiles of CO gas column density (N$_{\rm CO}$) for all five MAPS disks, based on four CO isotopologue transitions, i.e., (2-1) and (1-0) of \cc~ and \ct. For the HD 163296 and MWC 480 disks, we also present \nco profiles derived from the \cseven~(1-0) line, as this line is sufficiently strong in these two disks.  

In general, the \nco profiles are consistent among CO isotopologue lines, within a factor of 2. We also note that the spatial resolution of the (1-0) lines is a factor of two lower than the (2-1) lines, and therefore fewer substructures are seen in the \nco profiles from (1-0) lines. For the HD 163296 disk, the \nco derived~from \cseven~and \cc~lines are consistent outside 100\,au but  differ inside 100\,au by up to a factor of five. We discuss this discrepancy in detail in Section~\ref{sec:co_inside_100}. The derived \nco profiles show some small scale wiggles (10-20\% variation on the scale of 10\,au), which are numerical noise from our thermal chemical models. Here, we focus on larger scale features. 

The \nco profiles from different transitions are in good agreement. Therefore we use the \nco profile from the \cc~(2-1) line for general analysis since this line has a lower optical depth than \ct~lines, and has higher resolution and signal-to-noise ratio than the \cc~(1-0) lines. 

\begin{figure*}[htbp]
\centering
\vspace{-0.8cm}
\includegraphics[width=1\textwidth]{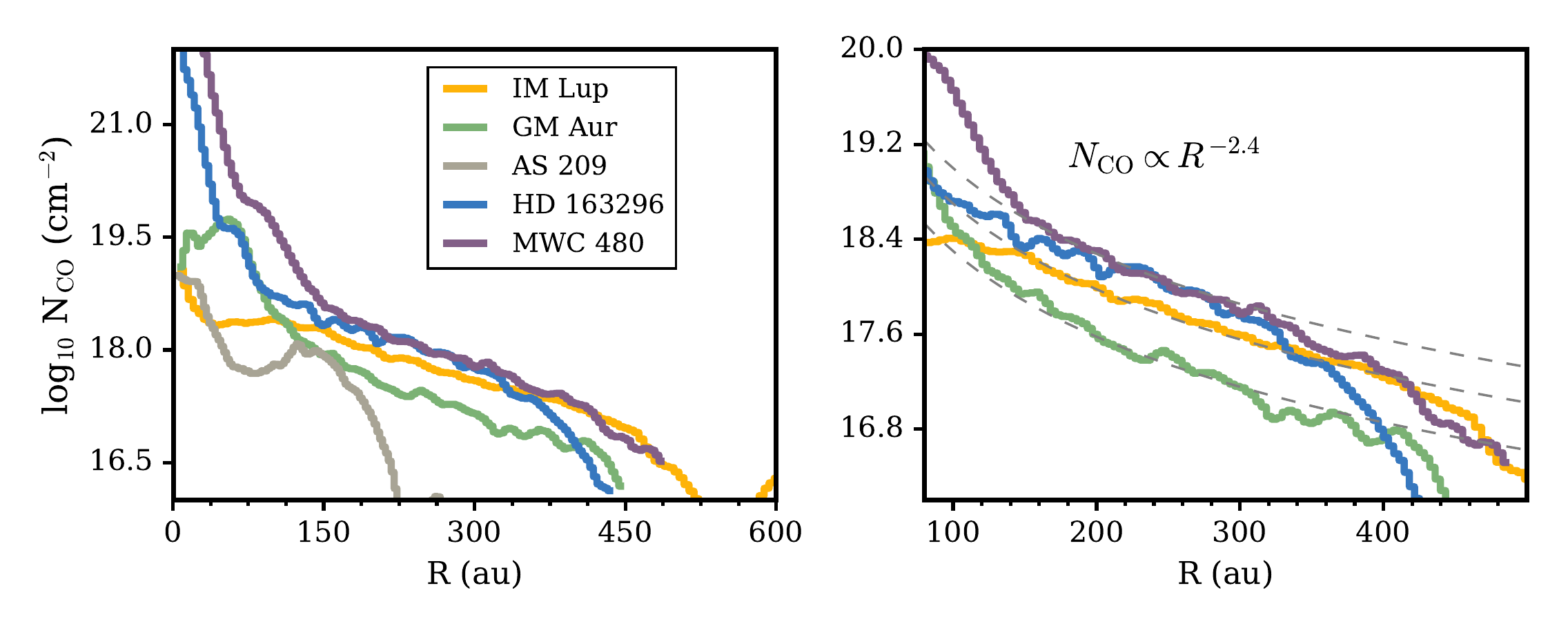}
\vspace{-1.2cm}
\caption{Left: Comparison of CO column density profiles among five MAPS disks. The \nco~here is based on \cc~(2-1) line in thermo-chemical models. Right: The CO column density of four disks between 100 and 500\,au. The dashed lines are power-law functions of $\Sigma(R)\propto~R^{-2.4}$. It shows that the CO profiles of these four disks have a similar slope between 150 and 400\,au. \label{fig:Nco_profiles_comparison}}
\end{figure*}

Among the five MAPS disks, their \nco distributions vary significantly in their general shapes and absolute values (see Figure~\ref{fig:Nco_profiles_comparison}). In the two disks around Herbig stars, the \nco profiles appear to be similar: the \nco shows a steep decrease with radius out to 100\,au, which is followed by a slow decrease out to 400\,au. In the three disks around T Tauri stars, however, the \nco profile varies significantly from one disk to another. For the IM Lup disk, the \nco shows a steep decrease with radius inside 20\,au and a slow decrease between 50-500\,au. The GM Aur disk shows a CO cavity inside 40\,au. Beyond 40\,au, its \nco profile is similar to that of the two disks around Herbig stars. For the AS 209 disk, \nco decreases with radius steeply inside 40\,au, shows a broad shallow gap between 60-120\,au, and a deeper gap around 240\,au. The AS 209 disk has the smallest CO gas disk with a steep decrease between 150-300\,au. Interestingly, the \nco of the remaining four disks show a very similar shallow slope between 150-400\,au, see Figure~\ref{fig:Nco_profiles_comparison}. The slope can be characterized by a power-law function of $\Sigma(R)\propto~R^{-2.4}$. 

Comparing the absolute \nco value of the five disks, the largest differences occur inside 150\,au, where the \nco in the two disks around Herbig stars are 1-3 orders of magnitude higher than that of the three disks around T Tauri stars.  For individual disks, the absolute value of \nco varies by 4-5 orders of magnitude throughout a disk. The IM Lup disk shows the smallest range of \nco changes throughout its disk, while the MWC 480 disk shows the largest variation.

\subsection{CO column density from empirical temperature structure}\label{sec:empirical_temperature}

The \nco profiles derived above depend on gas temperature structures in thermo-chemical models. To test the robustness of these results, we employ a second method to measure \nco using the empirical gas temperature structure measured by \citet{law20_maps_surface}. 

Gas temperature in the disk can be measured from the surface brightness of optically thick line emission provided the emission fills the beam \citep{Weaver18,Pinte18_imlup,Isella18}. Thanks to the high spatial resolution of MAPS data, the emission surfaces of lines can be directly measured (see our discussion in Section 3.1.6), which are a function of radius and vertical height ($r,z$). For optically thick line emission, the gas temperature at these ($r,z$) locations can be directly measured from their line surface brightness. \citet{law20_maps_surface} measured gas temperatures at the emission surfaces of CO and \ct~(2-1) lines. These empirical temperature measurements are for discrete locations in a given disk. To construct the 2D gas temperature structure, \citet{law20_maps_surface} used the two-layer model from \citet{Dullemond20} to fit the discrete temperature measurements.

\begin{figure}[htbp]
\centering
\includegraphics[width=0.52\textwidth]{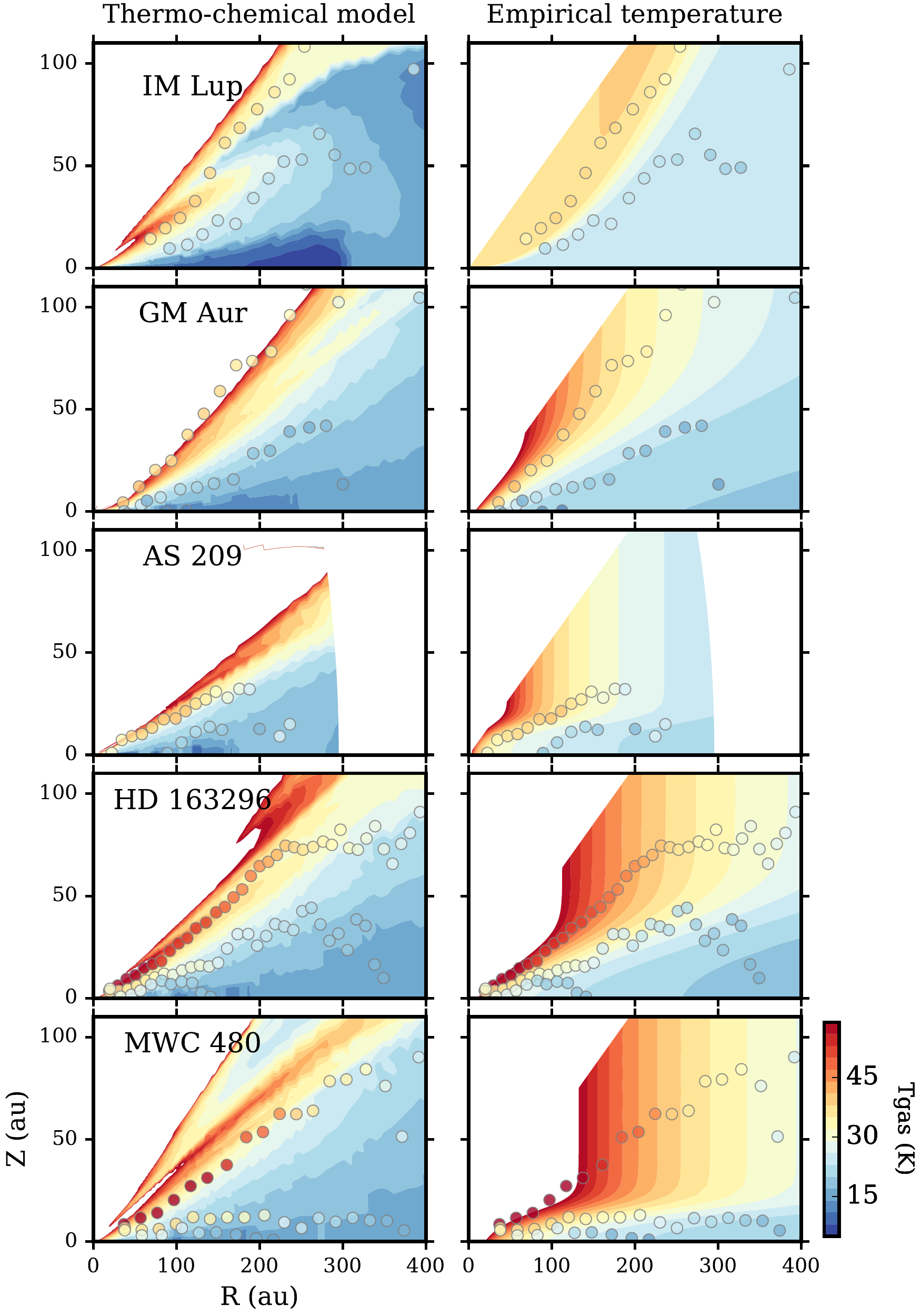}
\caption{Comparison between gas temperature structure of our thermo-chemical models (left) and the empirical temperature structure (right) from \citet{law20_maps_surface}. The circles are surface brightness temperature measurements from CO, $^{13}$CO, and \cc~(2-1) line observations. The emission surfaces of CO is the highest, $^{13}$CO is in the middle, and \cc~at the bottom, because of their relative abundances \citep{Qi11,zhang17}. All panels use the same color scale. \label{fig:tgas_comp}}
\end{figure}

In Figure~\ref{fig:tgas_comp}, we compare the gas temperatures between our thermo-chemical models and empirical temperature structure from \citet{law20_maps_surface}. In general, thermo-chemical models match well with the empirical temperatures in the 20-40\,K region. The temperature structures of IM Lup and AS 209 show the best match between these two temperature models. For the HD 163296 and MWC 480 disks, our thermo-chemical models underpredict the temperature at the CO (2-1) emission surfaces. The largest discrepancy is seen in the MWC 480 disk models: gas temperature in our thermo-chemical model is generally colder by 5-10\,K than the empirical temperature structure.
A caveat of our thermo-chemical models is that the gas temperature structure is calculated with the assumption of no-CO depletion. Because CO is an important coolant in the disk atmosphere, a strong depletion of CO can lead to a warmer atmosphere. \citet{Calahan20b} and \citet{Schwarz20} presented thermo-chemical models with spatially varying C/H elemental ratio for the HD 163296 and GM Aur disks, respectively. They found that the gas temperature becomes warmer at regions above Z/R $>$0.2 when accounting for CO depletion, but no significant changes were seen in the gas temperatures of the $^{13}$CO and \cc~emission surfaces. 

We replace the gas temperature distribution with the empirically-based temperature structure and then use the same approach described in Section~\ref{sec:co_column_thermo} to derive the CO column density. The dust temperature and CO abundance structure are kept the same as the thermo-chemical models. Figure~\ref{fig:co_profile_empi} in the Appendix shows our best-fitting models compared with the observed profiles.

\begin{figure*}[htbp]
\centering
\includegraphics[width=1\textwidth]{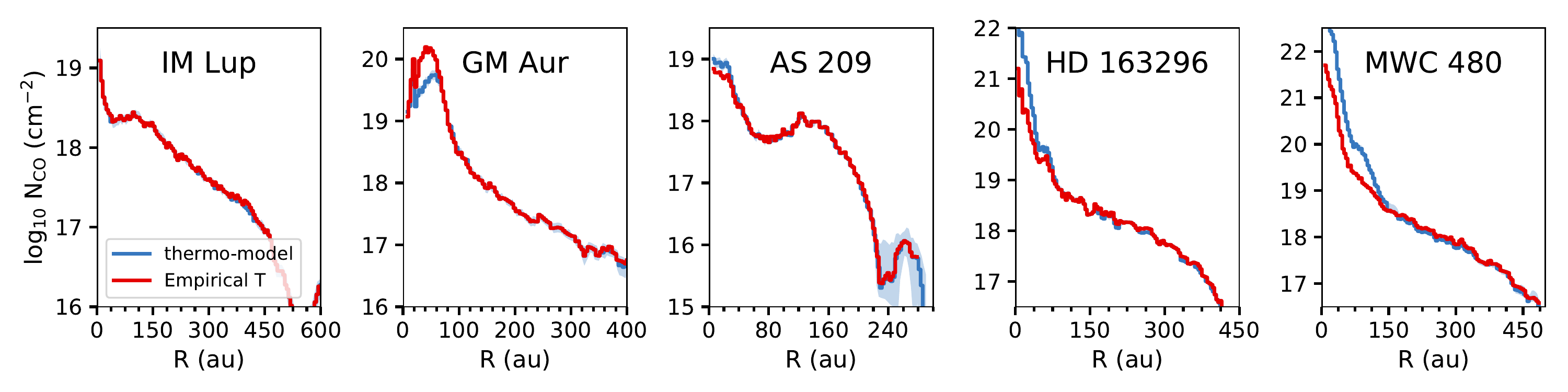}
\caption{Comparison of CO gas column density derived using thermo-chemical models and CO depletion profiles and empirical temperature structures. The column density shown here is based on \cc~(2-1) line. \label{fig:nco_comparison}}
\end{figure*}

Figure~\ref{fig:nco_comparison} shows the \nco profiles derived using the empirical temperature structures, compared with the profiles derived from our thermo-chemical models. The two N$_{\rm CO}$ profiles of each disk generally have excellent agreement for most of the disk regions. For the IM Lup and AS 209 disks, the two N$_{\rm CO}$ profiles are nearly identical. This is expected as these two disks have the best agreements in temperature structures between their thermo models and empirical temperature structures. For the GM Aur, AS 209, and HD 163296 disks, small differences are seen inside 100\,au. The largest difference is in the MWC 480 models, as temperature from the RAC2D model generally underpredicts the gas temperature inside 150\,au. The empirical temperature models also show a steep increase of \nco inside 100\,au in the HD 163296 and MWC 480 disks, although the absolute \nco~is smaller than the one based on the thermo-chemical models. These results indicate that exact \nco derived will depend on the temperature structure inside 100\,au of these disks. Based on current best constraints of the gas temperature structures in the two disks, a steep increase of \nco is needed to reproduce the observed line intensity profiles.

\subsection{CO column density from \ce{C^{17}O} hyperfine line fitting } \label{sec:c17o_col}

The third method we use to derive \nco is fitting \ce{C^{17}O} hyperfine structure transitions. This method is independent of the temperature and CO abundance structures adopted in the thermo-chemical models. 

The \ce{C^{17}O} $J$=1-0 rotational transition is split into three components which can be fitted to obtain constraints on the \ce{C^{17}O} column density, excitation temperature, and optical depth \citep{Mangum15}. This method has been used recently in disk studies of \ce{C_2H}, \ce{HCN}, and \ce{CN} \citep{Bergner2019, 2020ApJ...899..157T, bergner20, Guzman20}.  We present the first application of this method to \ce{C^{17}O} in protoplanetary disks. For our sample, only the \ce{C^{17}O} (1-0) images of the HD 163296 and MWC 480 disks have sufficient signal-to-noise ratio for this analysis.

The three components of \ce{C^{17}O} (1-0) line have rest frequencies of 
112.358777, 112.358982 and 112.360007~GHz \citep{Klapper_2003}, as listed in Table~2 of \citet{oberg20_maps}. The 0.5~\kms{} spectral resolution of the MAPS Band 3 data means that 2 of the components are always blended but the third component is $\sim$2\,\kms{} offset from the other two components,  which is enough for us to model and fit the individual components. 

Using the spectral shifting and stacking tool in \texttt{Gofish}, we generate spectra averaged over annuli across the disk. This technique removes most of the Keplerian broadening. This technique also increases the effective signal-to-noise of the data compared to a single-pixel spectrum. The HD~163296 and MWC~480 spectra were averaged over annuli 1/2~$\times$ the beam major axis resulting in radial bins of width 15 and 24~au, respectively. 

Following the same method outlined in \citet{bergner20}, we generate a model spectra for the \ce{C^{17}O} (1-0) hyperfine transitions
to fit the observed spectra. We assume LTE and that all three hyper-fine transitions have the same excitation temperature. The free parameters in the fit are: the total column density of \cseven~($N_T$, $\mathrm{cm^{-2}}$), the excitation temperature ($T_{ex}$, K), the observed line width ($\sigma_{\mathrm{obs}}$, \kms), and systematic velocity (V$_{lsrk}$, \kms). The $\sigma_{\mathrm{obs}}$ parameter is larger than the local thermal+turbulence broadening, likely due to a combination of (1) emission contribution from the back and front sides of a flared disk, which have different projected velocities, (2) beam smearing, in which a wide range of Keplerian velocities are incorporated within the same beam, (3) Hanning smoothing of the data which effectively reduces the spectral resolution \citep{bergner20}.

The bounds for the column density and excitation temperature were set to $10^{10} - 10^{22}$ $\mathrm{cm^{-2}}$ and $10 - 50$~K.
This encompasses the range of temperature for the \ce{C^{17}O} emitting region (see Section~\ref{sec:empirical_temperature}). 
$\sigma_{\mathrm{obs}}$ is set to range from 0.01 to 8~\kms~and the V$_{lsrk}$ is between 0 and 10~\kms.
The total line optical depth is the sum of the individual optical depths at the line center of each of the three transitions. 
This includes a correction factor to account for beam smearing 
that will artificially lower the inferred optical depth. 
For full details of the fitting procedure, see \citet{bergner20}. 

\begin{figure}[h!]
    \centering
    \vspace{0.3cm}
    \includegraphics[width=0.7\hsize]{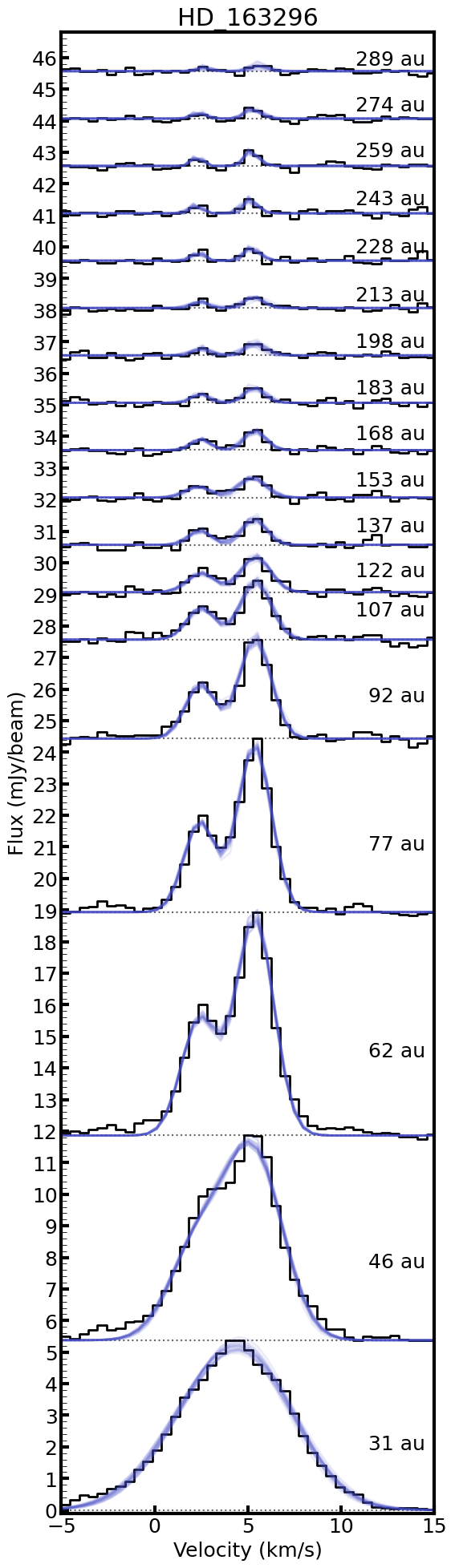}
    \caption{Radially averaged \ce{C^{17}O} (1-0) spectra (black) and the results from the hyperfine line fitting (blue) for the HD~163296 disk. The center radius of each bin is shown and the spectra are offset along the axis for clarity.}
    \label{fig:c17o_hd163296_spectra}
\end{figure}

We use the MCMC package \texttt{emcee} \citep{emcee} to sample the posterior distributions of the four fit parameters.
The radially-averaged spectra from \texttt{Gofish} and the model fits for HD~163296 are shown in Figure~\ref{fig:c17o_hd163296_spectra}. 
The same plot for the MWC~480 disk and the fits for N(\ce{C^{17}O}), $T_{ex}$, and $\tau$ for both disks are shown in the Appendix in Figure~\ref{fig:c17o_fits}.
The \ce{C^{17}O} line becomes optically thick ($\tau > 1$) within $\approx$55~au in the HD~163296 disk and $\approx$100~au in the MWC~480 disk. The $T_{ex}$ is poorly constrained as all of the transitions have the same upper-energy level but the derived column density is not sensitive to the assumed $T_{ex}$. We tested fixing the $T_{ex}$ to 20, 25, 30, 35 and 40~K for the HD~163296 disk and in the region where the line is optically thin, the column densities were well within the fit uncertainties. 

The N(\ce{C^{17}O}) measured from the hyperfine-line fitting method has an inner and outer  limits of the radius. Inside the inner 1.5$\times$ beams, the fit is poor due to the complete blending of the lines and beam smearing. Beyond $\sim$250~au the spectra are too noisy to obtain robust fits in the HD~163296 and MWC~480 disks.

The resulting N(\ce{C^{17}O}) radial profiles are then converted to total N(CO) radial profiles under the assumption 
that \ce{^{16}O}/\ce{^{18}O}$= 557$ and \ce{^{18}O}/\ce{^{17}O}$= 3.6$ \citep{wilson99}.

\begin{figure*}
    \centering
    \includegraphics[width=1.0\hsize]{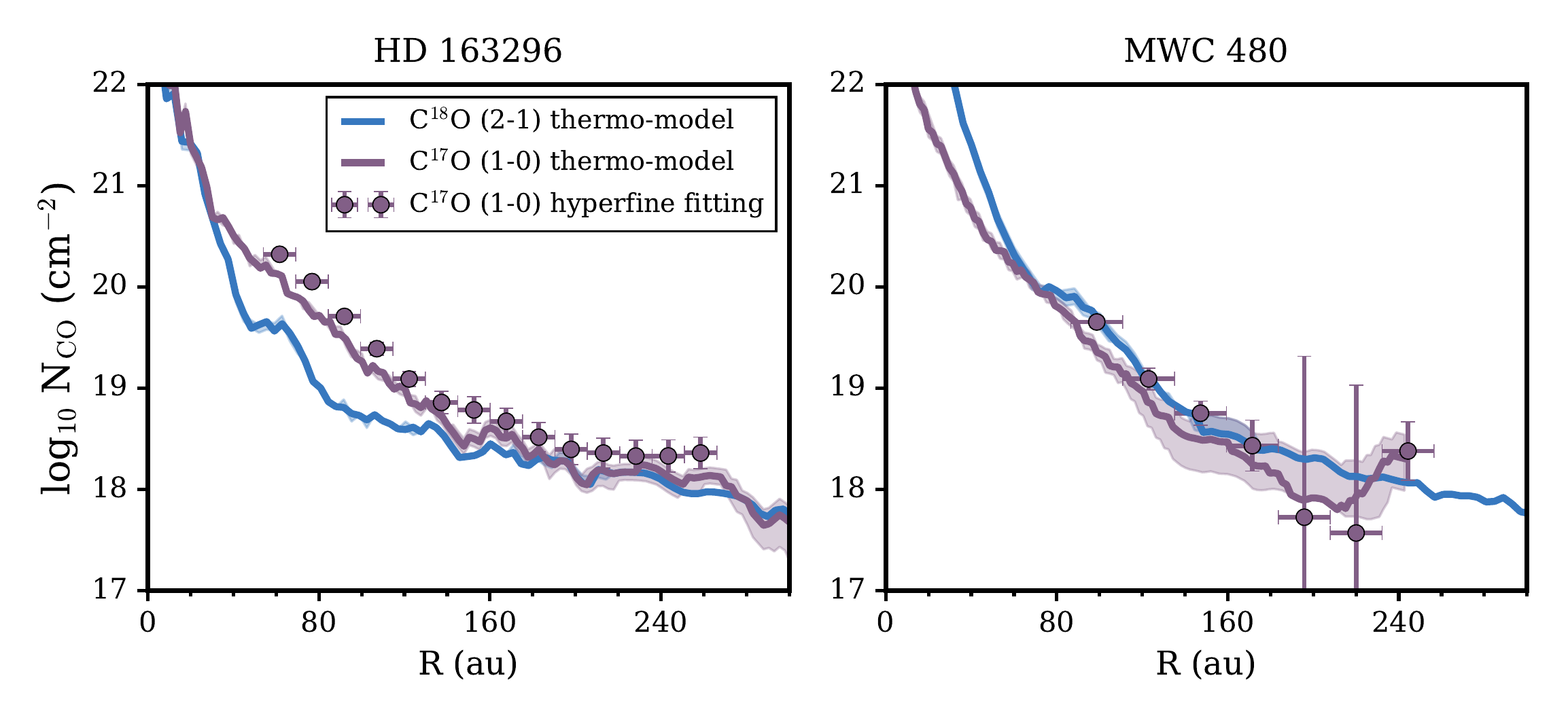}
    \vspace{-0.7cm}
    \caption{Comparison of CO column densities derived from hyperfine line models with that from thermo-chemical models of the HD~163296 and MWC~480 disks.  The blue and purple lines are \nco profiles from the \cc\,(2-1) and \cseven\,(1-0) lines, respectively, in thermo-chemical models (Section~\ref{sec:co_column_thermo}). The purple colored markers are the best-fitting results for the hyperfine lines of \cseven~ (1-0) (Section \ref{sec:c17o_col}). The x-scale error bars denote the size of the radial bin that the spectra are averaged over. In the results from hyperfine line fitting, we do not show points from the inner 1.5$\times$ the beam major axis, where the lines are too blended, or points from the outer disk, where the signal-to-noise of the spectra is too low.}
    \label{nc17o}
\end{figure*}

Figure~\ref{nc17o} presents the \nco profiles derived from the \ce{C^{17}O} fitting compared with the profiles derived in Section~\ref{sec:thermo-chemical models} for HD~163296 and MWC~480 from the \ce{C^{18}O} (2-1) and \ce{C^{17}O} (1-0) lines. 
In general, the profiles are consistent with each other, i.e., within a factor of a 2-3. 
For the HD 163296 disk, the \nco derived from \cseven~(1-0) using both methods suggests a higher CO column density between 30-150\,au than that of \cc~(2-1) line. See detailed discussions on the difference in the following section.

Overall, we have shown that with sufficiently high signal-to-noise data, fits to the \ce{C^{17}O} (1-0) hyperfine transitions can provide good constraints on the CO column density.

\subsection{CO column density inside 100\,au of HD 163296 and MWC 480} \label{sec:co_inside_100}

\begin{figure*}
    \centering
    \includegraphics[width=0.7\hsize]{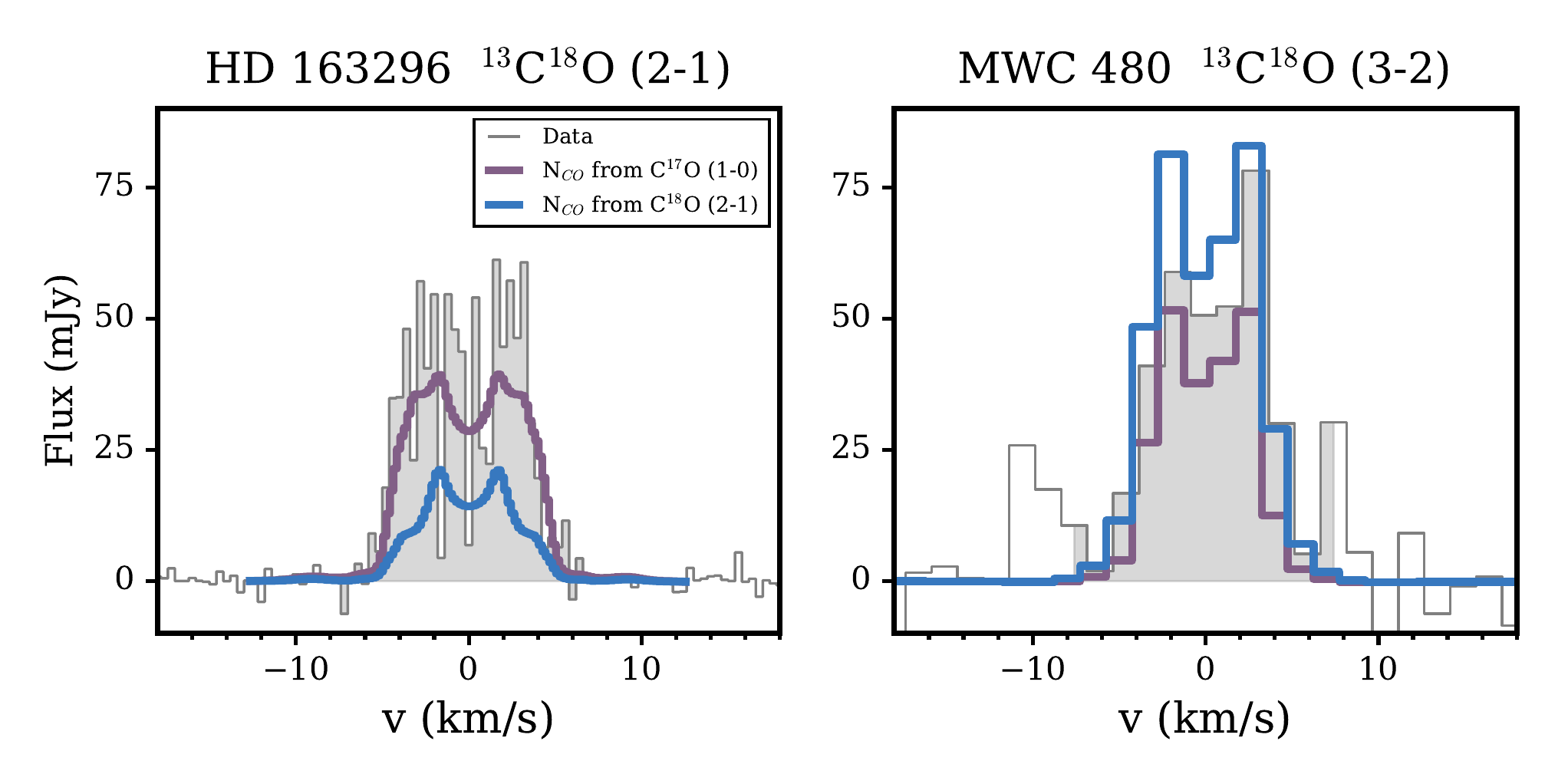}
    \caption{Comparison of $^{13}$C$^{18}$O spectra (in grey) with models based on \nco from \cc\,(2-1) and \cseven\,(1-0) lines in thermo-chemical models (Section~3.1). }
    \label{fig:13c18o}
\end{figure*}

 Inside 100\,au, the HD 163296 and MWC 480 disks have a CO column density that is 1-2 orders of magnitude higher than that of the other three disks around T Tauri stars.  The \cc~(2-1) line emission interior to $\sim$100\,au in these two disks is optically thick. In these regions, the \nco derived from \cc~thus relies on the robustness of the vertical temperature and CO abundance in models. 
The \nco derived from the \cseven~(1-0) line of HD 163296 shows a 2-6 times higher column density inside $\sim$100\,au, while the \nco inferred from the \cseven~and \cc~lines match well in the MWC 480 disk. Both disks have previous detections of rarer CO isotopologue lines, including $^{13}$C$^{18}$O and  $^{13}$C$^{17}$O lines \citep{Booth19,Zhang20_hd163,Loomis20_mwc480}, which provides constraints closer to the midplane in the two disks. Here we use the disk-integrated $^{13}$C$^{18}$O line spectra to test the robustness of the \nco results inside 100\,au for the HD 163296 and MWC 480 disks. We apply the CO depletion profiles from the \cc~and \cseven~lines to the CO abundance from our thermo-chemical models and generate simulated spectra for $^{13}$C$^{18}$O (2-1) or (3-2). The results are shown in Figure~\ref{fig:13c18o}.

For HD 163296, the \nco from \cseven~(1-0) line has a better match to the $^{13}$C$^{18}$O (2-1) line spectrum. The high \nco is consistent with previous results from   \citet{Zhang20_hd163}, who found that the CO abundance inside 70\,au in the HD 163296 disk is a factor of few larger than that in the outer disk region. As the $^{13}$C$^{18}$O~and \cc~(2-1) lines are at similar frequencies, the difference of \nco cannot be explained by optically thick dust. One possible explanation is that the CO-to-H$_2$ gas abundance ratio close to the mid-plane may be a factor of few higher than that of the disk atmosphere inside 70\,au. This enhanced CO abundance at the mid-plane is possible if a large amount of icy pebbles evaporate their CO ice mantle at the mid-plane \citep{Krijt18,Krijt20}. For the MWC 480 disk, the \nco discrepancy inside 100\,au is smaller and both results match the $^{13}$C$^{18}$O~(3-2) line within uncertainties. Higher spatial resolution observations of $^{13}$C$^{18}$O are necessary to better constrain the \nco inside 100\,au of these two disks.

\section{Discussion} \label{sec:discussion}

\subsection{Minimum gas disk masses}\label{sec:minimum_mass}
Despite the large uncertainty of CO abundance in disks, the amount of CO gas in disks provides a lower limit of total gas mass. Here we estimate minimum gas disk masses in two ways: (1) we calculate the total CO gas mass for each disk, and then scale it with a constant CO-to-H$_2$ abundance ratio of 10$^{-4}$ to get a total disk gas mass, $M_{gas1}$. This is a lower limit of disk gas mass, as it does not include gas mass in the CO freeze-out and photodissociation regions. (2) We scale the gas surface density in our disk models with the CO depletion profile, and calculate a total disk gas mass, $M_{gas2}$. This approach assumes that the difference between thermo-chemical models and observations is caused by gas depletion and thus our CO depletion factor is a gas depletion factor.  In this way, the gas masses are corrected for the CO freeze-out and photodissociation effects, but the result depends on the temperature and CO abundance structure in models. For the mass estimation, we use the CO depletion profiles based on the \cc~(2-1) line from  empirical temperature models (Section~\ref{sec:empirical_temperature}). The only exception is that for the inner 150\,au of the HD 163296 disk, we adopt the \nco from \cseven~(1-0) line, due to the discrepancy discussed in Section~\ref{sec:co_inside_100}. We note that our initial chemical condition in thermo-chemical models has a CO-to-H$_2$ abundance of 2.8$\times10^{-4}$ \citep{Lacy94}, and thus the undepleted CO abundance in warm molecular layer is between 1-2.8$\times10^{-4}$, slightly higher than the canonic assumption of 10$^{-4}$.  Due to the difference in assumed CO-to-H$_2$ abundance ratios, even $M_{gas2}$ includes corrections of freeze-out and photodissociation, in the IM Lup and MWC 480 disks, the $M_{gas2}$ are still slightly smaller or the same as $M_{gas1}$.  

The estimations of minimum gas disk masses are summarized in Table~\ref{tab:mini_mass}. The total dust masses are adopted from our best-fit thermo-chemical models (see Table~\ref{tab:disk_para}). These five disks have a lower limit of gas-to-dust ratio between 1 and 24, demonstrating that even if no CO depletion is assumed, these disks still have a gas-to-dust ratio $\ge$1. Based on the CO derived minimum gas masses, the two disks around Herbig stars have a higher gas-to-dust mass ratio than those around T Tauri stars. This is consistent with the expectation that disks around Herbig stars would have less depletion of CO, due to their warmer disk conditions \citep{Bosman18,Schwarz18,Schwarz19a}. Interestingly, the IM Lup disk has the lowest CO gas-to-dust mass ratio, despite the fact that it is the youngest disk in the sample. 

\begin{figure*}[htbp]
\centering
\includegraphics[width=1.0\textwidth]{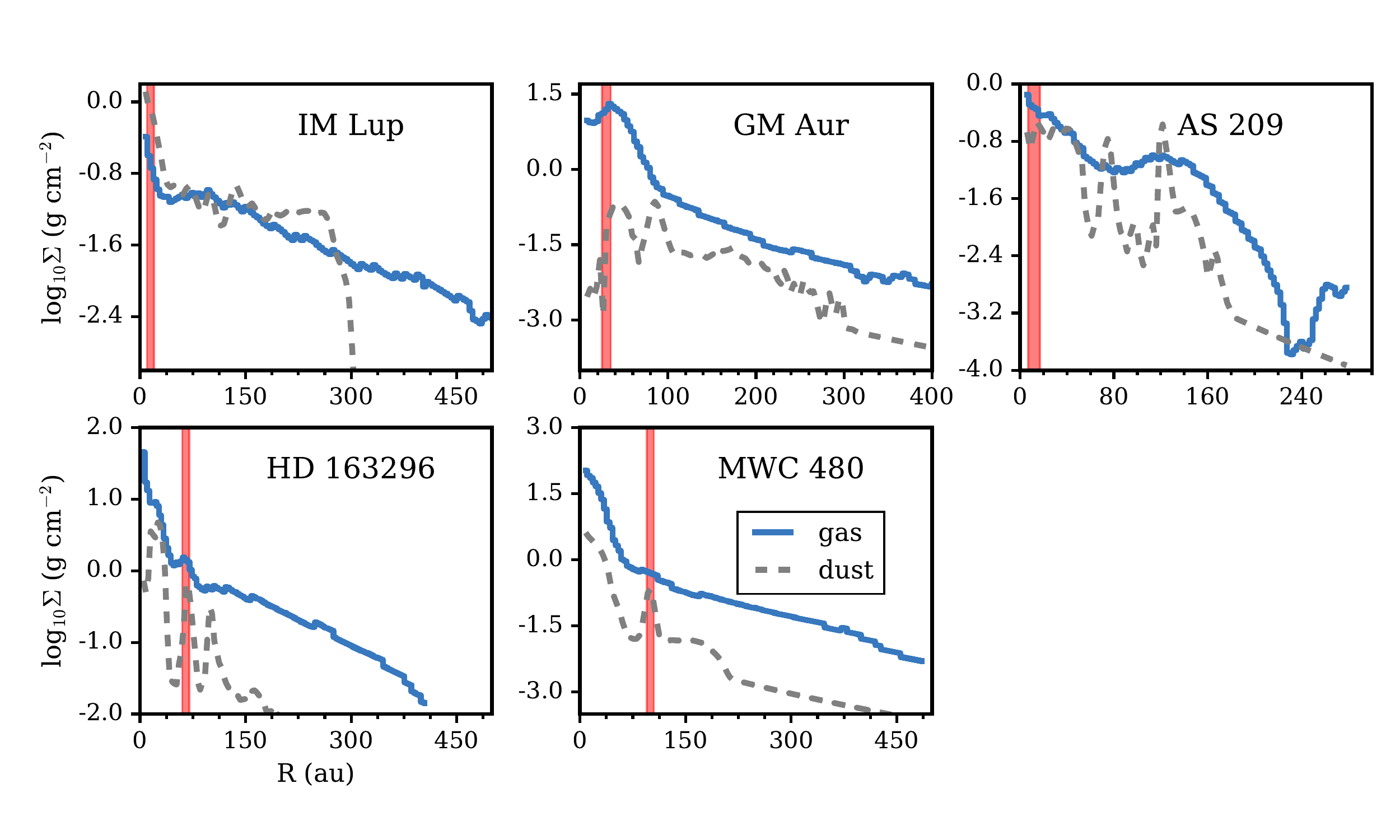}
\caption{Radial profiles of gas and dust in MAPS disks. The gas surface density is derived by scaling gas surface density of thermo-chemical models with CO depletion profiles (see details in method 2 in Section~\ref{sec:minimum_mass}). The dust surface density is from the best-fit thermo-chemical models. The red shaded region indicates the mid-plane CO snowline locations in our thermo-chemical models. \label{fig:Nco_vs_cont}} 
\end{figure*}

In Figure~\ref{fig:Nco_vs_cont}, we compare the radial distributions of CO derived gas mass (method 2 above) with dust mass in our models. It shows that if the CO intensity variation is due to gas depletion, the IM Lup disk would have a gas-to-dust ratio $\le$1 inside 200\,au. Analyses of other molecules in the IM Lup disk showed it is unlikely the case, based on the observations of N-carriers \citep{Cleeves18}. Therefore, the IM Lup disk likely has a low CO abundance rather than a gas-to-dust ratio of 1 in the disk.   

\begin{deluxetable*}{lcccccc}
\tablecaption{Minimum disk mass and snowline location\label{tab:mini_mass}}
\tablehead{
\colhead{Source} & \colhead{Mgas1} &\colhead{Mgas2} & \colhead{g2d1} &\colhead{g2d2} &\colhead{r$_{\rm CO}$} &\colhead{r$_{\rm N_2}$}\\
\colhead{} & \colhead{(10$^{-3}~{\rm M_\odot)}$} & \colhead{(10$^{-3}~{\rm M_{\odot}}$)} &\colhead{} &\colhead{}&\colhead{(au)}&\colhead{(au)}
}
\colnumbers
\startdata
IM Lup       &    1.9 &    1.9 &         1 &         1 &     15$\pm$5 &25$\pm$5\\
GM Aur       &    8.0 &   19.9 &        12 &        29 &     30$\pm$5  &70$\pm$5\\
AS 209       &    0.4 &    1.0 &         1 &         2 &     12$\pm$5 &20$\pm$5\\
HD 163296    &   31.3 &   42.7 &        14 &        19 &     65$\pm$5 &90$\pm$5\\
MWC 480      &   35.9 &   28.6 &        25 &        20 &    100$\pm$5 &150$\pm$5\\
\enddata

\tablecomments{(1) Source name (2) $M_{gas1}$: the minimum disk gas masses by scaling the total CO gas mass with n$_{\rm CO}$/n$_{\rm H_2}$ = 10$^{-4}$. (3) $M_{gas2}$: the gas disk mass by scaling the gas surface density at each radius by CO depletion profiles derived in Section~\ref{sec:empirical_temperature}.
(4) Minimum gas-to-dust mass ratio based on $M_{gas1}$/$M_{dust}$. (5) $M_{gas2}$/$M_{dust}$.(6) mid-plane CO snowline location in the our thermo-chemical model. (7) mid-plane N$_2$ snowline location.}
\end{deluxetable*}

\subsection{Radial dependence of CO depletion in MAPS disks}
\begin{figure*}
    \centering
    \includegraphics[width=1.0\hsize]{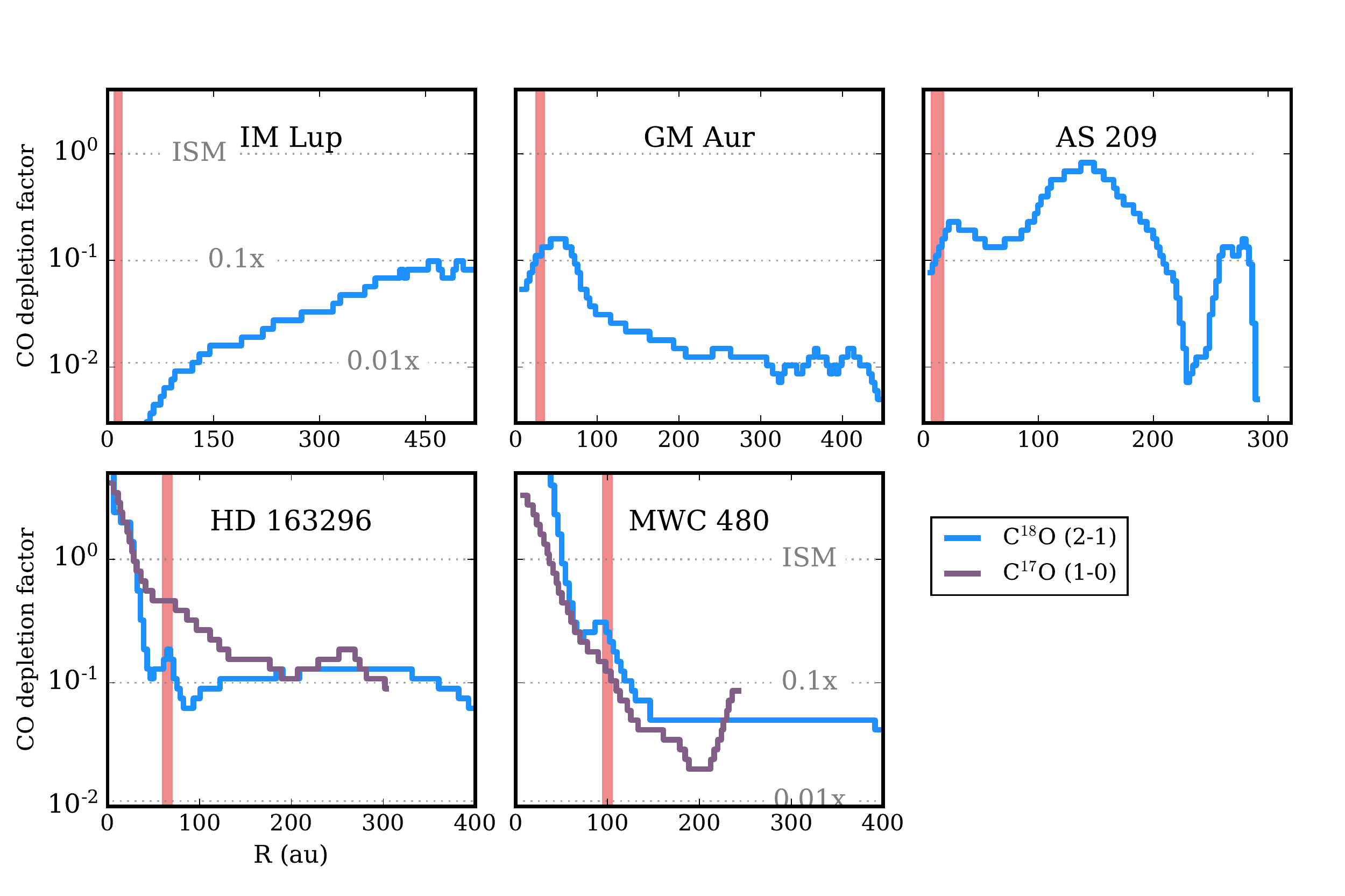}
    \caption{CO depletion profiles derived from \cc~(2-1) and \cseven~(1-0) line observations. We show both depletion profiles derived from thermo-chemical models (solid line, Section~\ref{sec:thermo-chemical models}).  The red shaded region indicates the mid-plane CO snowline location in our thermo-chemical models. The depletion profiles show that the CO abundance is depleted by a factor of 10-100 in the three disks around T Tauri stars. For the two disks around Herbig stars, the CO abundances are depleted by a factor of $\sim$10 outside 150\,au, but enhanced to the ISM level inside their mid-plane CO snowlines.  }
    \label{fig:co_dep_profiles}
\end{figure*}

In Figure~\ref{fig:co_dep_profiles}, we show the CO depletion profiles derived from thermo-chemical models.
Compared to our thermo-chemical models, observations require 1-2 orders of magnitude lower CO abundance in most of the disk region. This is consistent with previous findings that the disk-averaged CO gas may be heavily depleted in some gas-rich Class II disks \citep{schwarz16,mcclure16,zhang17,Zhang19}. 

Various chemical and physical processes can reduce CO gas abundance in the disk atmosphere. Chemical processes can reprocess CO into other carbon-carriers \citep{aikawa99_abundance, bergin14,Eistrup16,yu16,Schwarz18,Bosman18}. Additionally, dust growth and settling can sequester CO into the mid-plane and the radial drift of ice pebbles can bring CO into the inner disk region \citep{xu17,Krijt18,Booth19_pebble,Krijt20}. These processes do not occur at the same rate at each radius, and thus the radial and vertical variation of CO abundance in disks can provide useful constraints on the CO depletion processes in disks. 

\citet{Zhang19} analyzed the radial variation of CO depletion in five disks using \cc~(2-1) or (3-2) line observations. They found that the CO gas abundance is heavily depleted in the region just beyond the mid-plane CO snowline and the CO abundance tends to be higher at the outermost region of the gas disk. They noted that the CO abundance inside the CO snowline of the HD 163296 disk is one order of magnitude higher than the region outside, which is further confirmed by a follow-up study of $^{13}$C$^{18}$O (2-1) \citep{Zhang20_hd163}. 

The work here provides higher resolution constraints on the CO depletion profiles than previous observational studies.  
In Figure~\ref{fig:co_dep_profiles}, we show that the two disks around Herbig stars (HD 163296 and MWC 480) exhibit a very similar depletion pattern: outside the mid-plane CO snowline, CO is depleted by a factor of ten or more; inside the CO snowline, CO gas abundance increases rapidly. This pattern is consistent with the dust evolution picture: CO is sequestered to the mid-plane as icy dust grains settle, and icy pebbles drift inwards into the region inside the mid-plane CO snowline, where CO ice evaporates and enhances the CO abundance at the inner disk regions \citep{booth17,Booth19_pebble, Krijt18,Krijt20}. 

The three disks around T Tauri stars, however, do not show a clear trend. For the IM Lup disk, its CO abundance increases monotonically with radius. For the GM Aur disk, beyond its 40\,au cavity, its CO abundance has a sharp decrease out to 80\,au and then a slow decrease out to the edge of the disk. This profile is similar to that of the two disks around Herbig stars, but 80\,au is much further out than the mid-plane CO snowline of 30\,au in our chemical model of the GM Aur disk.  The AS 209 disk has two local peaks around 30 and 150\,au. In our models, the mid-plane CO snowlines of these T Tauri sources are comparable to or slightly smaller than our spatial resolution. But we do not see signals of CO enrichment like that in the two disks around Herbig stars.  

\subsection{CO column density distribution vs. mm-sized dust \label{sec:co_vs_cont}}

\subsubsection{CO gap properties}
\begin{figure*}[htbp]
\includegraphics[width=1\textwidth]{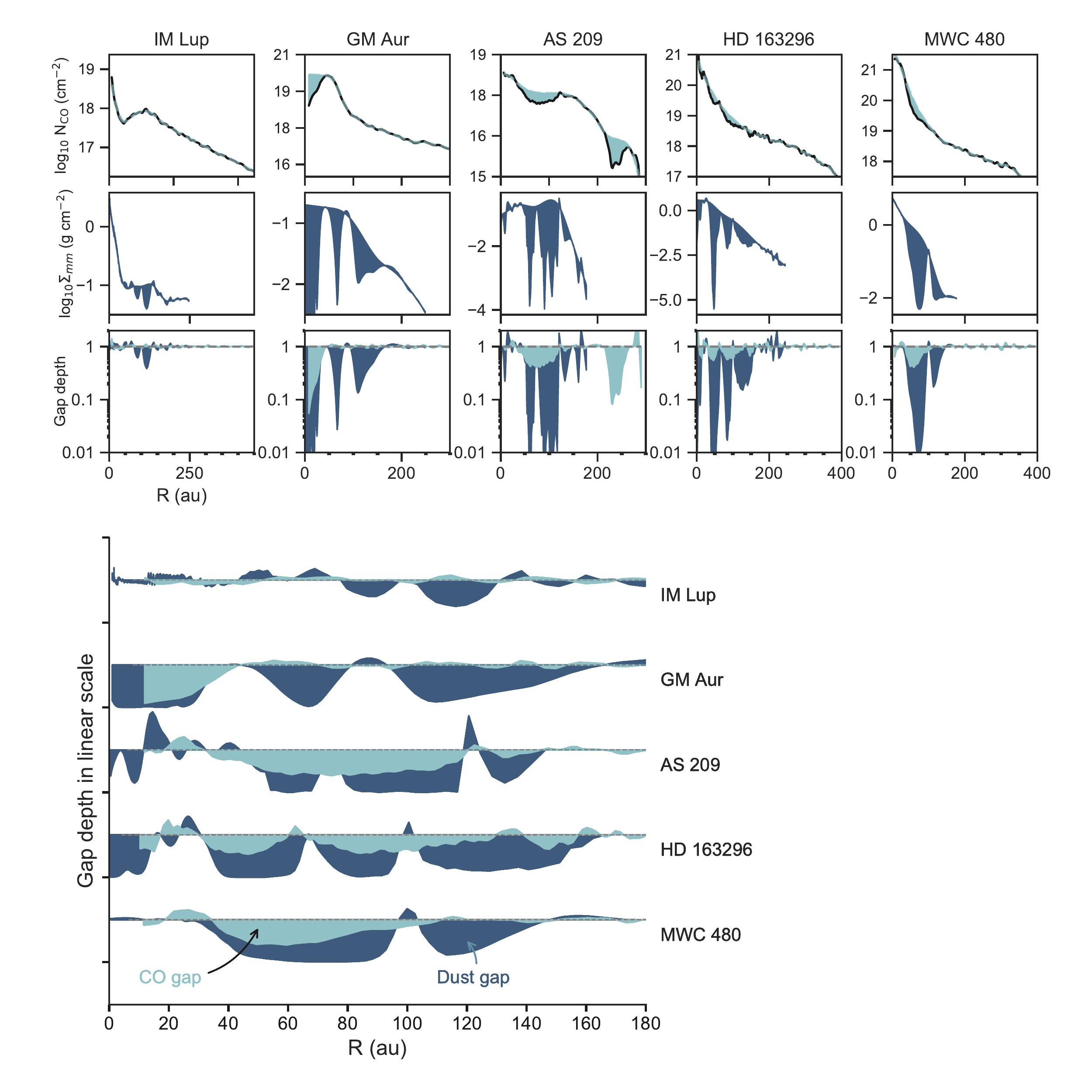}
\vspace{-0.8cm}
\caption{Comparison of CO gas and dust gaps. Top plot: row1, CO column density distribution (solid line). The regions filled in light blue color are CO gaps identified. row 2: mm-dust distributions. The regions filled in blue are dust gaps identified. row 3: Comparison between gas and dust gaps in logarithmic scale. The bottom plot: Cross-comparison of the locations and depths of the CO gas and dust gaps among the five MAPS disks (depth is shown in linear scale).  \label{fig:co_gaps}}
\end{figure*}

\begin{figure*}[htbp]
\centering
\includegraphics[width=1\textwidth]{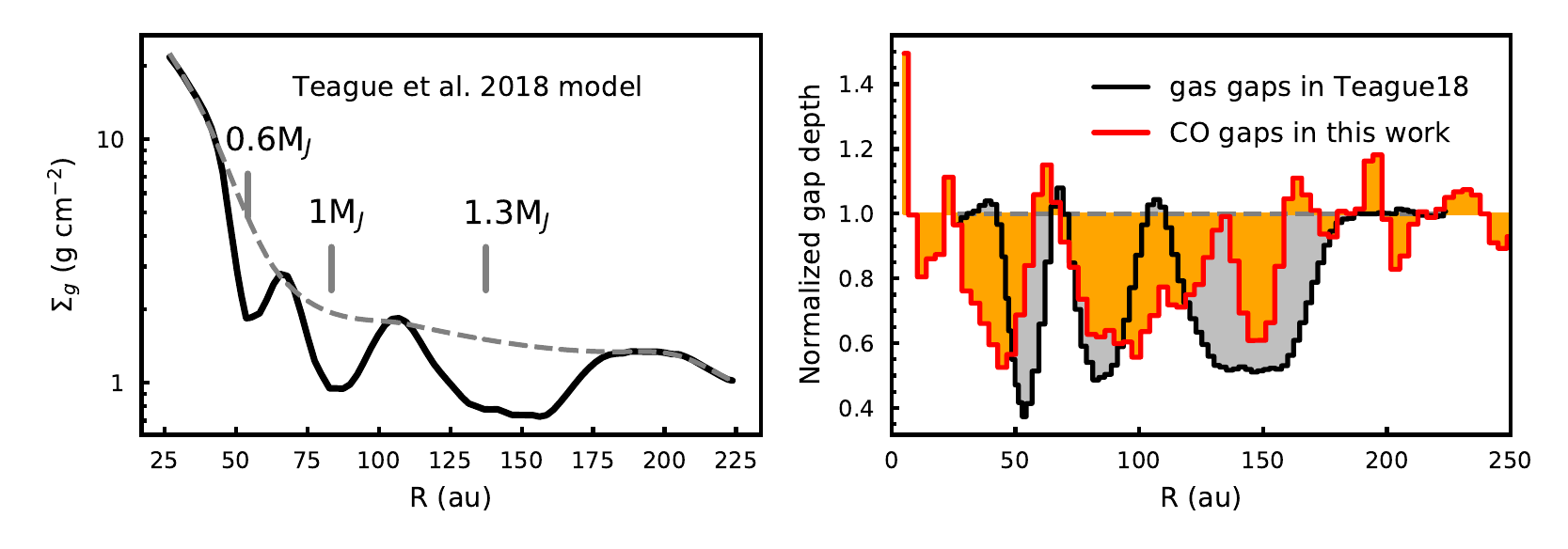}

\caption{CO gaps in the HD 163296 disk. Left: \citet{Teague18a} gas distribution model that reproduces the velocity deviations from a Keplerian rotation field in \cc~(2-1) line observations. The dashed line is the smooth function used to identify gap properties. Right: Comparison of gas gaps in \citet{Teague18a} with the CO gaps found in this work.  }  \label{fig:hd163296_gap}
\end{figure*}

\begin{figure*}[htbp]
\centering
\includegraphics[width=1\textwidth]{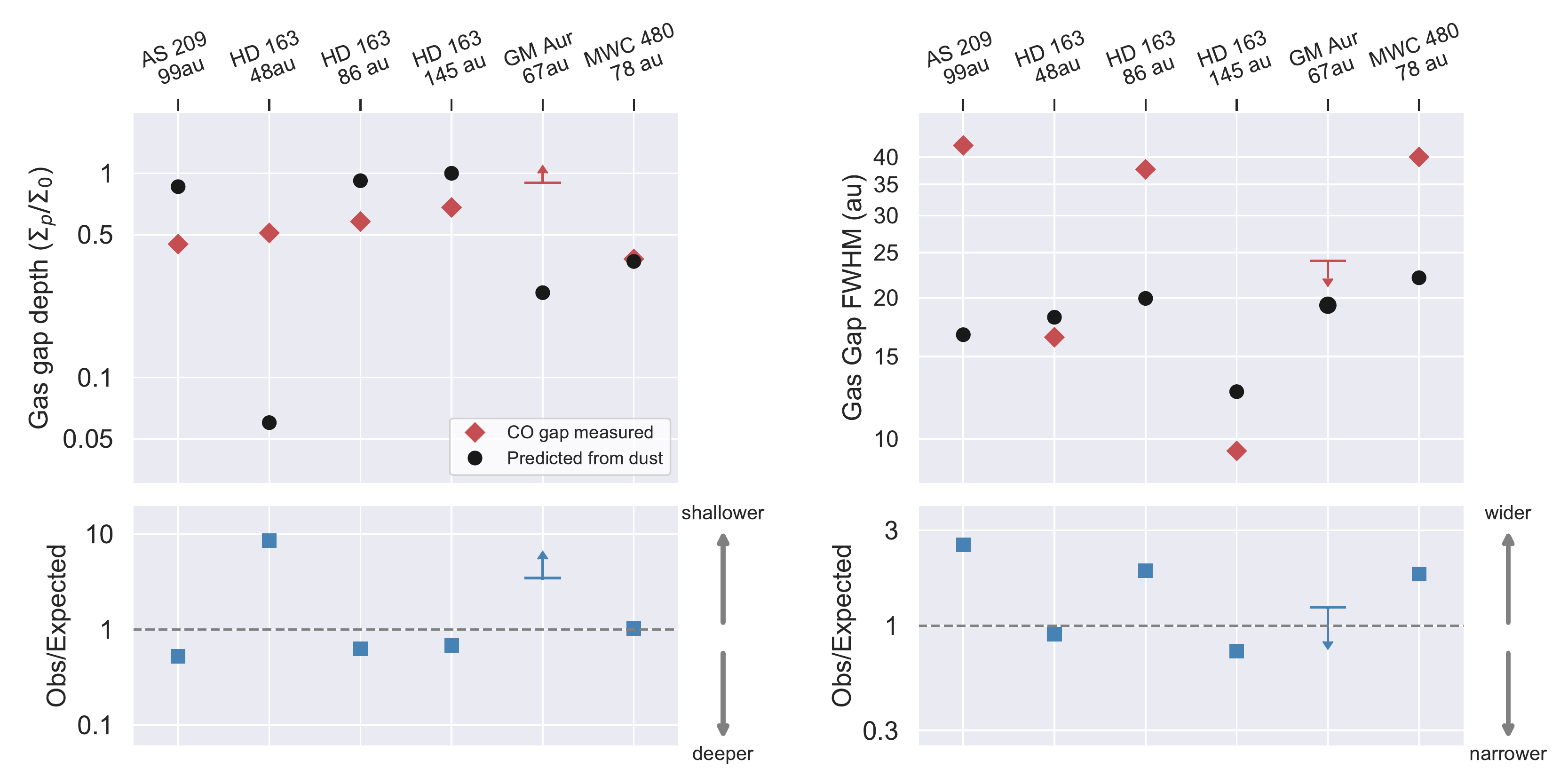}

\caption{Comparison of observed CO gap properties (depth and width) with predictions of planet-disk interactions. The top panels show the depths and widths of CO gaps (red diamonds) and predictions based on mm-continuum gaps (black circles). See detailed derivations in Section~\ref{sec:co_vs_cont}, and numbers are listed in the Appendix Table~\ref{tab:gap_comparison}. The bottom left panel: the ratios of CO gap depth over the gas gap depth predicted. It shows that the CO gap depths are generally consistent with predictions based on mm-continuum gaps, but there are also cases where CO gaps are 5-10 times shallower compared to the predictions. Bottom right panel: the same as the bottom left panel, but for the ratio of gap width. It shows that the observed widths are slightly larger or comparable to the expectations.  }  \label{fig:planet_properties}
\end{figure*}

In Figure~\ref{fig:co_gaps}, we show the CO gas profiles and mm-sized dust distributions in our best-fit models. 
At our spatial resolution (10-24\,au), a few substructures in \nco are noticeable. The most notable features are the AS 209 disk has a deep gap ($\sim$97\% depletion) around 240\,au, and GM Aur has a deep inner cavity inside 40\,au \footnote{We note that the \nco of the IM Lup disk appears to have a gap between 20-80\,au. But it is possibly just a result of the high dust optical depth inside 20\,au in our model (see Figure~\ref{fig:Nco_vs_cont}). For the IM Lup disk, all CO profiles show a depression inside 80\,au, and its 1.3\,mm continuum emission shows a steep increase inside 20\,au. Given the large uncertainty of optical depth inside 20\,au, we do not consider it as a real gas gap here.}  Except these features, other \nco substructures are more subtle compared to substructures seen in the dust distribution. To better characterize these substructures, we subtract a smooth function from the \nco profiles to identify significant features in the residuals. We then fit Gaussian functions to the residuals to estimate the widths and locations of gaps.   

 Table~\ref{tab:co_gap_properties} lists the CO gap properties. We note that the exact properties will likely vary with the particular choice of smooth functions. In general, when CO gaps are present in the pebble disk region, their gap locations roughly coincide with the gap locations of mm-sized dust, suggesting a correlation between these pairs. However, this is not a one-to-one correlation, as some of the deepest dust gaps do not have CO gap pairs. 
 
 The AS 209 disk shows a broad CO gap between 40-120\,au, spanning the range of three deep dust gaps. Our fitting requires two overlapping Gaussian functions at 61 and 94\,au to match this broad gap. The peaks of the two Gaussians are consistent with the locations of two deep dust gaps at 61 and 90\,au. The AS 209 disk also has a deep CO gap around 236\,au, which is beyond the pebble disk region. The GM Aur disk does not show clear substructures in its CO distribution beyond its central cavity, even though it has two deep dust gaps at 70 and 120\,au.  The HD 163296 disk has three CO gaps at 43, 94, and 148\,au, and these CO gaps are close to three dust gaps at 48, 86, and 145\,au. The MWC 480 disk has one broad gap at 63\,au, offset from its dust gap at 73\,au.  
In terms of the gap depth, CO gaps inside the pebble disk region only show a modest depression, where the \nco is 38-68\% of that at nearby regions. The deepest CO gaps in the sample is the 236\,au gap of the AS 209 disk, which is outside of the pebble disk. The density inside this gap is only $\sim$3\% of the background density. 

\begin{deluxetable}{lccc}[htbp]
\tablecaption{Properties of CO gaps \label{tab:co_gap_properties}}
\tablehead{
\colhead{Source} & \colhead{r$_{\rm gap}$} &\colhead{$\sigma$} & \colhead{depth}\\
\colhead{} & \colhead{(au)} & \colhead{(au)} &\colhead{}
}
\colnumbers
\startdata
GM Aur&   14$\pm$1 &    14$\pm$1 &     0.95$\pm$0.03 \\
AS 209&   59$\pm$9 &    13$\pm$6 &     0.47$\pm$0.27 \\
&   92$\pm$12 &    18$\pm$8 &     0.55$\pm$0.12 \\
&  238$\pm$1 &    13$\pm$1 &     0.97$\pm$0.08 \\
HD 163296&   44$\pm$1 &     7$\pm$1 &     0.49$\pm$0.04 \\
&   93$\pm$1 &    16$\pm$1 &     0.42$\pm$0.03 \\
&  148$\pm$1 &     4$\pm$1 &     0.32$\pm$0.05 \\
MWC 480&   63$\pm$1 &    17$\pm$1 &     0.62$\pm$0.03 \\
\enddata
\tablecomments{(1) Source name (2) CO gap location (3) gap width derived by fitting a Gaussian. $\sigma$ is the standard deviation of the best-fitting Gaussian function. (4) gap depth, defined as $(\Sigma_{\rm_{base}}-\Sigma_{\rm {gap}})/\Sigma_{\rm_{base}}$.  }
\end{deluxetable}

\subsubsection{Testing the planet scenario: can CO and dust gaps be explained by the same planet masses?}

It is still unclear what mechanisms cause these substructures in the CO gas distribution. Here we briefly discuss planet-disk interactions as one possible cause.

In our sample, the HD 163296 disk shows the strongest correlation between CO and dust gap locations. Various previous studies have suggested that the HD 163296 disk hosts multiple giant planets. One evidence is based on substructures in its 1.25\,mm continuum image, for which three giant planets between 0.1-4\,M$_J$ were needed to explain the dust gaps at 10, 48, and 86\,au \citep{Zhang_S18}. Another evidence is from local velocity deviations from Keplerian rotation \citep{Teague18a,Teague19Nat}. Hydro-dynamical models showed that three giant planets at 54, 83, 136\,au provide the best match to the velocity perturbations\footnote{The original planet locations reported in \citet{Teague18a} were based on a distance of 122\,pc for HD 163296. Here we corrected the planet locations using the latest GAIA distance of 101\,pc.}. In Figure~\ref{fig:hd163296_gap}, we compare our \nco gap locations and depths with gas surface density derived from velocity deviation of \citet{Teague18a}. The widths and depths of CO gaps are comparable to the gas gaps in \citet{Teague18a} model, but the peaks differ by up to 10~\,au  and the CO gap at 148\,au is 2-3 times narrower than the one in their model. In addition to these putative planets within the pebble disk, \citet{Pinte18_hd163} suggested a 2\,M$_J$ planet at 260\,au, using local deviation from Keplerian velocity. This velocity deviation has been further confirmed in the CO observations of the DSHARP and MAPS programs \citep{Pinte20_dsharp,Teague20_maps}. But we do not find any clear CO gaps around 260\,au. 

For the AS 209 disk, we identify two CO gaps. The inner gap between 50-120\,au  coincides with two mm-continuum gaps at 61 and 94\,au. \citet{Zhang_S18} attributed these two dust gaps to a single planet (0.2-1.3\,M$_J$) at 99\,au. \citet{Favre19} showed that a planet of 0.2-0.3\,M$_J$ at 100\,au is roughly consistent with their \cc~and \ct~(2-1) observations. However, using higher spatial resolution CO and C$_2$H observations, \citet{alarcon20_maps} showed that the CO gap is at least partially caused by a local depletion of CO abundance and thus ruled out the presence of $>0.2$\,M$_J$ planet around 100\,au.  The AS 209 disk also has a deep CO gap around 236\,au, which coincides with a local pressure minimum identified through velocity deviation \citep{Teague18_as209}. This gap location also potentially coincides with a ring in scattered light image of the AS 209 disk \citep{Avenhaus18_sphere}. The origin of the 236\,au gap requires further investigation. 

Figure~\ref{fig:co_gaps} shows that CO gaps are much shallower than the mm-continuum gaps. This is qualitatively consistent with the predictions of planet-disk interactions that gas gaps are shallower than gaps of pebbles \citep[e.g.,][]{Zhang_S18}. 
It is of particular interest to see whether the CO gas continuum gaps are quantitatively consistent with the expectation of planet-disk interactions. 

Here we do a quick check on whether the CO and continuum gaps can be explained by the same planet masses. 
Following the method developed by \citet{Zhang_S18}, we first measure continuum gap widths, and then combine these widths with scale height ratios and gas surface densities in our best-fit disk parameters to derive planet masses (assuming a viscosity parameter $\alpha$=10$^{-3}$). We then use the inferred planet masses and eqs.~(5-6) in \citet{Kanagawa15} to derive the expected gas gap depths\footnote{We note that we did not use \citet{Zhang_S18}'s eq.~(24) to link planet masses with gas gap depth, because their definition of gas depth does not work well for the steep \nco~profiles we derived in this work.}. We use eq.~(22) in \citet{Zhang_S18} to derive the expected gas gap width. The continuum gap widths, disk parameters, and estimated planet masses are listed in the Appendix in Table~\ref{tab:gap_comparison}. We note that these values are just used for a consistency check of the inferred planets by CO and continuum, and the results should not be considered rigorous derivations of planet masses. 

Figure~\ref{fig:planet_properties} shows the comparison of CO gas properties with expected properties of gas gaps. For gap depths, the CO gas depths are generally consistent with the expectations within a factor of 2. But there are two cases where the CO gaps are 5-10 times shallower than the predictions:  the 48\,au gap of HD 163296, and the 67\,au gap of GM Aur. For gap widths, the observed CO gap widths are consistent with the expected values within a factor of 3, but tend to be larger. This is partially due to our beam size of 15-24\,au, but the beam size difference is not sufficient to explain the two shallow gaps mentioned above. Nevertheless, our simple estimation of gap depths does not consider possible complications, such as CO abundance variations across a gap, the strength of viscosity, and impact of planet migration. Therefore, the discrepancy between the observed CO gap depths and the predictions in some disks does not necessarily rule out the possibility of planets at these places.
To further test the planet scenario, it will be beneficial to combine constraints from different methods, e.g., velocity deviations from Kepleration rotation \citep[e.g.,][]{Teague18_as209, Rab2020, Teague20_maps, alarcon20_maps}, and other chemical tracers like HCO$^+$ \citep[e.g.,][]{Smirnov-Pinchukov20}.

\subsection{Possible effects of isotope-selective photodissociation } \label{sec:photodissociation}

Our thermo-chemical models do not include isotope-selective photodissociation effects. Here we discuss how it might affect our results of the CO column densities.

CO is one of the few molecules that can self-shield itself from UV radiation, as its photodissociation occurs by absorbing UV photons at discrete wavelengths and being subsequently excited to predissociated bound states \citep[e.g.,][]{vanDishoeck88, visser09}. Due to the differences in abundances,  there is a transition zone where the ratios of $^{12}$CO/\cc~and $^{13}$CO/\cc~are higher than the elemental isotopic ratios. If the transition zone takes a significant fraction of the total CO gas column, the \nco derived from \cc~and an ISM element isotopic ratio would be lower than the true \nco \citep{miotello14}. Besides CO itself, H$_2$ and small dust particles in the disk provide additional attenuation to UV radiation.  In general, the more massive a disk is, the smaller the isotope-selective photodissociation effect for the total CO column density. For example, \citet{miotello14} showed that for an ISM ratio of CO-to-H$_2$ abundance the isotope-selective photodissociation is  unimportant for disks $>$10$^{-2}$\,M$_\odot$. 

For the five MAPS disks studied here, we expect the effects of isotope-selective photodissociation can only change our results up to a factor of 2-3. On the global disk scale, the CO column densities derived from $^{13}$CO and \cc~lines are consistent with a constant ratio of $^{13}$CO/C$^{18}$O=557/69, as far out in the disks that \cc~is detected. In contrast, models with significant isotope-selective photodissociation predict that $^{13}$CO/C$^{18}$O ratio is 3-10 times higher \citep[e.g.,][]{miotello14}.  The isotope-selective effect could also be important inside dust gaps, if the gas and dust densities are reduced significantly. However, as shown in Figure~\ref{fig:co_gaps}, the \cc~depletions inside gaps are only modest or even not detected inside the mm-continuum gap locations. It is unlikely that isotope-selective photodissociation is much stronger inside these gaps compared to the nearby regions. Further investigation of the UV intensity inside these gaps can be done with modeling of small dust particles (based on scattered light images), but it is beyond the scope of this work. In short, isotope-selective photodissociation effects would not significantly change our results.

\subsection{Gas distribution beyond 150\,au: evidence of viscous disks}
The global mass distribution of a disk profoundly affects planet formation and migration \citep[e.g.,][]{morbidelli16b}. Turbulent viscosity and disk winds are currently two leading mechanisms proposed to explain global disk mass transportation \citep{shakura73,Balbus91, bai13}. These two mechanisms can result in very different gas mass distributions, especially at the $<$10\,au and $>$100\,au regions of the disk. For example, current disk wind models generally predict steeper density profile beyond 100\,au compared to a pure viscous disk \citep{Bai16, suzuki16}. The difference in the outer disk is mainly because a viscous disk spreads with time as angular momentum moves to larger radii, while in the disk wind case the disk size needs not grow with time.  Global disk properties, such as accretion rates and disk sizes have been tried to constrain viscous disk evolution models but observations showed relatively large scatter \citep{Mulders17, Najita18_evolution, Hendler20}. The high resolution MAPS measurements of CO distribution now provide detailed constraints on the gas mass distribution in the outer disk region. Here we compare the \nco distributions with theoretical predictions of gas mass distribution at the outer disk region.

As shown in Figure~\ref{fig:Nco_profiles_comparison}, four out of five MAPS disks show similar \nco distributions between 150-400\,au. Except for the AS 209 disk, the remaining four disks all show a smooth long tail, which can be well characterized by a power-law function of N$_{\rm CO}\propto r^{-2.4\pm0.2}$. This \nco tail is a very shallow profile compared to current predictions of disk wind models \citep{Bai16, suzuki16} but comparable to that of viscously evolved disk models.

\begin{figure*}[htbp]
\centering
\includegraphics[width=1.\textwidth]{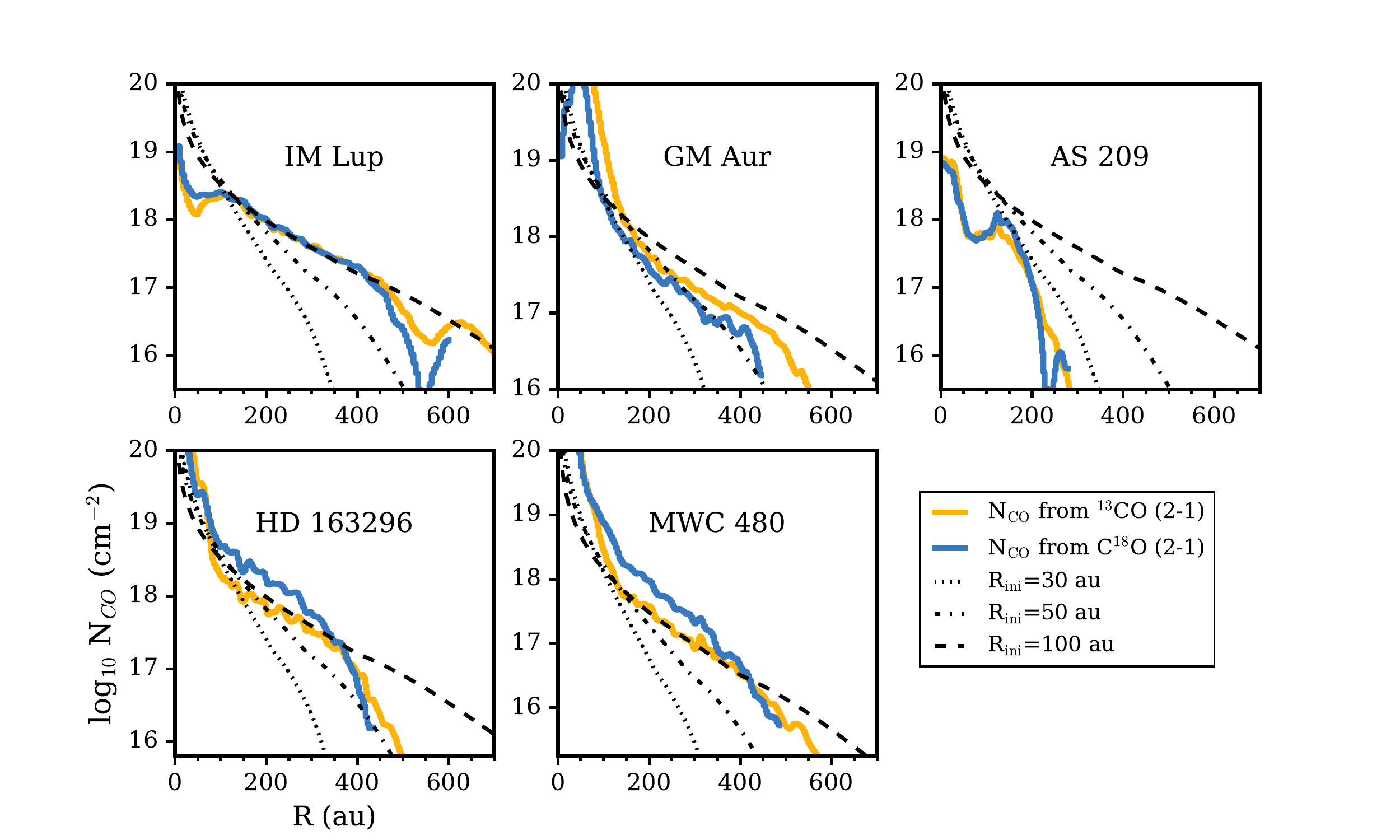}
\caption{Comparison of CO gas column density profiles of MAPS disks with that of viscously evolving disk models \citep{Trapman20_viscous}. The viscous disk models are for $\alpha=10^{-3}$ around a 1\,M$_\odot$ star after 1\,Myr evolution. The initial disk mass in the models is  0.06\,M$_\odot$ and the initial radii, R$_{\rm init}$, is 30, 50, and 100\,au, respectively. \citet{Trapman20_viscous} did not include CO depletion, and therefore we scaled their CO column density by a factor of 0.1 to better match our results. The CO distributions in the IM Lup, GM Aur, HD 163296, and MWC 480 disks match well with the viscous models of R$_{\rm init}$=50, 100\,au between the 150-400\,au region. The CO profile of the AS 209 disk is more consistent with the R$_{\rm init}$=30\,au model. \label{fig:nco_viscous}}
\end{figure*}

For illustrative purposes, we compare our \nco profiles with that from thermo-chemical models of viscous disks of \citet{Trapman20_viscous}. We adopt \nco profiles from three viscous disk models with $\alpha=10^{-3}$ around a 1\,M$_\odot$ star after 1\,Myr of evolution. These models have an initial disk mass of 0.06\,M$_\odot$ and their initial disk size, R$_{\rm init}$, is 30, 50, and 100\,au, respectively. These models predict that the slope of CO column density distribution is similar to the gas between 100-500\,au region (see Figure~\ref{fig:trapman20_models} in the Appendix).   We scale their absolute CO column density by a factor of 0.1 to match our \nco measurements. The differences are likely because these models do not consider CO depletion. Figure~\ref{fig:nco_viscous} shows the \nco~measured in this work with that of viscous models.  Between 150-400\,au, the measured \nco profiles of the IM Lup, HD 163296, and MWC 480 disks match well to the R$_{\rm init}$=100\,au model, and the GM Aur disk to the R$_{\rm init}$=50\,au model. The AS 209 disk is similar to the R$_{\rm init}$=30\,au model but likely requires an even smaller R$_{\rm init}$ for better match. Beyond 400\,au, the measured \nco profiles decrease faster compared to the models. The faster decreases might be caused by photodissociation of CO or real gas surface density changes due to photoevaporation.

In short, the \nco distributions in 150-400\,au of the five MAPS disk match well with predictions of viscously evolving disk models. The initial evidence shows the importance of spatially resolved \nco profiles to constrain current disk evolution theories. A systematic study of \nco profiles in a larger disk sample and comparison with both viscously evolving and wind-driven models are needed for further investigation.

\subsection{CO snowlines and continuum substructures}
Snowlines of major volatile species have been proposed as a possible cause of substructures seen in (sub)mm continuum of disks \citep{zhang15b,okuzumi16}. The expectation is that the dust size distribution and mass surface density have a significant discontinuity at the snowline locations, as predicted by current models of icy dust growth \citep[e.g.,][]{ros13, banzatti15, okuzumi16, Pinilla17, Ros19}. Here we compare the locations of CO snowline in our models with continuum substructures. 

In our thermo-chemical models, the mid-plane CO snowline is defined as the radius where the CO gas and ice abundance become equal at the disk mid-plane. The mid-plane snowline locations of IM Lup and AS 209 disks are slightly warmer than the CO condensation temperature ($\sim$20\,K), because in our models a significant fraction of CO ices at the mid-plane has been processed into other species after 1\,Myr).The snowline locations in our models are listed in Table~\ref{tab:mini_mass}. 
These snowline locations are consistent with the observations of \ce{C^{17}O} (1-0) line in the HD~163296 and MWC 480 disks, in which the \ce{C^{17}O} line becomes optically thick in the inner disks of at $\sim$60~au and 80~au, respectively. Interestingly, a bright continuum ring is seen at the CO snowline location of both disks, as shown in Figure~\ref{fig:Nco_vs_cont}. This bright continuum ring is consistent with a pile-up of dust grains due to the sintering effect at snowlines \citep{okuzumi16,banzatti15,Pinilla17}.

\section{Conclusions} \label{sec:summary}

In this work, we present high resolution (15-24\,au) observations and analysis of CO isotopologue lines from the MAPS ALMA Large Program. Our analysis employs \cc~and \ct~$J$=(2-1), (1-0), and C$^{17}$O (1-0) lines of five protoplanetary disks around IM Lup, GM Aur, AS 209, HD 163296, and MWC 480. Our findings are summarized as follows:

\begin{itemize}

    \item We retrieve CO gas density distributions, using three independent assumptions for the underlying gas temperature: (1) a thermo-chemical modeling framework based on the CO data, the broadband spectral energy distribution, and the mm continuum emission; (2) an empirical distribution based on optically thick CO emission lines; and (3) a direct estimate based on fits to the \cseven~hyperfine lines. Results from all three methods generally show excellent agreement. 

   \item By comparing model temperature structures with empirical measurements from optically thick lines, we show that current thermo-chemical models can provide reliable temperature structures within the 15-40\,K region.   
   
    \item We show that fitting \ce{C^{17}O} hyperfine structure components provides reliable constraints on the CO gas column density distribution in disks. The results are comparable to those derived from thermo-chemical models. The hyperfine analysis is independent of the disk model, or any previous knowledge about the temperature structure. 
    
    \item Assuming a gas-to-dust mass ratio of 100, we find that all five disks have a CO-to-H$_2$ abundance of 10-100 times lower than the ISM ratio of 10$^{-4}$. The radial distribution of CO gas column density show significant variations in terms of the absolute value and the radial profile among the five disks. 
    
    \item The MWC 480 and HD 163296 disks show a steep increase of CO gas column density inside their CO snowlines compared to predictions of thermo-chemical models, suggesting their CO abundance inside the CO snowline is significantly higher than the warm molecular layer in the outer disk region. This enhancement of CO abundance inside the CO snowline is consistent with predictions of a large amount of icy pebble drift into the inner disk regions. 
    
   \item Four of the five disks have remarkably similar surface density profiles between 150-400\,au, which can be well characterized as $\Sigma_{\rm CO}\propto R^{-2.4}$. 
   We show that the \nco profiles of all five disks between 150-400\,au are consistent with predictions of viscously evolving disks with an initial disk size between 30-100\,au. 
    
    \item We find that when CO gaps are present in the pebble disk region, their locations are correlated with locations of dust gaps. But some of the deepest dust gaps do not have a corresponding CO gap. These CO gas gaps are depleted by a factor of 30-62\% of compared to nearby regions, a much shallower depression than the 1-2 orders of depletion seen in dust gaps. The relative depths of CO gaps and mm-continuum gaps are generally consistent with predictions of planet-disk interactions. But there are also cases where CO gaps are 5-10 times shallower than predictions based on continuum observations.
    
\end{itemize}

In summary, the MAPS observations show that high spatial resolution observations of CO isotopologue lines can provide direct constraints on fundamental disk properties, such as gas thermal structure and gas mass distributions. These constraints are key tests of current theories of planet formation, including dust evolution, planet-disk interaction, and global evolution of disk structures. Similar CO observations for a larger sample of protoplanetary disks will be essential for our understanding of typical environment and processes of planet formation.

\acknowledgments
This paper makes use of the following ALMA data: ADS/JAO.ALMA\#2018.1.01055.L. ALMA is a partnership of ESO (representing its member states), NSF (USA) and NINS (Japan), together with NRC (Canada), MOST and ASIAA (Taiwan), and KASI (Re- public of Korea), in cooperation with the Republic of Chile. The Joint ALMA Observatory is operated by ESO, AUI/NRAO and NAOJ. The National Radio Astronomy Observatory is a facility of the National Science Foundation operated under cooperative agreement by Associated Universities, Inc.

K.Z. acknowledges the support of the Office of the Vice Chancellor for Research and Graduate Education at the University of Wisconsin – Madison with funding from the Wisconsin Alumni Research Foundation. K.Z., K.R.S., J.H., J.B., J.B.B., and I.C. acknowledge the support of NASA through Hubble Fellowship grant HST-HF2-51401.001, HST-HF2-51419.001, HST-HF2-51460.001-A, HST-HF2-51427.001-A, HST-HF2-51429.001-A, and HST-HF2-51405.001-A awarded by the Space Telescope Science Institute, which is operated by the Association of Universities for Research in Astronomy, Inc., for NASA, under contract NAS5-26555. C.J.L. acknowledges funding from the National Science Foundation Graduate Research Fellowship under Grant DGE1745303. A.D.B., E.A.B., and A.F. acknowledges support from NSF AAG Grant \#1907653. V.V.G. acknowledges support from FONDECYT Iniciaci\'on 11180904 and ANID project Basal AFB-170002. Y.L. acknowledges the financial support by the Natural Science Foundation of China (Grant No. 11973090) K.I.\"O acknowledges support from the Simons Foundation (SCOL \#321183) and an NSF AAG Grant (\#1907653). S.M.A. and J.H. acknowledge funding support from the National Aeronautics and Space Administration under Grant No. 17-XRP17 2-0012 issued through the Exoplanets Research Program. J.D.I. acknowledges support from the Science and Technology Facilities Council of the United Kingdom (STFC) under ST/T000287/1. C.W.~acknowledges financial support from the University of Leeds and from the Science and Technology Facilities Council (grant numbers ST/R000549/1 and ST/T000287/1). R.T. and F.L. acknowledges support from the Smithsonian Institution as a Submillimeter Array (SMA) Fellow. M.L.R.H. acknowledges support from the Michigan Society of Fellows. Y. A. and G. C. acknowledge support by NAOJ ALMA Scientific Research grant code 2019-13B. Y.A. acknowledges support by Grant-in-Aid for Scientific Research (S) 18H05222, and Grant-in-Aid for Transformative Research Areas (A) 20H05844 and 20H05847. J.K.C. acknowledges support from the National Science Foundation Graduate Research Fellowship under Grant No. DGE 1256260 and the National Aeronautics and Space Administration FINESST grant, under Grant no. 80NSSC19K1534. A.S. acknowledges support from ANID/CONICYT Programa de Astronom\'ia Fondo ALMA-CONICYT 2018 31180052. F.M. acknowledges support from ANR of France under contract ANR-16-CE31-0013 (Planet-Forming-Disks)  and ANR-15-IDEX-02 (through CDP "Origins of Life"). R.L.G. acknowledges support from a CNES fellowship grant. H.N. acknowledges support by NAOJ ALMA Scientific Research grant code 2018-10B and Grant-in-Aid for Scientific Research 18H05441. Y.Y. is supported by IGPEES, WINGS Program, the University of Tokyo. L.M.P.\ acknowledges support from ANID project Basal AFB-170002 and from ANID FONDECYT Iniciaci\'on project \#11181068. T.T. is supported by JSPS KAKENHI Grant Numbers JP17K14244 and JP20K04017.

\vspace{5cm}

\newpage
\facilities{ALMA}

\software{Astropy \citep{astropy},  
          CASA \citep{CASA}, 
          RAC2D \citep{du14},
          RADMC3D \citep{radmc3d},
          \texttt{Gofish} \citep{GoFish},
          \texttt{emcee} \citep{emcee},
          \texttt{dsharp\_opac} \citep{Birnstiel18}.
          }

\appendix


\begin{figure*}[htbp]
    \centering
    \includegraphics[width=\hsize]{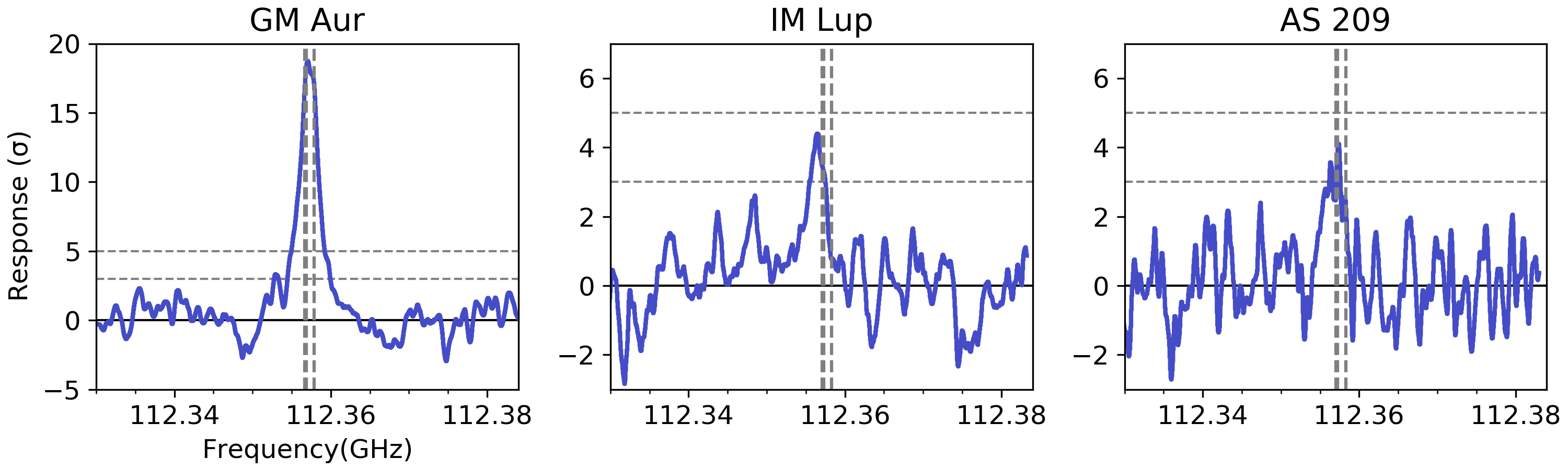}
    \caption{Matched filter responses of the GM~Aur, IM Lup and AS~209 disks using a 100~au Keplerian mask. Horizontal dashed lines mark the 3 and 5 $\sigma$ levels, and the vertical lines mark the frequencies of the \ce{C^{17}O} (1-0) hyperfine structure transitions after correction for the velocities of the sources.}
    \label{c17o_filter}
\end{figure*}

\begin{figure*}[htbp]
\centering
\includegraphics[width=0.95\textwidth]{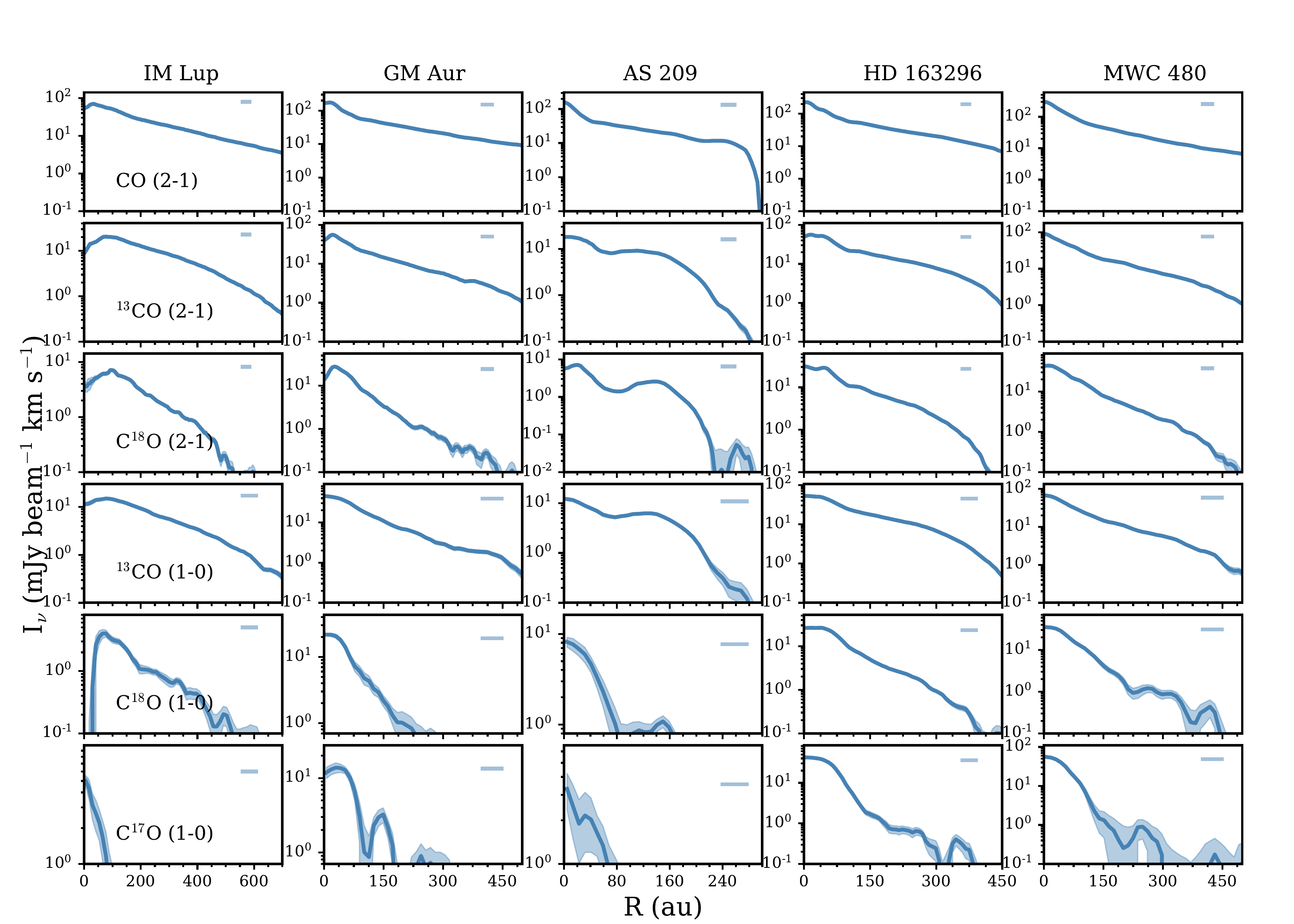}
\caption{Deprojected radial intensity profiles of CO lines, for the MAPS sample, ordered by increasing stellar mass from left to right. The same as Figure~\ref{fig:co_obs}, but on a logarithmic scale to highlight the outer disk emission. The beam sizes are plotted at the upper right corners. } \label{fig:co_obs_log}
\end{figure*}

\begin{figure*}[htbp]
\centering
\includegraphics[width=1\textwidth]{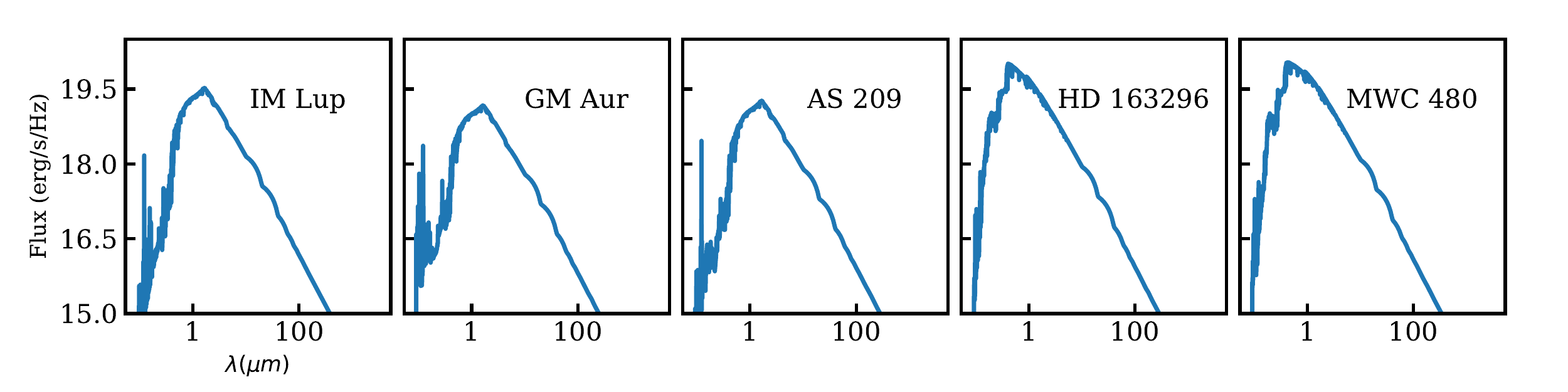}
\includegraphics[width=1\textwidth]{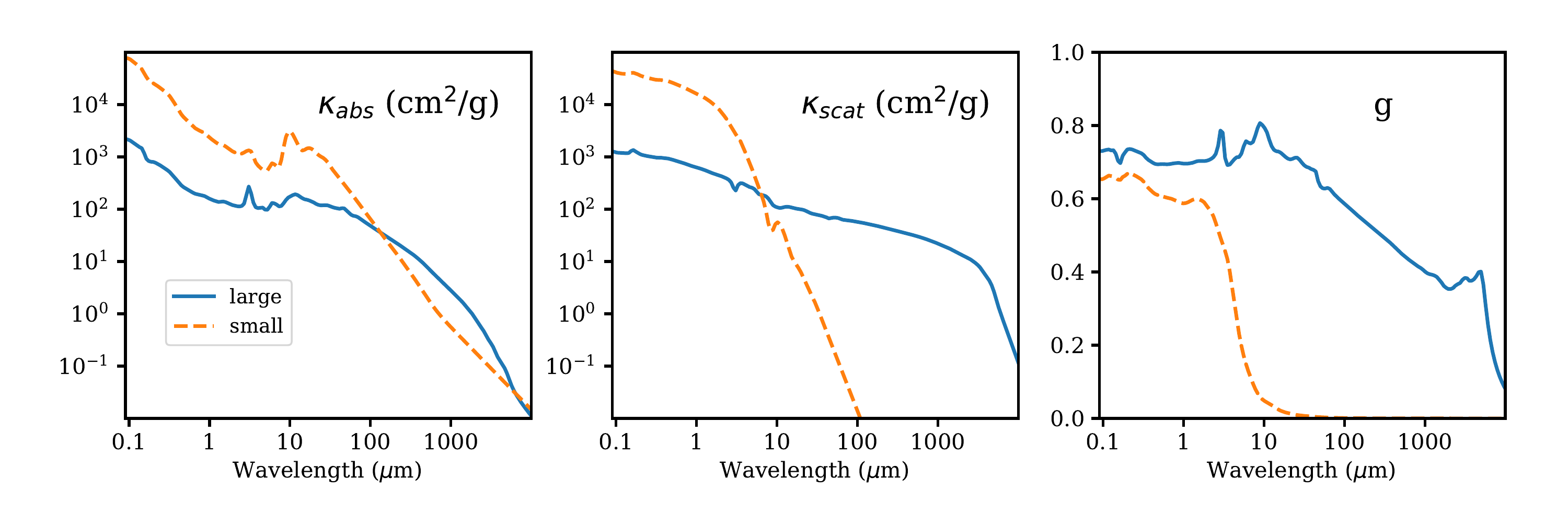}
\caption{Stellar spectra (top row) and dust absorption and scattering coefficients} (bottom row) used in thermo-chemical models \label{fig:stellar_spec}
\end{figure*}

\section{Robustness test of the CO retrieval method}
\label{sec:app:co_retrieval_test}

\begin{figure*}[htbp]
\centering
\includegraphics[width=1.0\textwidth]{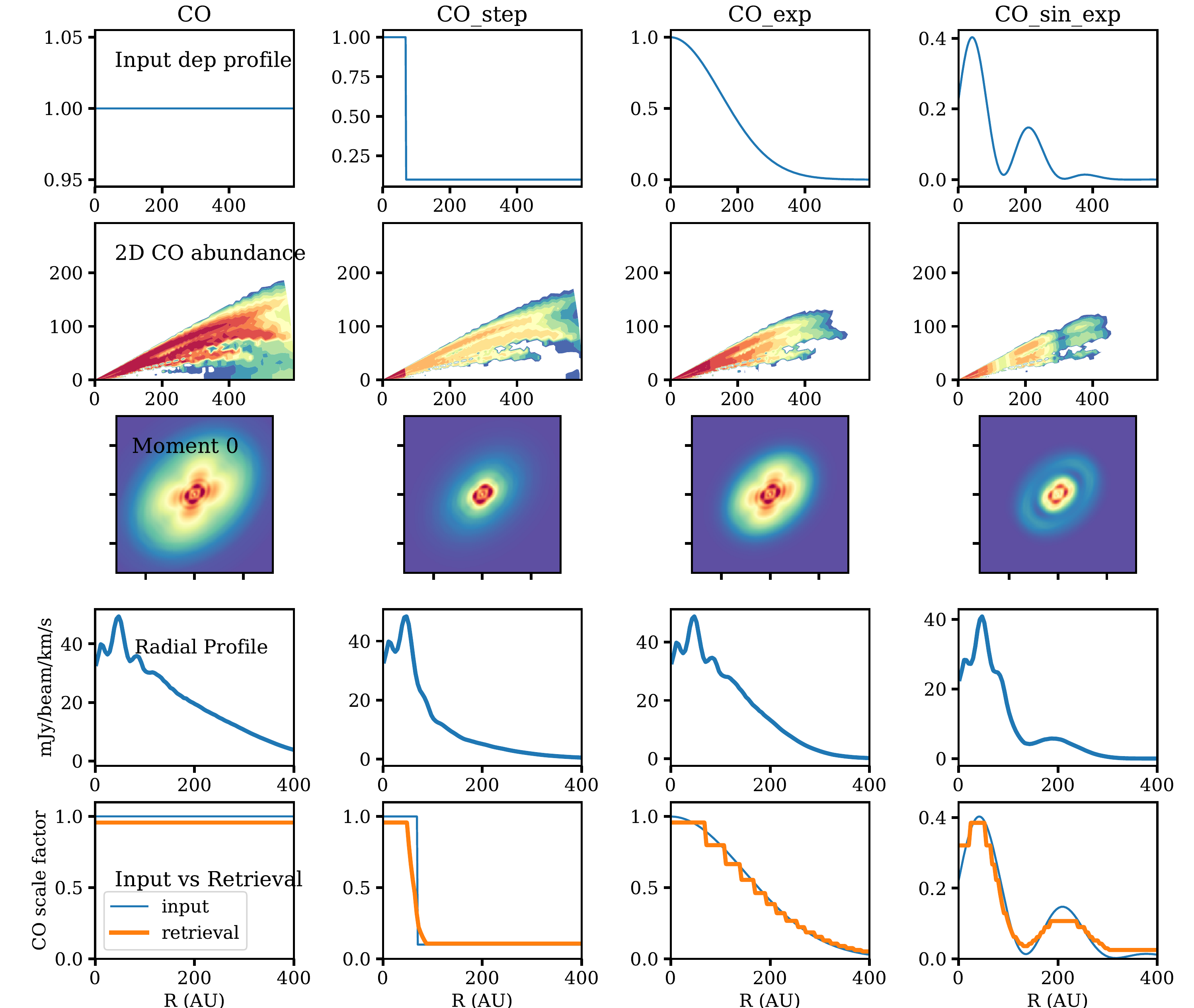}
\caption{Robustness test results of the CO abundance retrieval method. Row 1 shows the profiles of CO depletion used in the test. Row 2 shows the 2-dimensional CO abundance after scaling by the profiles. Row 3 and 4 show the result moment zero maps of the \ct~(2-1) line images and deprojected radial profiles. Row 5 shows the comparison between the input depletion profile and the retrieved CO depletion profile based on a grid of CO models with constant depletion throughout the whole disk. \label{fig:retrieval_test}} 
\end{figure*}

In Section~\ref{sec:thermo-chemical models}, we compare a grid of CO models with the observed CO radial intensity profiles to retrieve CO column density distribution. Here we test the robustness of this approach, using the HD 163296 model as an example. We modify the initial CO abundance structure with three radial depletion profiles, a step function, an exponentially decreasing function, and a sinusoidal function (see Figure~\ref{fig:retrieval_test}).  Using the modified CO abundance structures, we then generate simulated CO observations and compare simulated radial profiles with our grid of CO models to retrieve radial CO depletion profiles. In Figure~\ref{fig:retrieval_test}, we show that our method robustly recovered all three input depletion profiles, with an uncertainty less than 20\%, consistent with the step size of our logarithmic grid.

\begin{figure*}
\centering
\includegraphics[width=\textwidth]{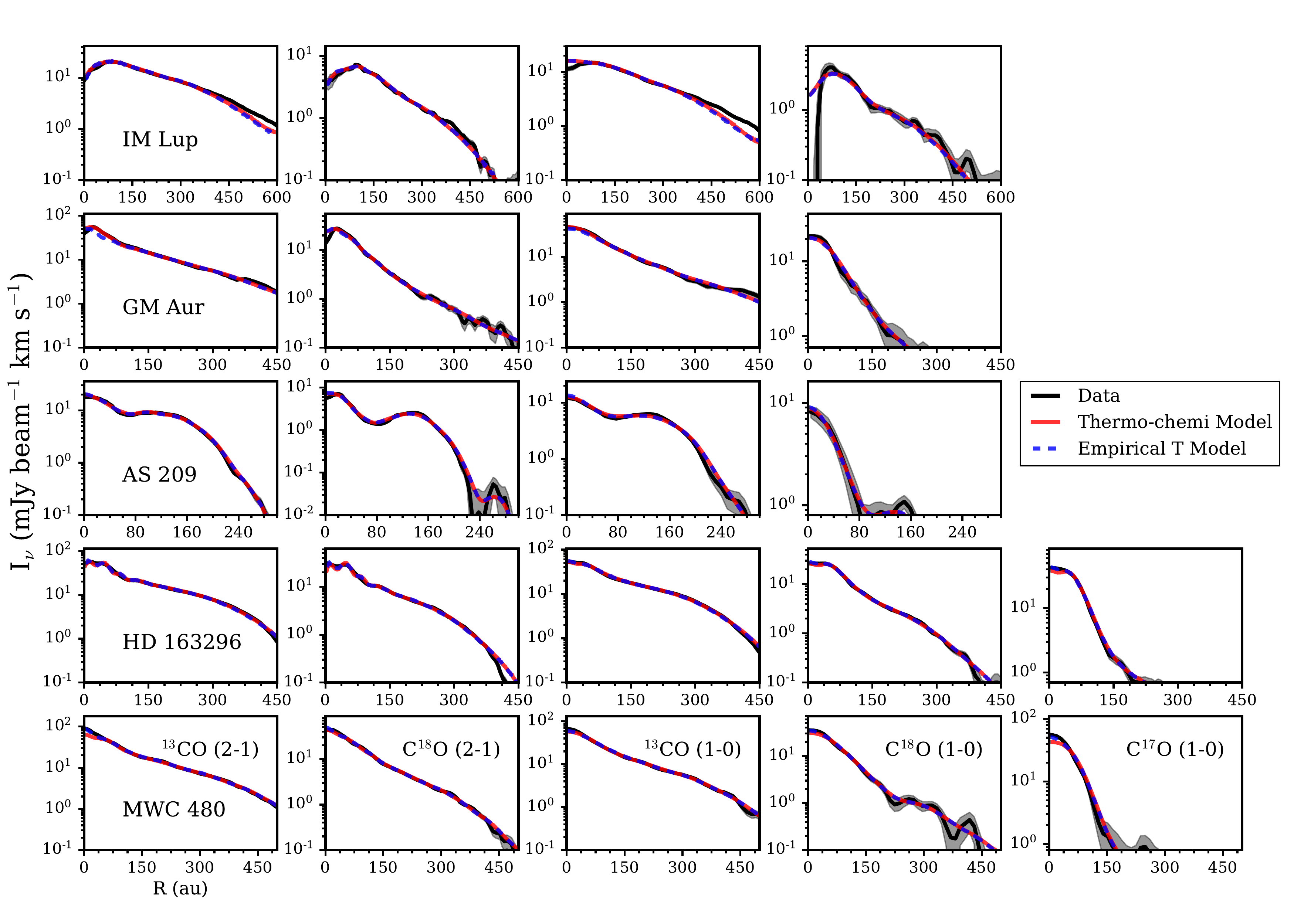}
\caption{Comparison of radial profiles of the CO isotopologue lines and best-fit results from thermo-chemical models (red solid line, and see model details in Section~\ref{sec:co_column_thermo}) and empirical temperature models (blue dash line, and see model details in Section~\ref{sec:empirical_temperature}).}    \label{fig:co_profile_empi} 
\end{figure*}

\begin{figure*}
\begin{minipage}{.23\textwidth}
        \centering
        \includegraphics[width=\textwidth]{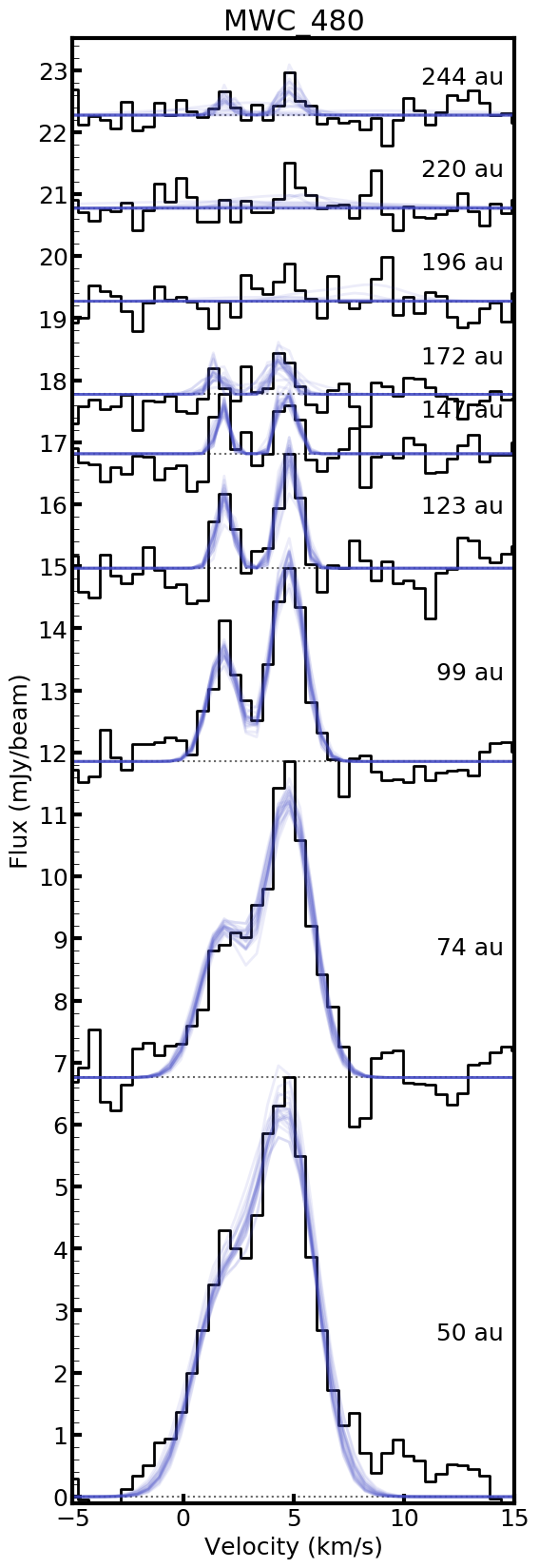}
    \end{minipage}%
    \begin{minipage}{0.72\textwidth}
        \centering
         \includegraphics[width=\hsize]{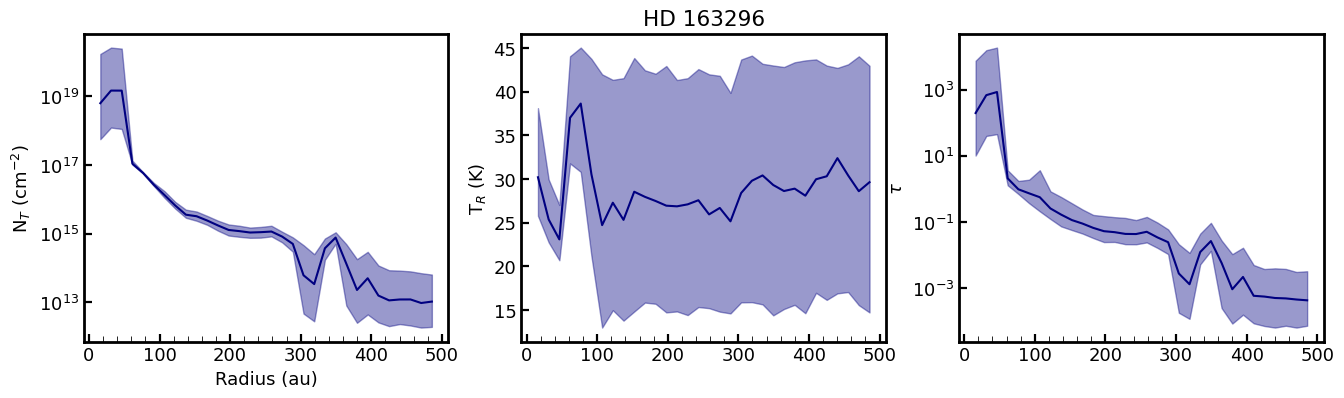}
        \includegraphics[width=\hsize]{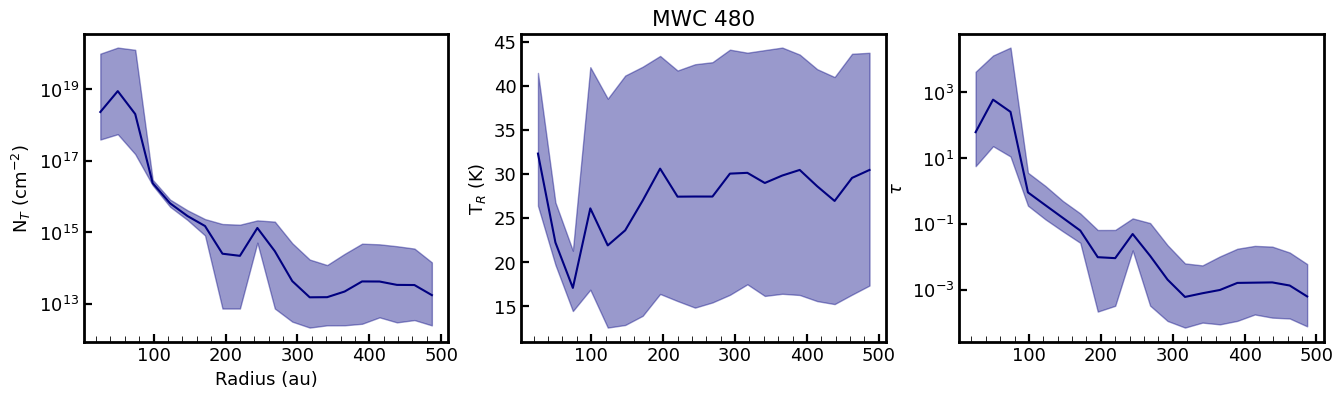}
        \caption{Left: radially averaged \ce{C^{17}O} spectra (black) and the results from the hyperfine line fitting (blue) for the MWC 480 disk. The center radius of each bin is shown and the spectra are offset along the axis for clarity. Right: results from MCMC fit to the \ce{C^{17}O} spectra.
    From left to right are the \ce{C^{17}O} column density, excitation temperature ($T_{ex}$) and optical depth ($\tau$).}
        \label{fig:c17o_fits}
    \end{minipage}
\end{figure*}

\begin{deluxetable*}{cccccccccccccc}
\tablecaption{Comparison of the CO gap properties with properties of gas gaps caused by planet-disk interactions \label{tab:gap_comparison}}
\tablehead{
 \colhead{Source}& \colhead{M$_\star$}& \colhead{r$_{\rm CO}$}& \colhead{r$_{\rm mm}$}& \colhead{h/r}& \colhead{$\Sigma_g$} & \colhead{$\Delta_{\rm mm}$}& \colhead{$\delta_{\rm CO}$}& \colhead{CO$_{\rm fwhm}$}& \colhead{m$_{p4}$}& \colhead{m$_{p3}$}& \colhead{m$_{p2}$}& \colhead{depth}& \colhead{width} \\
 \colhead{}& \colhead{(M$_\odot$)}& \colhead{(au)}& \colhead{(au)}& \colhead{}& \colhead{(g cm$^{-2}$)}& \colhead{}&\colhead{}& \colhead{(au)}& \colhead{(m$_{\rm Jup}$)}& \colhead{(m$_{\rm Jup}$)}& \colhead{(m$_{\rm Jup}$)}& \colhead{}& \colhead{(au)}
}
\colnumbers
\startdata
 AS 209     &   1.2 &     94 &     99 &   0.06 &   0.24 &   0.31 &     0.45 &     42.4 &     0.03 &     0.07 &     0.14 &     0.86 &     16.7 \\
 HD 163296  &   2.0 &     43 &     48 &   0.08 &  19.17 &   0.34 &     0.51 &     16.5 &     1.11 &     2.26 &     4.61 &     0.06 &     18.2 \\
 HD 163296  &   2.0 &     98 &     86 &   0.08 &   9.30 &   0.17 &     0.58 &     37.7 &     0.10 &     0.20 &     0.40 &     0.92 &     20.0 \\
 HD 163296  &   2.0 &    145 &    143 &   0.09 &   4.23 &   0.09 &     0.68 &      9.4 &     0.005 &     0.01 &     0.02 &     1.00 &     12.6 \\
 GM Aur     &   1.1 &     67 &     68 &   0.07 &  17.06 &   0.24 &    $>$0.90 &     \nodata &     0.17 &     0.34 &     0.70 &     0.26 &     19.3 \\
 MWC 480    &   2.1 &     63 &     73 &   0.10 &  10.98 &   0.29 &     0.38 &     40.0 &     0.66 &     1.35 &     2.75 &     0.37 &     22.1 \\
\enddata
\tablecomments{ (1) Source name; (2) Stellar mass; (3) CO gas gap location; (4) 1.3\,mm continuum gap location; (5) Scale height-to-radius ratio at r$_{\rm mm}$, based on model parameters in Table~\ref{tab:disk_para}; (6) Gas surface density at r$_{\rm mm}$, based on model parameters in Table~\ref{tab:disk_para}; (7) $\Delta_{\rm mm}$: a width parameter measured from mm continuum radial profiles, defined as (r$_{\rm out}-$r$_{\rm in}$)/r$_{\rm out}$. See details in eq. (21) of \citet{Zhang_S18}; (8) CO gap depth measured in this work, see Table~\ref{tab:co_gap_properties}. Here $\delta_{\rm CO}$ = $\Sigma_{\rm gap}/\Sigma_{0}$; (9) FWHM of CO gap measured in this work, adopted from Table~\ref{tab:co_gap_properties}; (10) Planet masses derived from \{$\Delta_{\rm mm}$, h/r, $\Sigma_g$, and M$_\star$\}, using Table~1 of \citet{Zhang_S18} and $\alpha=10^{-4}$; (11) the same as (10), but for $\alpha=10^{-3}$; (12) the same as (10), but for $\alpha=10^{-2}$; (13) expected gas gap depth, based on m$_{p3}$ and  $\alpha=10^{-3}$, using \citet{Kanagawa15} eq.~(5); (14) expected gas gap width based on m$_{p3}$ and $\alpha=10^{-3}$, using Table~2 of \citet{Zhang_S18}. }
\end{deluxetable*}

\begin{figure*}[htbp]
\centering
\includegraphics[width=1.0\textwidth]{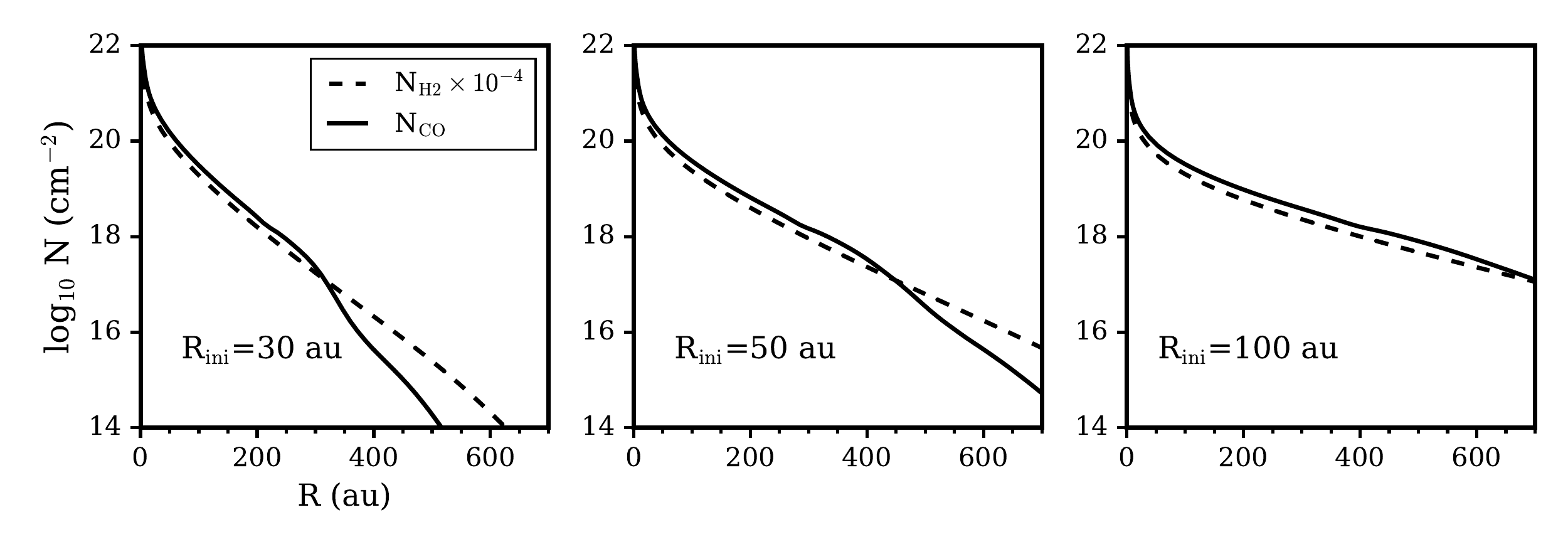}
\caption{Gas and CO column density profiles of 
viscously evolving disk models from \citet{Trapman20_viscous}. These models are for $\alpha=10^{-3}$ around a 1\,M$_\odot$ star after 1\,Myr of evolution. The initial disk mass in the models is  0.06\,M$_\odot$ and the initial radii, R$_{\rm init}$, is 30, 50, and 100\,au, respectively. These models show that between 150-400\,au, the CO column density profile is similar to the gas surface density profile.   \label{fig:trapman20_models}} 
\end{figure*}

\bibliography{lib}{}

\begin{thebibliography}{}
\expandafter\ifx\csname natexlab\endcsname\relax\def\natexlab#1{#1}\fi
\providecommand{\url}[1]{\href{#1}{#1}}
\providecommand{\dodoi}[1]{doi:~\href{http://doi.org/#1}{\nolinkurl{#1}}}
\providecommand{\doeprint}[1]{\href{http://ascl.net/#1}{\nolinkurl{http://ascl.net/#1}}}
\providecommand{\doarXiv}[1]{\href{https://arxiv.org/abs/#1}{\nolinkurl{https://arxiv.org/abs/#1}}}

\bibitem[{{Aikawa} {et~al.}(1998){Aikawa}, {Umebayashi}, {Nakano}, \&
  {Miyama}}]{aikawa98}
{Aikawa}, Y., {Umebayashi}, T., {Nakano}, T., \& {Miyama}, S. 1998, Faraday
  Discussions, 109, 281, \dodoi{10.1039/a800258d}

\bibitem[{{Aikawa} {et~al.}(1999){Aikawa}, {Umebayashi}, {Nakano}, \&
  {Miyama}}]{aikawa99_abundance}
{Aikawa}, Y., {Umebayashi}, T., {Nakano}, T., \& {Miyama}, S.~M. 1999, \apj,
  519, 705, \dodoi{10.1086/307400}

\bibitem[{{Alarc{\'o}n} {et~al.}(2020){Alarc{\'o}n}, {Teague}, {Zhang},
  {Bergin}, \& {Barraza-Alfaro}}]{Alarcon20}
{Alarc{\'o}n}, F., {Teague}, R., {Zhang}, K., {Bergin}, E.~A., \&
  {Barraza-Alfaro}, M. 2020, \apj, 905, 68, \dodoi{10.3847/1538-4357/abc1d6}

\bibitem[{{Alarc{\'o}n} {et~al.}(2021){Alarc{\'o}n}, {Bosman}, {Bergin},
  {Zhang}, {Teague}, {Bae}, {Aikawa}, {Andrews}, {Booth}, {Calahan}, {Cataldi},
  {Czekala}, {Huang}, {Ilee}, {Law}, {Le Gal}, {Liu}, {Long}, {Loomis},
  {M{\'e}nard}, {{\"O}berg}, {Schwarz}, {Van't Hoff}, {Walsh}, \&
  {Wilner}}]{alarcon20_maps}
{Alarc{\'o}n}, F., {Bosman}, A., {Bergin}, E., {et~al.} 2021, arXiv e-prints,
  arXiv:2109.06263.
\newblock \doarXiv{2109.06263}

\bibitem[{{Alcal{\'a}} {et~al.}(2017){Alcal{\'a}}, {Manara}, {Natta}, {Frasca},
  {Testi}, {Nisini}, {Stelzer}, {Williams}, {Antoniucci}, {Biazzo}, {Covino},
  {Esposito}, {Getman}, \& {Rigliaco}}]{Alcala17}
{Alcal{\'a}}, J.~M., {Manara}, C.~F., {Natta}, A., {et~al.} 2017, \aap, 600,
  A20, \dodoi{10.1051/0004-6361/201629929}

\bibitem[{{Anderson} {et~al.}(2013){Anderson}, {Bergin}, {Maret}, \&
  {Wakelam}}]{anderson13}
{Anderson}, D.~E., {Bergin}, E.~A., {Maret}, S., \& {Wakelam}, V. 2013, \apj,
  779, 141, \dodoi{10.1088/0004-637X/779/2/141}

\bibitem[{{Andrews} {et~al.}(2011){Andrews}, {Wilner}, {Espaillat}, {Hughes},
  {Dullemond}, {McClure}, {Qi}, \& {Brown}}]{andrews11}
{Andrews}, S.~M., {Wilner}, D.~J., {Espaillat}, C., {et~al.} 2011, \apj, 732,
  42, \dodoi{10.1088/0004-637X/732/1/42}

\bibitem[{{Andrews} {et~al.}(2018){Andrews}, {Huang}, {P{\'e}rez}, {Isella},
  {Dullemond}, {Kurtovic}, {Guzm{\'a}n}, {Carpenter}, {Wilner}, {Zhang}, {Zhu},
  {Birnstiel}, {Bai}, {Benisty}, {Hughes}, {{\"O}berg}, \&
  {Ricci}}]{Andrews18b}
{Andrews}, S.~M., {Huang}, J., {P{\'e}rez}, L.~M., {et~al.} 2018, \apj, 869,
  L41, \dodoi{10.3847/2041-8213/aaf741}

\bibitem[{{Ansdell} {et~al.}(2016){Ansdell}, {Williams}, {van der Marel},
  {Carpenter}, {Guidi}, {Hogerheijde}, {Mathews}, {Manara}, {Miotello},
  {Natta}, {Oliveira}, {Tazzari}, {Testi}, {van Dishoeck}, \& {van
  Terwisga}}]{ansdell16}
{Ansdell}, M., {Williams}, J.~P., {van der Marel}, N., {et~al.} 2016, \apj,
  828, 46, \dodoi{10.3847/0004-637X/828/1/46}

\bibitem[{{Astropy Collaboration} {et~al.}(2013){Astropy Collaboration},
  {Robitaille}, {Tollerud}, {Greenfield}, {Droettboom}, {Bray}, {Aldcroft},
  {Davis}, {Ginsburg}, {Price-Whelan}, {Kerzendorf}, {Conley}, {Crighton},
  {Barbary}, {Muna}, {Ferguson}, {Grollier}, {Parikh}, {Nair}, {Unther},
  {Deil}, {Woillez}, {Conseil}, {Kramer}, {Turner}, {Singer}, {Fox}, {Weaver},
  {Zabalza}, {Edwards}, {Azalee Bostroem}, {Burke}, {Casey}, {Crawford},
  {Dencheva}, {Ely}, {Jenness}, {Labrie}, {Lim}, {Pierfederici}, {Pontzen},
  {Ptak}, {Refsdal}, {Servillat}, \& {Streicher}}]{astropy}
{Astropy Collaboration}, {Robitaille}, T.~P., {Tollerud}, E.~J., {et~al.} 2013,
  \aap, 558, A33, \dodoi{10.1051/0004-6361/201322068}

\bibitem[{{Avenhaus} {et~al.}(2018){Avenhaus}, {Quanz}, {Garufi}, {Perez},
  {Casassus}, {Pinte}, {Bertrang}, {Caceres}, {Benisty}, \&
  {Dominik}}]{Avenhaus18_sphere}
{Avenhaus}, H., {Quanz}, S.~P., {Garufi}, A., {et~al.} 2018, \apj, 863, 44,
  \dodoi{10.3847/1538-4357/aab846}

\bibitem[{{Bai}(2016)}]{Bai16}
{Bai}, X.-N. 2016, \apj, 821, 80, \dodoi{10.3847/0004-637X/821/2/80}

\bibitem[{{Bai} \& {Stone}(2013)}]{bai13}
{Bai}, X.-N., \& {Stone}, J.~M. 2013, \apj, 769, 76,
  \dodoi{10.1088/0004-637X/769/1/76}

\bibitem[{{Balbus} \& {Hawley}(1991)}]{Balbus91}
{Balbus}, S.~A., \& {Hawley}, J.~F. 1991, \apj, 376, 214,
  \dodoi{10.1086/170270}

\bibitem[{{Banzatti} {et~al.}(2015){Banzatti}, {Pinilla}, {Ricci},
  {Pontoppidan}, {Birnstiel}, \& {Ciesla}}]{banzatti15}
{Banzatti}, A., {Pinilla}, P., {Ricci}, L., {et~al.} 2015, \apj, 815, L15,
  \dodoi{10.1088/2041-8205/815/1/L15}

\bibitem[{{Bergin} {et~al.}(2007){Bergin}, {Aikawa}, {Blake}, \& {van
  Dishoeck}}]{Bergin07}
{Bergin}, E.~A., {Aikawa}, Y., {Blake}, G.~A., \& {van Dishoeck}, E.~F. 2007,
  in Protostars and Planets V, ed. B.~{Reipurth}, D.~{Jewitt}, \& K.~{Keil},
  751.
\newblock \doarXiv{astro-ph/0603358}

\bibitem[{{Bergin} {et~al.}(2014){Bergin}, {Cleeves}, {Crockett}, \&
  {Blake}}]{bergin14}
{Bergin}, E.~A., {Cleeves}, L.~I., {Crockett}, N., \& {Blake}, G.~A. 2014,
  Faraday Discussions, 168, 61, \dodoi{10.1039/C4FD00003J}

\bibitem[{{Bergin} \& {Williams}(2017)}]{bergin17}
{Bergin}, E.~A., \& {Williams}, J.~P. 2017, {The Determination of
  Protoplanetary Disk Masses}, Vol. 445 (Astrophysics and Space Science
  Library), 1, \dodoi{10.1007/978-3-319-60609-5\_1}

\bibitem[{{Bergin} {et~al.}(2013){Bergin}, {Cleeves}, {Gorti}, {Zhang},
  {Blake}, {Green}, {Andrews}, {Evans}, {Henning}, {{\"O}berg}, {Pontoppidan},
  {Qi}, {Salyk}, \& {van Dishoeck}}]{bergin13}
{Bergin}, E.~A., {Cleeves}, L.~I., {Gorti}, U., {et~al.} 2013, \nat, 493, 644,
  \dodoi{10.1038/nature11805}

\bibitem[{{Bergner} {et~al.}(2019){Bergner}, {{\"O}berg}, {Bergin}, {Loomis},
  {Pegues}, \& {Qi}}]{Bergner2019}
{Bergner}, J.~B., {{\"O}berg}, K.~I., {Bergin}, E.~A., {et~al.} 2019, \apj,
  876, 25, \dodoi{10.3847/1538-4357/ab141e}

\bibitem[{{Bergner} {et~al.}(2020){Bergner}, {{\"O}berg}, {Bergin}, {Andrews},
  {Blake}, {Carpenter}, {Cleeves}, {Guzm{\'a}n}, {Huang}, {J{\o}rgensen}, {Qi},
  {Schwarz}, {Williams}, \& {Wilner}}]{Bergner_20evolution}
---. 2020, \apj, 898, 97, \dodoi{10.3847/1538-4357/ab9e71}

\bibitem[{{Bergner} {et~al.}(2021){Bergner}, {Oberg}, {Guzman}, {Law},
  {Loomis}, {Cataldi}, {Bosman}, {Aikawa}, {Andrews}, {Bergin}, {Booth},
  {Cleeves}, {Czekala}, {Huang}, {Ilee}, {Le Gal}, {Long}, {Nomura}, {Menard},
  {Qi}, {Schwarz}, {Teague}, {Tsukagoshi}, {Walsh}, {Wilner}, \&
  {Yamato}}]{bergner20}
{Bergner}, J.~B., {Oberg}, K.~I., {Guzman}, V.~V., {et~al.} 2021, arXiv
  e-prints, arXiv:2109.06694.
\newblock \doarXiv{2109.06694}

\bibitem[{{Birnstiel} {et~al.}(2018){Birnstiel}, {Dullemond}, {Zhu}, {Andrews},
  {Bai}, {Wilner}, {Carpenter}, {Huang}, {Isella}, {Benisty}, {P{\'e}rez}, \&
  {Zhang}}]{Birnstiel18}
{Birnstiel}, T., {Dullemond}, C.~P., {Zhu}, Z., {et~al.} 2018, \apjl, 869, L45,
  \dodoi{10.3847/2041-8213/aaf743}

\bibitem[{{Booth} {et~al.}(2019){Booth}, {Walsh}, {Ilee}, {Notsu}, {Qi},
  {Nomura}, \& {Akiyama}}]{Booth19}
{Booth}, A.~S., {Walsh}, C., {Ilee}, J.~D., {et~al.} 2019, \apjl, 882, L31,
  \dodoi{10.3847/2041-8213/ab3645}

\bibitem[{{Booth} {et~al.}(2017){Booth}, {Clarke}, {Madhusudhan}, \&
  {Ilee}}]{booth17}
{Booth}, R.~A., {Clarke}, C.~J., {Madhusudhan}, N., \& {Ilee}, J.~D. 2017,
  \mnras, 469, 3994, \dodoi{10.1093/mnras/stx1103}

\bibitem[{{Booth} \& {Ilee}(2019)}]{Booth19_pebble}
{Booth}, R.~A., \& {Ilee}, J.~D. 2019, \mnras, 487, 3998,
  \dodoi{10.1093/mnras/stz1488}

\bibitem[{{Bosman} {et~al.}(2018){Bosman}, {Walsh}, \& {van
  Dishoeck}}]{Bosman18}
{Bosman}, A.~D., {Walsh}, C., \& {van Dishoeck}, E.~F. 2018, \aap, 618, A182,
  \dodoi{10.1051/0004-6361/201833497}

\bibitem[{{Calahan} {et~al.}(2021{\natexlab{a}}){Calahan}, {Bergin}, {Zhang},
  {Teague}, {Cleeves}, {Bergner}, {Blake}, {Cazzoletti}, {Guzm{\'a}n},
  {Hogerheijde}, {Huang}, {Kama}, {Loomis}, {{\"O}berg}, {Qi}, {van Dishoeck},
  {Terwisscha van Scheltinga}, {Walsh}, \& {Wilner}}]{Calahan21_twhya}
{Calahan}, J.~K., {Bergin}, E., {Zhang}, K., {et~al.} 2021{\natexlab{a}}, \apj,
  908, 8, \dodoi{10.3847/1538-4357/abd255}

\bibitem[{{Calahan} {et~al.}(2021{\natexlab{b}}){Calahan}, {Bergin}, {Zhang},
  {Schwarz}, {Oberg}, {Guzman}, {Walsh}, {Aikawa}, {Alarcon}, {Andrews}, {Bae},
  {Bergner}, {Booth}, {Bosman}, {Cataldi}, {Czekala}, {Huang}, {Ilee}, {Law},
  {Le Gal}, {Long}, {Loomis}, {Menard}, {Nomura}, {Qi}, {Teague}, {van'T Hoff},
  {Wilner}, \& {Yamato}}]{Calahan20b}
{Calahan}, J.~K., {Bergin}, E.~A., {Zhang}, K., {et~al.} 2021{\natexlab{b}},
  arXiv e-prints, arXiv:2109.06202.
\newblock \doarXiv{2109.06202}

\bibitem[{{Cieza} {et~al.}(2021){Cieza}, {Gonz{\'a}lez-Ruilova}, {Hales},
  {Pinilla}, {Ru{\'\i}z-Rodr{\'\i}guez}, {Zurlo}, {Casassus}, {P{\'e}rez},
  {C{\'a}novas}, {Arce-Tord}, {Flock}, {Kurtovic}, {Marino}, {Nogueira},
  {Perez}, {Price}, {Principe}, \& {Williams}}]{Cieza21_ODISEAIII}
{Cieza}, L.~A., {Gonz{\'a}lez-Ruilova}, C., {Hales}, A.~S., {et~al.} 2021,
  \mnras, 501, 2934, \dodoi{10.1093/mnras/staa3787}

\bibitem[{{Cleeves} {et~al.}(2015){Cleeves}, {Bergin}, {Qi}, {Adams}, \&
  {{\"O}berg}}]{cleeves15}
{Cleeves}, L.~I., {Bergin}, E.~A., {Qi}, C., {Adams}, F.~C., \& {{\"O}berg},
  K.~I. 2015, \apj, 799, 204, \dodoi{10.1088/0004-637X/799/2/204}

\bibitem[{{Cleeves} {et~al.}(2016){Cleeves}, {{\"O}berg}, {Wilner}, {Huang},
  {Loomis}, {Andrews}, \& {Czekala}}]{cleeves16}
{Cleeves}, L.~I., {{\"O}berg}, K.~I., {Wilner}, D.~J., {et~al.} 2016, \apj,
  832, 110, \dodoi{10.3847/0004-637X/832/2/110}

\bibitem[{{Cleeves} {et~al.}(2018){Cleeves}, {{\"O}berg}, {Wilner}, {Huang},
  {Loomis}, {Andrews}, \& {Guzman}}]{Cleeves18}
---. 2018, \apj, 865, 155, \dodoi{10.3847/1538-4357/aade96}

\bibitem[{{Czekala} {et~al.}(2021){Czekala}, {Loomis}, {Teague}, {Booth},
  {Huang}, {Cataldi}, {Ilee}, {Law}, {Walsh}, {Bosman}, {Guzm{\'a}n}, {Le Gal},
  {{\"O}berg}, {Yamato}, {Aikawa}, {Andrews}, {Bae}, {Bergin}, {Bergner},
  {Cleeves}, {Kurtovic}, {M{\'e}nard}, {Nomura}, {P{\'e}rez}, {Qi}, {Schwarz},
  {Tsukagoshi}, {Waggoner}, {Wilner}, \& {Zhang}}]{Czekala20_maps}
{Czekala}, I., {Loomis}, R.~A., {Teague}, R., {et~al.} 2021, arXiv e-prints,
  arXiv:2109.06188.
\newblock \doarXiv{2109.06188}

\bibitem[{{Dionatos} {et~al.}(2019){Dionatos}, {Woitke}, {G{\"u}del},
  {Degroote}, {Liebhart}, {Anthonioz}, {Antonellini}, {Baldovin-Saavedra},
  {Carmona}, {Dominik}, {Greaves}, {Ilee}, {Kamp}, {M{\'e}nard}, {Min},
  {Pinte}, {Rab}, {Rigon}, {Thi}, \& {Waters}}]{Dionatos19}
{Dionatos}, O., {Woitke}, P., {G{\"u}del}, M., {et~al.} 2019, \aap, 625, A66,
  \dodoi{10.1051/0004-6361/201832860}

\bibitem[{{Dong} {et~al.}(2015){Dong}, {Zhu}, \& {Whitney}}]{dong15}
{Dong}, R., {Zhu}, Z., \& {Whitney}, B. 2015, \apj, 809, 93,
  \dodoi{10.1088/0004-637X/809/1/93}

\bibitem[{{Du} \& {Bergin}(2014)}]{du14}
{Du}, F., \& {Bergin}, E.~A. 2014, \apj, 792, 2,
  \dodoi{10.1088/0004-637X/792/1/2}

\bibitem[{{Du} {et~al.}(2017){Du}, {Bergin}, {Hogerheijde}, {van Dishoeck},
  {Blake}, {Bruderer}, {Cleeves}, {Dominik}, {Fedele}, {Lis}, {Melnick},
  {Neufeld}, {Pearson}, \& {Y{\i}ld{\i}z}}]{du17}
{Du}, F., {Bergin}, E.~A., {Hogerheijde}, M., {et~al.} 2017, \apj, 842, 98,
  \dodoi{10.3847/1538-4357/aa70ee}

\bibitem[{{Dubrulle} {et~al.}(1995){Dubrulle}, {Morfill}, \&
  {Sterzik}}]{dubrulle95}
{Dubrulle}, B., {Morfill}, G., \& {Sterzik}, M. 1995, \icarus, 114, 237,
  \dodoi{10.1006/icar.1995.1058}

\bibitem[{{Dullemond} \& {Dominik}(2004)}]{dullemond04_settling}
{Dullemond}, C.~P., \& {Dominik}, C. 2004, \aap, 421, 1075,
  \dodoi{10.1051/0004-6361:20040284}

\bibitem[{{Dullemond} {et~al.}(2020){Dullemond}, {Isella}, {Andrews},
  {Skobleva}, \& {Dzyurkevich}}]{Dullemond20}
{Dullemond}, C.~P., {Isella}, A., {Andrews}, S.~M., {Skobleva}, I., \&
  {Dzyurkevich}, N. 2020, \aap, 633, A137, \dodoi{10.1051/0004-6361/201936438}

\bibitem[{{Dullemond} {et~al.}(2012){Dullemond}, {Juhasz}, {Pohl}, {Sereshti},
  {Shetty}, {Peters}, {Commercon}, \& {Flock}}]{radmc3d}
{Dullemond}, C.~P., {Juhasz}, A., {Pohl}, A., {et~al.} 2012, {RADMC-3D: A
  multi-purpose radiative transfer tool}, Astrophysics Source Code Library.
\newblock \doeprint{1202.015}

\bibitem[{{Eistrup} {et~al.}(2016){Eistrup}, {Walsh}, \& {van
  Dishoeck}}]{Eistrup16}
{Eistrup}, C., {Walsh}, C., \& {van Dishoeck}, E.~F. 2016, \aap, 595, A83,
  \dodoi{10.1051/0004-6361/201628509}

\bibitem[{{Espaillat} {et~al.}(2010){Espaillat}, {D'Alessio}, {Hern{\'a}ndez},
  {Nagel}, {Luhman}, {Watson}, {Calvet}, {Muzerolle}, \&
  {McClure}}]{Espaillat10}
{Espaillat}, C., {D'Alessio}, P., {Hern{\'a}ndez}, J., {et~al.} 2010, \apj,
  717, 441, \dodoi{10.1088/0004-637X/717/1/441}

\bibitem[{{Facchini} {et~al.}(2018){Facchini}, {Pinilla}, {van Dishoeck}, \&
  {de Juan Ovelar}}]{Facchini18}
{Facchini}, S., {Pinilla}, P., {van Dishoeck}, E.~F., \& {de Juan Ovelar}, M.
  2018, \aap, 612, A104, \dodoi{10.1051/0004-6361/201731390}

\bibitem[{{Fairlamb} {et~al.}(2015){Fairlamb}, {Oudmaijer}, {Mendigut{\'\i}a},
  {Ilee}, \& {van den Ancker}}]{Fairlamb15}
{Fairlamb}, J.~R., {Oudmaijer}, R.~D., {Mendigut{\'\i}a}, I., {Ilee}, J.~D., \&
  {van den Ancker}, M.~E. 2015, \mnras, 453, 976, \dodoi{10.1093/mnras/stv1576}

\bibitem[{{Favre} {et~al.}(2013){Favre}, {Cleeves}, {Bergin}, {Qi}, \&
  {Blake}}]{Favre13}
{Favre}, C., {Cleeves}, L.~I., {Bergin}, E.~A., {Qi}, C., \& {Blake}, G.~A.
  2013, \apjl, 776, L38, \dodoi{10.1088/2041-8205/776/2/L38}

\bibitem[{{Favre} {et~al.}(2019){Favre}, {Fedele}, {Maud}, {Booth}, {Tazzari},
  {Miotello}, {Testi}, {Semenov}, \& {Bruderer}}]{Favre19}
{Favre}, C., {Fedele}, D., {Maud}, L., {et~al.} 2019, \apj, 871, 107,
  \dodoi{10.3847/1538-4357/aaf80c}

\bibitem[{{Fedele} {et~al.}(2017){Fedele}, {Carney}, {Hogerheijde}, {Walsh},
  {Miotello}, {Klaassen}, {Bruderer}, {Henning}, \& {van Dishoeck}}]{Fedele17}
{Fedele}, D., {Carney}, M., {Hogerheijde}, M.~R., {et~al.} 2017, \aap, 600,
  A72, \dodoi{10.1051/0004-6361/201629860}

\bibitem[{{Fedele} {et~al.}(2018){Fedele}, {Tazzari}, {Booth}, {Testi},
  {Clarke}, {Pascucci}, {Kospal}, {Semenov}, {Bruderer}, {Henning}, \&
  {Teague}}]{Fedele18}
{Fedele}, D., {Tazzari}, M., {Booth}, R., {et~al.} 2018, \aap, 610, A24,
  \dodoi{10.1051/0004-6361/201731978}

\bibitem[{{Flaherty} {et~al.}(2020){Flaherty}, {Hughes}, {Simon}, {Qi}, {Bai},
  {Bulatek}, {Andrews}, {Wilner}, \& {K{\'o}sp{\'a}l}}]{Flaherty20}
{Flaherty}, K., {Hughes}, A.~M., {Simon}, J.~B., {et~al.} 2020, \apj, 895, 109,
  \dodoi{10.3847/1538-4357/ab8cc5}

\bibitem[{{Flaherty} {et~al.}(2015){Flaherty}, {Hughes}, {Rosenfeld},
  {Andrews}, {Chiang}, {Simon}, {Kerzner}, \& {Wilner}}]{flaherty15}
{Flaherty}, K.~M., {Hughes}, A.~M., {Rosenfeld}, K.~A., {et~al.} 2015, \apj,
  813, 99, \dodoi{10.1088/0004-637X/813/2/99}

\bibitem[{{Flaherty} {et~al.}(2017){Flaherty}, {Hughes}, {Rose}, {Simon}, {Qi},
  {Andrews}, {K{\'o}sp{\'a}l}, {Wilner}, {Chiang}, {Armitage}, \&
  {Bai}}]{flaherty17}
{Flaherty}, K.~M., {Hughes}, A.~M., {Rose}, S.~C., {et~al.} 2017, \apj, 843,
  150, \dodoi{10.3847/1538-4357/aa79f9}

\bibitem[{{Flock} {et~al.}(2015){Flock}, {Ruge}, {Dzyurkevich}, {Henning},
  {Klahr}, \& {Wolf}}]{Flock15}
{Flock}, M., {Ruge}, J.~P., {Dzyurkevich}, N., {et~al.} 2015, \aap, 574, A68,
  \dodoi{10.1051/0004-6361/201424693}

\bibitem[{{Foreman-Mackey} {et~al.}(2013){Foreman-Mackey}, {Hogg}, {Lang}, \&
  {Goodman}}]{emcee}
{Foreman-Mackey}, D., {Hogg}, D.~W., {Lang}, D., \& {Goodman}, J. 2013, \pasp,
  125, 306, \dodoi{10.1086/670067}

\bibitem[{{Gaia Collaboration} {et~al.}(2018){Gaia Collaboration}, {Brown, A.
  G. A.}, {Vallenari, A.}, {Prusti, T.}, {de Bruijne, J. H. J.}, {Babusiaux,
  C.}, {Bailer-Jones, C. A. L.}, {Biermann, M.}, {Evans, D. W.}, {Eyer, L.},
  {Jansen, F.}, {Jordi, C.}, {Klioner, S. A.}, {Lammers, U.}, {Lindegren, L.},
  {Luri, X.}, {Mignard, F.}, {Panem, C.}, {Pourbaix, D.}, {Randich, S.},
  {Sartoretti, P.}, {Siddiqui, H. I.}, {Soubiran, C.}, {van Leeuwen, F.},
  {Walton, N. A.}, {Arenou, F.}, {Bastian, U.}, {Cropper, M.}, {Drimmel, R.},
  {Katz, D.}, {Lattanzi, M. G.}, {Bakker, J.}, {Cacciari, C.}, {Casta\~neda,
  J.}, {Chaoul, L.}, {Cheek, N.}, {De Angeli, F.}, {Fabricius, C.}, {Guerra,
  R.}, {Holl, B.}, {Masana, E.}, {Messineo, R.}, {Mowlavi, N.}, {Nienartowicz,
  K.}, {Panuzzo, P.}, {Portell, J.}, {Riello, M.}, {Seabroke, G. M.}, {Tanga,
  P.}, {Th\'evenin, F.}, {Gracia-Abril, G.}, {Comoretto, G.},
  {Garcia-Reinaldos, M.}, {Teyssier, D.}, {Altmann, M.}, {Andrae, R.}, {Audard,
  M.}, {Bellas-Velidis, I.}, {Benson, K.}, {Berthier, J.}, {Blomme, R.},
  {Burgess, P.}, {Busso, G.}, {Carry, B.}, {Cellino, A.}, {Clementini, G.},
  {Clotet, M.}, {Creevey, O.}, {Davidson, M.}, {De Ridder, J.}, {Delchambre,
  L.}, {Dell\'{}Oro, A.}, {Ducourant, C.}, {Fern\'andez-Hern\'andez, J.},
  {Fouesneau, M.}, {Fr\'emat, Y.}, {Galluccio, L.}, {Garc\'{\i}a-Torres, M.},
  {Gonz\'alez-N\'u\~nez, J.}, {Gonz\'alez-Vidal, J. J.}, {Gosset, E.}, {Guy, L.
  P.}, {Halbwachs, J.-L.}, {Hambly, N. C.}, {Harrison, D. L.}, {Hern\'andez,
  J.}, {Hestroffer, D.}, {Hodgkin, S. T.}, {Hutton, A.}, {Jasniewicz, G.},
  {Jean-Antoine-Piccolo, A.}, {Jordan, S.}, {Korn, A. J.}, {Krone-Martins, A.},
  {Lanzafame, A. C.}, {Lebzelter, T.}, {L\"offler, W.}, {Manteiga, M.},
  {Marrese, P. M.}, {Mart\'{\i}n-Fleitas, J. M.}, {Moitinho, A.}, {Mora, A.},
  {Muinonen, K.}, {Osinde, J.}, {Pancino, E.}, {Pauwels, T.}, {Petit, J.-M.},
  {Recio-Blanco, A.}, {Richards, P. J.}, {Rimoldini, L.}, {Robin, A. C.},
  {Sarro, L. M.}, {Siopis, C.}, {Smith, M.}, {Sozzetti, A.}, {S\"uveges, M.},
  {Torra, J.}, {van Reeven, W.}, {Abbas, U.}, {Abreu Aramburu, A.}, {Accart,
  S.}, {Aerts, C.}, {Altavilla, G.}, {\'Alvarez, M. A.}, {Alvarez, R.}, {Alves,
  J.}, {Anderson, R. I.}, {Andrei, A. H.}, {Anglada Varela, E.}, {Antiche, E.},
  {Antoja, T.}, {Arcay, B.}, {Astraatmadja, T. L.}, {Bach, N.}, {Baker, S. G.},
  {Balaguer-N\'u\~nez, L.}, {Balm, P.}, {Barache, C.}, {Barata, C.}, {Barbato,
  D.}, {Barblan, F.}, {Barklem, P. S.}, {Barrado, D.}, {Barros, M.}, {Barstow,
  M. A.}, {Bartholom\'e Mu\~noz, S.}, {Bassilana, J.-L.}, {Becciani, U.},
  {Bellazzini, M.}, {Berihuete, A.}, {Bertone, S.}, {Bianchi, L.}, {Bienaym\'e,
  O.}, {Blanco-Cuaresma, S.}, {Boch, T.}, {Boeche, C.}, {Bombrun, A.},
  {Borrachero, R.}, {Bossini, D.}, {Bouquillon, S.}, {Bourda, G.}, {Bragaglia,
  A.}, {Bramante, L.}, {Breddels, M. A.}, {Bressan, A.}, {Brouillet, N.},
  {Br\"usemeister, T.}, {Brugaletta, E.}, {Bucciarelli, B.}, {Burlacu, A.},
  {Busonero, D.}, {Butkevich, A. G.}, {Buzzi, R.}, {Caffau, E.}, {Cancelliere,
  R.}, {Cannizzaro, G.}, {Cantat-Gaudin, T.}, {Carballo, R.}, {Carlucci, T.},
  {Carrasco, J. M.}, {Casamiquela, L.}, {Castellani, M.}, {Castro-Ginard, A.},
  {Charlot, P.}, {Chemin, L.}, {Chiavassa, A.}, {Cocozza, G.}, {Costigan, G.},
  {Cowell, S.}, {Crifo, F.}, {Crosta, M.}, {Crowley, C.}, {Cuypers+, J.},
  {Dafonte, C.}, {Damerdji, Y.}, {Dapergolas, A.}, {David, P.}, {David, M.},
  {de Laverny, P.}, {De Luise, F.}, {De March, R.}, {de Martino, D.}, {de
  Souza, R.}, {de Torres, A.}, {Debosscher, J.}, {del Pozo, E.}, {Delbo, M.},
  {Delgado, A.}, {Delgado, H. E.}, {Di Matteo, P.}, {Diakite, S.}, {Diener,
  C.}, {Distefano, E.}, {Dolding, C.}, {Drazinos, P.}, {Dur\'an, J.},
  {Edvardsson, B.}, {Enke, H.}, {Eriksson, K.}, {Esquej, P.}, {Eynard Bontemps,
  G.}, {Fabre, C.}, {Fabrizio, M.}, {Faigler, S.}, {Falc\~ao, A. J.}, {Farr\`as
  Casas, M.}, {Federici, L.}, {Fedorets, G.}, {Fernique, P.}, {Figueras, F.},
  {Filippi, F.}, {Findeisen, K.}, {Fonti, A.}, {Fraile, E.}, {Fraser, M.},
  {Fr\'ezouls, B.}, {Gai, M.}, {Galleti, S.}, {Garabato, D.},
  {Garc\'{\i}a-Sedano, F.}, {Garofalo, A.}, {Garralda, N.}, {Gavel, A.},
  {Gavras, P.}, {Gerssen, J.}, {Geyer, R.}, {Giacobbe, P.}, {Gilmore, G.},
  {Girona, S.}, {Giuffrida, G.}, {Glass, F.}, {Gomes, M.}, {Granvik, M.},
  {Gueguen, A.}, {Guerrier, A.}, {Guiraud, J.}, {Guti\'errez-S\'anchez, R.},
  {Haigron, R.}, {Hatzidimitriou, D.}, {Hauser, M.}, {Haywood, M.}, {Heiter,
  U.}, {Helmi, A.}, {Heu, J.}, {Hilger, T.}, {Hobbs, D.}, {Hofmann, W.},
  {Holland, G.}, {Huckle, H. E.}, {Hypki, A.}, {Icardi, V.}, {Jan\ss{}en, K.},
  {Jevardat de Fombelle, G.}, {Jonker, P. G.}, {Juh\'asz, \'A. L.}, {Julbe,
  F.}, {Karampelas, A.}, {Kewley, A.}, {Klar, J.}, {Kochoska, A.}, {Kohley,
  R.}, {Kolenberg, K.}, {Kontizas, M.}, {Kontizas, E.}, {Koposov, S. E.},
  {Kordopatis, G.}, {Kostrzewa-Rutkowska, Z.}, {Koubsky, P.}, {Lambert, S.},
  {Lanza, A. F.}, {Lasne, Y.}, {Lavigne, J.-B.}, {Le Fustec, Y.}, {Le
  Poncin-Lafitte, C.}, {Lebreton, Y.}, {Leccia, S.}, {Leclerc, N.},
  {Lecoeur-Taibi, I.}, {Lenhardt, H.}, {Leroux, F.}, {Liao, S.}, {Licata, E.},
  {Lindstr\o{}m, H. E. P.}, {Lister, T. A.}, {Livanou, E.}, {Lobel, A.},
  {L\'opez, M.}, {Managau, S.}, {Mann, R. G.}, {Mantelet, G.}, {Marchal, O.},
  {Marchant, J. M.}, {Marconi, M.}, {Marinoni, S.}, {Marschalk\'o, G.},
  {Marshall, D. J.}, {Martino, M.}, {Marton, G.}, {Mary, N.}, {Massari, D.},
  {Matijevic, G.}, {Mazeh, T.}, {McMillan, P. J.}, {Messina, S.}, {Michalik,
  D.}, {Millar, N. R.}, {Molina, D.}, {Molinaro, R.}, {Moln\'ar, L.},
  {Montegriffo, P.}, {Mor, R.}, {Morbidelli, R.}, {Morel, T.}, {Morris, D.},
  {Mulone, A. F.}, {Muraveva, T.}, {Musella, I.}, {Nelemans, G.}, {Nicastro,
  L.}, {Noval, L.}, {O\'{}Mullane, W.}, {Ord\'enovic, C.}, {Ord\'o\~nez-Blanco,
  D.}, {Osborne, P.}, {Pagani, C.}, {Pagano, I.}, {Pailler, F.}, {Palacin, H.},
  {Palaversa, L.}, {Panahi, A.}, {Pawlak, M.}, {Piersimoni, A. M.}, {Pineau,
  F.-X.}, {Plachy, E.}, {Plum, G.}, {Poggio, E.}, {Poujoulet, E.}, {Prsa, A.},
  {Pulone, L.}, {Racero, E.}, {Ragaini, S.}, {Rambaux, N.}, {Ramos-Lerate, M.},
  {Regibo, S.}, {Reyl\'e, C.}, {Riclet, F.}, {Ripepi, V.}, {Riva, A.}, {Rivard,
  A.}, {Rixon, G.}, {Roegiers, T.}, {Roelens, M.}, {Romero-G\'omez, M.},
  {Rowell, N.}, {Royer, F.}, {Ruiz-Dern, L.}, {Sadowski, G.}, {Sagrist\`a
  Sell\'es, T.}, {Sahlmann, J.}, {Salgado, J.}, {Salguero, E.}, {Sanna, N.},
  {Santana-Ros, T.}, {Sarasso, M.}, {Savietto, H.}, {Schultheis, M.}, {Sciacca,
  E.}, {Segol, M.}, {Segovia, J. C.}, {S\'egransan, D.}, {Shih, I-C.},
  {Siltala, L.}, {Silva, A. F.}, {Smart, R. L.}, {Smith, K. W.}, {Solano, E.},
  {Solitro, F.}, {Sordo, R.}, {Soria Nieto, S.}, {Souchay, J.}, {Spagna, A.},
  {Spoto, F.}, {Stampa, U.}, {Steele, I. A.}, {Steidelm\"uller, H.},
  {Stephenson, C. A.}, {Stoev, H.}, {Suess, F. F.}, {Surdej, J.}, {Szabados,
  L.}, {Szegedi-Elek, E.}, {Tapiador, D.}, {Taris, F.}, {Tauran, G.}, {Taylor,
  M. B.}, {Teixeira, R.}, {Terrett, D.}, {Teyssandier, P.}, {Thuillot, W.},
  {Titarenko, A.}, {Torra Clotet, F.}, {Turon, C.}, {Ulla, A.}, {Utrilla, E.},
  {Uzzi, S.}, {Vaillant, M.}, {Valentini, G.}, {Valette, V.}, {van Elteren,
  A.}, {Van Hemelryck, E.}, {van Leeuwen, M.}, {Vaschetto, M.}, {Vecchiato,
  A.}, {Veljanoski, J.}, {Viala, Y.}, {Vicente, D.}, {Vogt, S.}, {von Essen,
  C.}, {Voss, H.}, {Votruba, V.}, {Voutsinas, S.}, {Walmsley, G.}, {Weiler,
  M.}, {Wertz, O.}, {Wevers, T.}, {Wyrzykowski, L.}, {Yoldas, A.}, {Zerjal,
  M.}, {Ziaeepour, H.}, {Zorec, J.}, {Zschocke, S.}, {Zucker, S.}, {Zurbach,
  C.}, \& {Zwitter, T.}}]{gaia2}
{Gaia Collaboration}, {Brown, A. G. A.}, {Vallenari, A.}, {et~al.} 2018, A\&A,
  616, A1, \dodoi{10.1051/0004-6361/201833051}

\bibitem[{{Guzm{\'a}n} {et~al.}(2021){Guzm{\'a}n}, {Bergner}, {Law}, {Oberg},
  {Walsh}, {Cataldi}, {Aikawa}, {Bergin}, {Czekala}, {Huang}, {Andrews},
  {Loomis}, {Zhang}, {Le Gal}, {Alarc{\'o}n}, {Ilee}, {Teague}, {Cleeves},
  {Wilner}, {Long}, {Schwarz}, {Bosman}, {P{\'e}rez}, {M{\'e}nard}, \&
  {Liu}}]{Guzman20}
{Guzm{\'a}n}, V.~V., {Bergner}, J.~B., {Law}, C.~J., {et~al.} 2021, arXiv
  e-prints, arXiv:2109.06391.
\newblock \doarXiv{2109.06391}

\bibitem[{{Hartmann} {et~al.}(2016){Hartmann}, {Herczeg}, \&
  {Calvet}}]{Hartmann16_ARAA}
{Hartmann}, L., {Herczeg}, G., \& {Calvet}, N. 2016, \araa, 54, 135,
  \dodoi{10.1146/annurev-astro-081915-023347}

\bibitem[{{Hauschildt} {et~al.}(1999){Hauschildt}, {Allard}, {Ferguson},
  {Baron}, \& {Alexander}}]{Nextgen_giant}
{Hauschildt}, P.~H., {Allard}, F., {Ferguson}, J., {Baron}, E., \& {Alexander},
  D.~R. 1999, \apj, 525, 871, \dodoi{10.1086/307954}

\bibitem[{{Hendler} {et~al.}(2020){Hendler}, {Pascucci}, {Pinilla}, {Tazzari},
  {Carpenter}, {Malhotra}, \& {Testi}}]{Hendler20}
{Hendler}, N., {Pascucci}, I., {Pinilla}, P., {et~al.} 2020, \apj, 895, 126,
  \dodoi{10.3847/1538-4357/ab70ba}

\bibitem[{{Herczeg} {et~al.}(2002){Herczeg}, {Linsky}, {Valenti},
  {Johns-Krull}, \& {Wood}}]{Herczeg02}
{Herczeg}, G.~J., {Linsky}, J.~L., {Valenti}, J.~A., {Johns-Krull}, C.~M., \&
  {Wood}, B.~E. 2002, \apj, 572, 310, \dodoi{10.1086/339731}

\bibitem[{{Herczeg} {et~al.}(2004){Herczeg}, {Wood}, {Linsky}, {Valenti}, \&
  {Johns-Krull}}]{Herczeg04}
{Herczeg}, G.~J., {Wood}, B.~E., {Linsky}, J.~L., {Valenti}, J.~A., \&
  {Johns-Krull}, C.~M. 2004, \apj, 607, 369, \dodoi{10.1086/383340}

\bibitem[{{Hogerheijde} {et~al.}(2011){Hogerheijde}, {Bergin}, {Brinch},
  {Cleeves}, {Fogel}, {Blake}, {Dominik}, {Lis}, {Melnick}, {Neufeld},
  {Pani{\'c}}, {Pearson}, {Kristensen}, {Y{\i}ld{\i}z}, \& {van
  Dishoeck}}]{hogerheijde11_water}
{Hogerheijde}, M.~R., {Bergin}, E.~A., {Brinch}, C., {et~al.} 2011, Science,
  334, 338, \dodoi{10.1126/science.1208931}

\bibitem[{{Huang} {et~al.}(2017){Huang}, {{\"O}berg}, {Qi}, {Aikawa},
  {Andrews}, {Furuya}, {Guzm{\'a}n}, {Loomis}, {van Dishoeck}, \&
  {Wilner}}]{Huang17}
{Huang}, J., {{\"O}berg}, K.~I., {Qi}, C., {et~al.} 2017, \apj, 835, 231,
  \dodoi{10.3847/1538-4357/835/2/231}

\bibitem[{{Huang} {et~al.}(2018){Huang}, {Andrews}, {Dullemond}, {Isella},
  {P{\'e}rez}, {Guzm{\'a}n}, {{\"O}berg}, {Zhu}, {Zhang}, {Bai}, {Benisty},
  {Birnstiel}, {Carpenter}, {Hughes}, {Ricci}, {Weaver}, \&
  {Wilner}}]{Huang18b}
{Huang}, J., {Andrews}, S.~M., {Dullemond}, C.~P., {et~al.} 2018, \apj, 869,
  L42, \dodoi{10.3847/2041-8213/aaf740}

\bibitem[{{Huang} {et~al.}(2020){Huang}, {Andrews}, {Dullemond}, {{\"O}berg},
  {Qi}, {Zhu}, {Birnstiel}, {Carpenter}, {Isella}, {Mac{\'\i}as}, {McClure},
  {P{\'e}rez}, {Teague}, {Wilner}, \& {Zhang}}]{Huang20}
---. 2020, \apj, 891, 48, \dodoi{10.3847/1538-4357/ab711e}

\bibitem[{{Ingleby} {et~al.}(2015){Ingleby}, {Espaillat}, {Calvet}, {Sitko},
  {Russell}, \& {Champney}}]{Ingleby15}
{Ingleby}, L., {Espaillat}, C., {Calvet}, N., {et~al.} 2015, \apj, 805, 149,
  \dodoi{10.1088/0004-637X/805/2/149}

\bibitem[{{Isella} {et~al.}(2016){Isella}, {Guidi}, {Testi}, {Liu}, {Li}, {Li},
  {Weaver}, {Boehler}, {Carperter}, {De Gregorio-Monsalvo}, {Manara}, {Natta},
  {P{\'e}rez}, {Ricci}, {Sargent}, {Tazzari}, \& {Turner}}]{Isella16}
{Isella}, A., {Guidi}, G., {Testi}, L., {et~al.} 2016, Physical Review Letters,
  117, 251101, \dodoi{10.1103/PhysRevLett.117.251101}

\bibitem[{{Isella} {et~al.}(2018){Isella}, {Huang}, {Andrews}, {Dullemond},
  {Birnstiel}, {Zhang}, {Zhu}, {Guzm{\'a}n}, {P{\'e}rez}, {Bai}, {Benisty},
  {Carpenter}, {Ricci}, \& {Wilner}}]{Isella18}
{Isella}, A., {Huang}, J., {Andrews}, S.~M., {et~al.} 2018, \apjl, 869, L49,
  \dodoi{10.3847/2041-8213/aaf747}

\bibitem[{{Kama} {et~al.}(2016){Kama}, {Bruderer}, {van Dishoeck},
  {Hogerheijde}, {Folsom}, {Miotello}, {Fedele}, {Belloche}, {G{\"u}sten}, \&
  {Wyrowski}}]{kama16}
{Kama}, M., {Bruderer}, S., {van Dishoeck}, E.~F., {et~al.} 2016, \aap, 592,
  A83, \dodoi{10.1051/0004-6361/201526991}

\bibitem[{{Kanagawa} {et~al.}(2015){Kanagawa}, {Muto}, {Tanaka}, {Tanigawa},
  {Takeuchi}, {Tsukagoshi}, \& {Momose}}]{Kanagawa15}
{Kanagawa}, K.~D., {Muto}, T., {Tanaka}, H., {et~al.} 2015, \apjl, 806, L15,
  \dodoi{10.1088/2041-8205/806/1/L15}

\bibitem[{Klapper {et~al.}(2003)Klapper, Surin, Lewen, Muller, Pak, \&
  Winnewisser}]{Klapper_2003}
Klapper, G., Surin, L., Lewen, F., {et~al.} 2003, The Astrophysical Journal,
  582, 262, \dodoi{10.1086/344615}

\bibitem[{{Krijt} {et~al.}(2020){Krijt}, {Bosman}, {Zhang}, {Schwarz},
  {Ciesla}, \& {Bergin}}]{Krijt20}
{Krijt}, S., {Bosman}, A.~D., {Zhang}, K., {et~al.} 2020, \apj, 899, 134,
  \dodoi{10.3847/1538-4357/aba75d}

\bibitem[{{Krijt} {et~al.}(2016){Krijt}, {Ciesla}, \& {Bergin}}]{krijt16}
{Krijt}, S., {Ciesla}, F.~J., \& {Bergin}, E.~A. 2016, \apj, 833, 285,
  \dodoi{10.3847/1538-4357/833/2/285}

\bibitem[{{Krijt} {et~al.}(2018){Krijt}, {Schwarz}, {Bergin}, \&
  {Ciesla}}]{Krijt18}
{Krijt}, S., {Schwarz}, K.~R., {Bergin}, E.~A., \& {Ciesla}, F.~J. 2018, \apj,
  864, 78, \dodoi{10.3847/1538-4357/aad69b}

\bibitem[{{Lacy} {et~al.}(1994){Lacy}, {Knacke}, {Geballe}, \&
  {Tokunaga}}]{Lacy94}
{Lacy}, J.~H., {Knacke}, R., {Geballe}, T.~R., \& {Tokunaga}, A.~T. 1994,
  \apjl, 428, L69, \dodoi{10.1086/187395}

\bibitem[{{Law} {et~al.}(2021{\natexlab{a}}){Law}, {Loomis}, {Teague},
  {{\"O}berg}, {Czekala}, {Andrews}, {Huang}, {Aikawa}, {Alarc{\'o}n}, {Bae},
  {Bergin}, {Bergner}, {Boehler}, {Booth}, {Bosman}, {Calahan}, {Cataldi},
  {Cleeves}, {Furuya}, {Guzm{\'a}n}, {Ilee}, {Le Gal}, {Liu}, {Long},
  {M{\'e}nard}, {Nomura}, {Qi}, {Schwarz}, {Sierra}, {Tsukagoshi}, {Yamato},
  {van't Hoff}, {Walsh}, {Wilner}, \& {Zhang}}]{law_maps_radial}
{Law}, C.~J., {Loomis}, R.~A., {Teague}, R., {et~al.} 2021{\natexlab{a}}, arXiv
  e-prints, arXiv:2109.06210.
\newblock \doarXiv{2109.06210}

\bibitem[{{Law} {et~al.}(2021{\natexlab{b}}){Law}, {Teague}, {Loomis}, {Bae},
  {{\"O}berg}, {Czekala}, {Andrews}, {Aikawa}, {Alarc{\'o}n}, {Bergin},
  {Bergner}, {Booth}, {Bosman}, {Calahan}, {Cataldi}, {Cleeves}, {Furuya},
  {Guzm{\'a}n}, {Huang}, {Ilee}, {Le Gal}, {Liu}, {Long}, {M{\'e}nard},
  {Nomura}, {P{\'e}rez}, {Qi}, {Schwarz}, {Soto}, {Tsukagoshi}, {Yamato},
  {van't Hoff}, {Walsh}, {Wilner}, \& {Zhang}}]{law20_maps_surface}
{Law}, C.~J., {Teague}, R., {Loomis}, R.~A., {et~al.} 2021{\natexlab{b}}, arXiv
  e-prints, arXiv:2109.06217.
\newblock \doarXiv{2109.06217}

\bibitem[{{Liu} {et~al.}(2019){Liu}, {Dipierro}, {Ragusa}, {Lodato}, {Herczeg},
  {Long}, {Harsono}, {Boehler}, {Menard}, {Johnstone}, {Pascucci}, {Pinilla},
  {Salyk}, {van der Plas}, {Cabrit}, {Fischer}, {Hendler}, {Manara}, {Nisini},
  {Rigliaco}, {Avenhaus}, {Banzatti}, \& {Gully-Santiago}}]{Liu19}
{Liu}, Y., {Dipierro}, G., {Ragusa}, E., {et~al.} 2019, \aap, 622, A75,
  \dodoi{10.1051/0004-6361/201834157}

\bibitem[{{Long} {et~al.}(2017){Long}, {Herczeg}, {Pascucci}, {Drabek-Maunder},
  {Mohanty}, {Testi}, {Apai}, {Hendler}, {Henning}, {Manara}, \&
  {Mulders}}]{long17}
{Long}, F., {Herczeg}, G.~J., {Pascucci}, I., {et~al.} 2017, \apj, 844, 99,
  \dodoi{10.3847/1538-4357/aa78fc}

\bibitem[{{Long} {et~al.}(2018){Long}, {Pinilla}, {Herczeg}, {Harsono},
  {Dipierro}, {Pascucci}, {Hendler}, {Tazzari}, {Ragusa}, {Salyk}, {Edwards},
  {Lodato}, {van de Plas}, {Johnstone}, {Liu}, {Boehler}, {Cabrit}, {Manara},
  {Menard}, {Mulders}, {Nisini}, {Fischer}, {Rigliaco}, {Banzatti}, {Avenhaus},
  \& {Gully-Santiago}}]{long18}
{Long}, F., {Pinilla}, P., {Herczeg}, G.~J., {et~al.} 2018, \apj, 869, 17,
  \dodoi{10.3847/1538-4357/aae8e1}

\bibitem[{{Loomis} {et~al.}(2018){Loomis}, {{\"O}berg}, {Andrews}, {Walsh},
  {Czekala}, {Huang}, \& {Rosenfeld}}]{Loomis2018}
{Loomis}, R.~A., {{\"O}berg}, K.~I., {Andrews}, S.~M., {et~al.} 2018, \aj, 155,
  182, \dodoi{10.3847/1538-3881/aab604}

\bibitem[{{Loomis} {et~al.}(2020){Loomis}, {{\"O}berg}, {Andrews}, {Bergin},
  {Bergner}, {Blake}, {Cleeves}, {Czekala}, {Huang}, {Le Gal}, {M{\'e}nard},
  {Pegues}, {Qi}, {Walsh}, {Williams}, \& {Wilner}}]{Loomis20_mwc480}
---. 2020, \apj, 893, 101, \dodoi{10.3847/1538-4357/ab7cc8}

\bibitem[{{Lynden-Bell} \& {Pringle}(1974)}]{lynden-bell74}
{Lynden-Bell}, D., \& {Pringle}, J.~E. 1974, \mnras, 168, 603,
  \dodoi{10.1093/mnras/168.3.603}

\bibitem[{{Mac{\'\i}as} {et~al.}(2018){Mac{\'\i}as}, {Espaillat}, {Ribas},
  {Schwarz}, {Anglada}, {Osorio}, {Carrasco-Gonz{\'a}lez}, {G{\'o}mez}, \&
  {Robinson}}]{Macias18}
{Mac{\'\i}as}, E., {Espaillat}, C.~C., {Ribas}, {\'A}., {et~al.} 2018, \apj,
  865, 37, \dodoi{10.3847/1538-4357/aad811}

\bibitem[{{Manara} {et~al.}(2016){Manara}, {Rosotti}, {Testi}, {Natta},
  {Alcal{\'a}}, {Williams}, {Ansdell}, {Miotello}, {van der Marel}, {Tazzari},
  {Carpenter}, {Guidi}, {Mathews}, {Oliveira}, {Prusti}, \& {van
  Dishoeck}}]{manara16}
{Manara}, C.~F., {Rosotti}, G., {Testi}, L., {et~al.} 2016, \aap, 591, L3,
  \dodoi{10.1051/0004-6361/201628549}

\bibitem[{{Mangum} \& {Shirley}(2015)}]{Mangum15}
{Mangum}, J.~G., \& {Shirley}, Y.~L. 2015, \pasp, 127, 266,
  \dodoi{10.1086/680323}

\bibitem[{{Mathis} {et~al.}(1977){Mathis}, {Rumpl}, \& {Nordsieck}}]{Mathis77}
{Mathis}, J.~S., {Rumpl}, W., \& {Nordsieck}, K.~H. 1977, \apj, 217, 425,
  \dodoi{10.1086/155591}

\bibitem[{{McClure} {et~al.}(2016){McClure}, {Bergin}, {Cleeves}, {van
  Dishoeck}, {Blake}, {Evans}, {Green}, {Henning}, {{\"O}berg}, {Pontoppidan},
  \& {Salyk}}]{mcclure16}
{McClure}, M.~K., {Bergin}, E.~A., {Cleeves}, L.~I., {et~al.} 2016, \apj, 831,
  167, \dodoi{10.3847/0004-637X/831/2/167}

\bibitem[{{McMullin} {et~al.}(2007){McMullin}, {Waters}, {Schiebel}, {Young},
  \& {Golap}}]{CASA}
{McMullin}, J.~P., {Waters}, B., {Schiebel}, D., {Young}, W., \& {Golap}, K.
  2007, in Astronomical Society of the Pacific Conference Series, Vol. 376,
  Astronomical Data Analysis Software and Systems XVI, ed. R.~A. {Shaw},
  F.~{Hill}, \& D.~J. {Bell}, 127

\bibitem[{{Mendigut{\'\i}a} {et~al.}(2013){Mendigut{\'\i}a}, {Brittain},
  {Eiroa}, {Meeus}, {Montesinos}, {Mora}, {Muzerolle}, {Oudmaijer}, \&
  {Rigliaco}}]{Mendigutia13}
{Mendigut{\'\i}a}, I., {Brittain}, S., {Eiroa}, C., {et~al.} 2013, \apj, 776,
  44, \dodoi{10.1088/0004-637X/776/1/44}

\bibitem[{{Miotello} {et~al.}(2014){Miotello}, {Bruderer}, \& {van
  Dishoeck}}]{miotello14}
{Miotello}, A., {Bruderer}, S., \& {van Dishoeck}, E.~F. 2014, \aap, 572, A96,
  \dodoi{10.1051/0004-6361/201424712}

\bibitem[{{Miotello} {et~al.}(2017){Miotello}, {van Dishoeck}, {Williams},
  {Ansdell}, {Guidi}, {Hogerheijde}, {Manara}, {Tazzari}, {Testi}, {van der
  Marel}, \& {van Terwisga}}]{miotello17}
{Miotello}, A., {van Dishoeck}, E.~F., {Williams}, J.~P., {et~al.} 2017, \aap,
  599, A113, \dodoi{10.1051/0004-6361/201629556}

\bibitem[{{Molyarova} {et~al.}(2017){Molyarova}, {Akimkin}, {Semenov},
  {Henning}, {Vasyunin}, \& {Wiebe}}]{Molyarova17}
{Molyarova}, T., {Akimkin}, V., {Semenov}, D., {et~al.} 2017, \apj, 849, 130,
  \dodoi{10.3847/1538-4357/aa9227}

\bibitem[{{Montesinos} {et~al.}(2009){Montesinos}, {Eiroa}, {Mora}, \&
  {Mer{\'\i}n}}]{Montesinos09}
{Montesinos}, B., {Eiroa}, C., {Mora}, A., \& {Mer{\'\i}n}, B. 2009, \aap, 495,
  901, \dodoi{10.1051/0004-6361:200810623}

\bibitem[{{Morbidelli} \& {Raymond}(2016)}]{morbidelli16b}
{Morbidelli}, A., \& {Raymond}, S.~N. 2016, Journal of Geophysical Research
  (Planets), 121, 1962, \dodoi{10.1002/2016JE005088}

\bibitem[{{Mulders} {et~al.}(2017){Mulders}, {Pascucci}, {Manara}, {Testi},
  {Herczeg}, {Henning}, {Mohanty}, \& {Lodato}}]{Mulders17}
{Mulders}, G.~D., {Pascucci}, I., {Manara}, C.~F., {et~al.} 2017, \apj, 847,
  31, \dodoi{10.3847/1538-4357/aa8906}

\bibitem[{{Najita} \& {Bergin}(2018)}]{Najita18_evolution}
{Najita}, J.~R., \& {Bergin}, E.~A. 2018, \apj, 864, 168,
  \dodoi{10.3847/1538-4357/aad80c}

\bibitem[{{Nakagawa} {et~al.}(1986){Nakagawa}, {Sekiya}, \&
  {Hayashi}}]{Nakagawa86}
{Nakagawa}, Y., {Sekiya}, M., \& {Hayashi}, C. 1986, \icarus, 67, 375,
  \dodoi{10.1016/0019-1035(86)90121-1}

\bibitem[{{Nomura} {et~al.}(2016){Nomura}, {Tsukagoshi}, {Kawabe}, {Ishimoto},
  {Okuzumi}, {Muto}, {Kanagawa}, {Ida}, {Walsh}, {Millar}, \& {Bai}}]{nomura16}
{Nomura}, H., {Tsukagoshi}, T., {Kawabe}, R., {et~al.} 2016, \apjl, 819, L7,
  \dodoi{10.3847/2041-8205/819/1/L7}

\bibitem[{{Oberg} {et~al.}(2021){Oberg}, {Guzman}, {Walsh}, {Aikawa}, {Bergin},
  {Law}, {Loomis}, {Alarcon}, {Andrews}, {Bae}, {Bergner}, {Boehler}, {Booth},
  {Bosman}, {Calahan}, {Cataldi}, {Cleeves}, {Czekala}, {Furuya}, {Huang},
  {Ilee}, {Kurtovic}, {Le Gal}, {Liu}, {Long}, {Menard}, {Nomura}, {Perez},
  {Qi}, {Schwarz}, {Sierra}, {Teague}, {Tsukagoshi}, {Yamato}, {van 't Hoff},
  {Waggoner}, {Wilner}, \& {Zhang}}]{oberg20_maps}
{Oberg}, K.~I., {Guzman}, V.~V., {Walsh}, C., {et~al.} 2021, arXiv e-prints,
  arXiv:2109.06268.
\newblock \doarXiv{2109.06268}

\bibitem[{{Okuzumi} {et~al.}(2016){Okuzumi}, {Momose}, {Sirono}, {Kobayashi},
  \& {Tanaka}}]{okuzumi16}
{Okuzumi}, S., {Momose}, M., {Sirono}, S.-i., {Kobayashi}, H., \& {Tanaka}, H.
  2016, \apj, 821, 82, \dodoi{10.3847/0004-637X/821/2/82}

\bibitem[{{Pi{\'e}tu} {et~al.}(2007){Pi{\'e}tu}, {Dutrey}, \&
  {Guilloteau}}]{pietu07}
{Pi{\'e}tu}, V., {Dutrey}, A., \& {Guilloteau}, S. 2007, \aap, 467, 163,
  \dodoi{10.1051/0004-6361:20066537}

\bibitem[{{Pinilla} {et~al.}(2017){Pinilla}, {Pohl}, {Stammler}, \&
  {Birnstiel}}]{Pinilla17}
{Pinilla}, P., {Pohl}, A., {Stammler}, S.~M., \& {Birnstiel}, T. 2017, \apj,
  845, 68, \dodoi{10.3847/1538-4357/aa7edb}

\bibitem[{{Pinte} {et~al.}(2018{\natexlab{a}}){Pinte}, {M{\'e}nard},
  {Duch{\^e}ne}, {Hill}, {Dent}, {Woitke}, {Maret}, {van der Plas}, {Hales},
  {Kamp}, {Thi}, {de Gregorio-Monsalvo}, {Rab}, {Quanz}, {Avenhaus}, {Carmona},
  \& {Casassus}}]{Pinte18_imlup}
{Pinte}, C., {M{\'e}nard}, F., {Duch{\^e}ne}, G., {et~al.} 2018{\natexlab{a}},
  \aap, 609, A47, \dodoi{10.1051/0004-6361/201731377}

\bibitem[{{Pinte} {et~al.}(2018{\natexlab{b}}){Pinte}, {Price}, {M{\'e}nard},
  {Duch{\^e}ne}, {Dent}, {Hill}, {de Gregorio- Monsalvo}, {Hales}, \&
  {Mentiplay}}]{Pinte18_hd163}
{Pinte}, C., {Price}, D.~J., {M{\'e}nard}, F., {et~al.} 2018{\natexlab{b}},
  \apj, 860, L13, \dodoi{10.3847/2041-8213/aac6dc}

\bibitem[{{Pinte} {et~al.}(2020){Pinte}, {Price}, {M{\'e}nard}, {Duch{\^e}ne},
  {Christiaens}, {Andrews}, {Huang}, {Hill}, {van der Plas}, {Perez}, {Isella},
  {Boehler}, {Dent}, {Mentiplay}, \& {Loomis}}]{Pinte20_dsharp}
---. 2020, \apjl, 890, L9, \dodoi{10.3847/2041-8213/ab6dda}

\bibitem[{Qi {et~al.}(2011)Qi, D'Alessio, Oberg, Wilner, Hughes, Andrews, \&
  Ayala}]{Qi11}
Qi, C., D'Alessio, P., Oberg, K.~I., {et~al.} 2011, \apj, 740, 84

\bibitem[{{Qi} {et~al.}(2006){Qi}, {Wilner}, {Calvet}, {Bourke}, {Blake},
  {Hogerheijde}, {Ho}, \& {Bergin}}]{Qi06}
{Qi}, C., {Wilner}, D.~J., {Calvet}, N., {et~al.} 2006, \apjl, 636, L157,
  \dodoi{10.1086/500241}

\bibitem[{{Rab} {et~al.}(2020){Rab}, {Kamp}, {Dominik}, {Ginski}, {Muro-Arena},
  {Thi}, {Waters}, \& {Woitke}}]{Rab2020}
{Rab}, C., {Kamp}, I., {Dominik}, C., {et~al.} 2020, \aap, 642, A165,
  \dodoi{10.1051/0004-6361/202038712}

\bibitem[{{Riols} \& {Lesur}(2019)}]{Riols19}
{Riols}, A., \& {Lesur}, G. 2019, \aap, 625, A108,
  \dodoi{10.1051/0004-6361/201834813}

\bibitem[{{Ros} \& {Johansen}(2013)}]{ros13}
{Ros}, K., \& {Johansen}, A. 2013, \aap, 552, A137,
  \dodoi{10.1051/0004-6361/201220536}

\bibitem[{{Ros} {et~al.}(2019){Ros}, {Johansen}, {Riipinen}, \&
  {Schlesinger}}]{Ros19}
{Ros}, K., {Johansen}, A., {Riipinen}, I., \& {Schlesinger}, D. 2019, \aap,
  629, A65, \dodoi{10.1051/0004-6361/201834331}

\bibitem[{{Rosotti} {et~al.}(2021){Rosotti}, {Ilee}, {Facchini}, {Tazzari},
  {Booth}, {Clarke}, \& {Kama}}]{Rosotti21}
{Rosotti}, G.~P., {Ilee}, J.~D., {Facchini}, S., {et~al.} 2021, \mnras, 501,
  3427, \dodoi{10.1093/mnras/staa3869}

\bibitem[{{Rosotti} {et~al.}(2016){Rosotti}, {Juhasz}, {Booth}, \&
  {Clarke}}]{Rosotti16}
{Rosotti}, G.~P., {Juhasz}, A., {Booth}, R.~A., \& {Clarke}, C.~J. 2016,
  \mnras, 459, 2790, \dodoi{10.1093/mnras/stw691}

\bibitem[{{Salyk} {et~al.}(2013){Salyk}, {Herczeg}, {Brown}, {Blake},
  {Pontoppidan}, \& {van Dishoeck}}]{salyk13}
{Salyk}, C., {Herczeg}, G.~J., {Brown}, J.~M., {et~al.} 2013, \apj, 769, 21,
  \dodoi{10.1088/0004-637X/769/1/21}

\bibitem[{{Schwarz} {et~al.}(2016){Schwarz}, {Bergin}, {Cleeves}, {Blake},
  {Zhang}, {{\"O}berg}, {van Dishoeck}, \& {Qi}}]{schwarz16}
{Schwarz}, K.~R., {Bergin}, E.~A., {Cleeves}, L.~I., {et~al.} 2016, \apj, 823,
  91, \dodoi{10.3847/0004-637X/823/2/91}

\bibitem[{{Schwarz} {et~al.}(2018){Schwarz}, {Bergin}, {Cleeves}, {Zhang},
  {{\"O}berg}, {Blake}, \& {Anderson}}]{Schwarz18}
---. 2018, \apj, 856, 85, \dodoi{10.3847/1538-4357/aaae08}

\bibitem[{{Schwarz} {et~al.}(2019){Schwarz}, {Bergin}, {Cleeves}, {Zhang},
  {{\"O}berg}, {Blake}, \& {Anderson}}]{Schwarz19a}
---. 2019, \apj, 877, 131, \dodoi{10.3847/1538-4357/ab1c5e}

\bibitem[{{Schwarz} {et~al.}(2021){Schwarz}, {Calahan}, {Zhang}, {Alarc{\'o}n},
  {Aikawa}, {Andrews}, {Bae}, {Bergin}, {Booth}, {Bosman}, {Cataldi},
  {Cleeves}, {Czekala}, {Huang}, {Ilee}, {Law}, {Le Gal}, {Liu}, {Long},
  {Loomis}, {Mac{\'\i}as}, {McClure}, {M{\'e}nard}, {{\"O}berg}, {Teague}, {van
  Dishoeck}, {Walsh}, \& {Wilner}}]{Schwarz20}
{Schwarz}, K.~R., {Calahan}, J.~K., {Zhang}, K., {et~al.} 2021, arXiv e-prints,
  arXiv:2109.06228.
\newblock \doarXiv{2109.06228}

\bibitem[{{Shakura} \& {Sunyaev}(1973)}]{shakura73}
{Shakura}, N.~I., \& {Sunyaev}, R.~A. 1973, \aap, 24, 337

\bibitem[{{Sierra} {et~al.}(2021){Sierra}, {P{\'e}rez}, {Zhang}, {Law},
  {Guzm{\'a}n}, {Qi}, {Bosman}, {{\"O}berg}, {Andrews}, {Long}, {Teague},
  {Booth}, {Walsh}, {Wilner}, {M{\'e}nard}, {Cataldi}, {Czekala}, {Bae},
  {Huang}, {Bergner}, {Ilee}, {Benisty}, {Le Gal}, {Loomis}, {Tsukagoshi},
  {Liu}, {Yamato}, \& {Aikawa}}]{Sierra20}
{Sierra}, A., {P{\'e}rez}, L.~M., {Zhang}, K., {et~al.} 2021, arXiv e-prints,
  arXiv:2109.06433.
\newblock \doarXiv{2109.06433}

\bibitem[{{Simon} {et~al.}(2019){Simon}, {Guilloteau}, {Beck}, {Chapillon}, {Di
  Folco}, {Dutrey}, {Feiden}, {Grosso}, {Pi{\'e}tu}, {Prato}, \&
  {Schaefer}}]{Simon19}
{Simon}, M., {Guilloteau}, S., {Beck}, T.~L., {et~al.} 2019, \apj, 884, 42,
  \dodoi{10.3847/1538-4357/ab3e3b}

\bibitem[{{Smirnov-Pinchukov} {et~al.}(2020){Smirnov-Pinchukov}, {Semenov},
  {Akimkin}, \& {Henning}}]{Smirnov-Pinchukov20}
{Smirnov-Pinchukov}, G.~V., {Semenov}, D.~A., {Akimkin}, V.~V., \& {Henning},
  T. 2020, \aap, 644, A4, \dodoi{10.1051/0004-6361/202038572}

\bibitem[{{Suriano} {et~al.}(2018){Suriano}, {Li}, {Krasnopolsky}, \&
  {Shang}}]{Suriano18}
{Suriano}, S.~S., {Li}, Z.-Y., {Krasnopolsky}, R., \& {Shang}, H. 2018, \mnras,
  477, 1239, \dodoi{10.1093/mnras/sty717}

\bibitem[{{Suzuki} {et~al.}(2016){Suzuki}, {Ogihara}, {Morbidelli}, {Crida}, \&
  {Guillot}}]{suzuki16}
{Suzuki}, T.~K., {Ogihara}, M., {Morbidelli}, A., {Crida}, A., \& {Guillot}, T.
  2016, \aap, 596, A74, \dodoi{10.1051/0004-6361/201628955}

\bibitem[{Teague(2019)}]{GoFish}
Teague, R. 2019, The Journal of Open Source Software, 4, 1632,
  \dodoi{10.21105/joss.01632}

\bibitem[{{Teague} {et~al.}(2019){Teague}, {Bae}, \& {Bergin}}]{Teague19Nat}
{Teague}, R., {Bae}, J., \& {Bergin}, E.~A. 2019, \nat, 574, 378,
  \dodoi{10.1038/s41586-019-1642-0}

\bibitem[{{Teague} {et~al.}(2018{\natexlab{a}}){Teague}, {Bae}, {Bergin},
  {Birnstiel}, \& {Foreman-Mackey}}]{Teague18a}
{Teague}, R., {Bae}, J., {Bergin}, E.~A., {Birnstiel}, T., \& {Foreman-Mackey},
  D. 2018{\natexlab{a}}, \apj, 860, L12, \dodoi{10.3847/2041-8213/aac6d7}

\bibitem[{{Teague} {et~al.}(2018{\natexlab{b}}){Teague}, {Bae}, {Birnstiel}, \&
  {Bergin}}]{Teague18_as209}
{Teague}, R., {Bae}, J., {Birnstiel}, T., \& {Bergin}, E.~A.
  2018{\natexlab{b}}, \apj, 868, 113, \dodoi{10.3847/1538-4357/aae836}

\bibitem[{{Teague} \& {Loomis}(2020)}]{2020ApJ...899..157T}
{Teague}, R., \& {Loomis}, R. 2020, \apj, 899, 157,
  \dodoi{10.3847/1538-4357/aba956}

\bibitem[{{Teague} {et~al.}(2016){Teague}, {Guilloteau}, {Semenov}, {Henning},
  {Dutrey}, {Pi{\'e}tu}, {Birnstiel}, {Chapillon}, {Hollenbach}, \&
  {Gorti}}]{teague16}
{Teague}, R., {Guilloteau}, S., {Semenov}, D., {et~al.} 2016, \aap, 592, A49,
  \dodoi{10.1051/0004-6361/201628550}

\bibitem[{{Teague} {et~al.}(2021){Teague}, {Bae}, {Aikawa}, {Andrews},
  {Bergin}, {Bergner}, {Boehler}, {Booth}, {Bosman}, {Cataldi}, {Czekala},
  {Guzm{\'a}n}, {Huang}, {Ilee}, {Law}, {Le Gal}, {Long}, {Loomis},
  {M{\'e}nard}, {{\"O}berg}, {P{\'e}rez}, {Schwarz}, {Sierra}, {Walsh},
  {Wilner}, {Yamato}, \& {Zhang}}]{Teague20_maps}
{Teague}, R., {Bae}, J., {Aikawa}, Y., {et~al.} 2021, arXiv e-prints,
  arXiv:2109.06218.
\newblock \doarXiv{2109.06218}

\bibitem[{{Trapman} {et~al.}(2017){Trapman}, {Miotello}, {Kama}, {van
  Dishoeck}, \& {Bruderer}}]{Trapman17}
{Trapman}, L., {Miotello}, A., {Kama}, M., {van Dishoeck}, E.~F., \&
  {Bruderer}, S. 2017, \aap, 605, A69, \dodoi{10.1051/0004-6361/201630308}

\bibitem[{{Trapman} {et~al.}(2020){Trapman}, {Rosotti}, {Bosman},
  {Hogerheijde}, \& {van Dishoeck}}]{Trapman20_viscous}
{Trapman}, L., {Rosotti}, G., {Bosman}, A.~D., {Hogerheijde}, M.~R., \& {van
  Dishoeck}, E.~F. 2020, \aap, 640, A5, \dodoi{10.1051/0004-6361/202037673}

\bibitem[{{Umebayashi} \& {Nakano}(1981)}]{Umebayashi81}
{Umebayashi}, T., \& {Nakano}, T. 1981, Publications of the Astronomical
  Society of Japan, 33, 617

\bibitem[{{van der Marel} {et~al.}(2015){van der Marel}, {van Dishoeck},
  {Bruderer}, {P{\'e}rez}, \& {Isella}}]{vanderMarel15}
{van der Marel}, N., {van Dishoeck}, E.~F., {Bruderer}, S., {P{\'e}rez}, L., \&
  {Isella}, A. 2015, \aap, 579, A106, \dodoi{10.1051/0004-6361/201525658}

\bibitem[{{van der Marel} {et~al.}(2018){van der Marel}, {Williams}, \&
  {Bruderer}}]{vanderMarel18}
{van der Marel}, N., {Williams}, J.~P., \& {Bruderer}, S. 2018, \apjl, 867,
  L14, \dodoi{10.3847/2041-8213/aae88e}

\bibitem[{{van Dishoeck} \& {Black}(1988)}]{vanDishoeck88}
{van Dishoeck}, E.~F., \& {Black}, J.~H. 1988, \apj, 334, 771,
  \dodoi{10.1086/166877}

\bibitem[{{Visser} {et~al.}(2009){Visser}, {van Dishoeck}, {Doty}, \&
  {Dullemond}}]{visser09}
{Visser}, R., {van Dishoeck}, E.~F., {Doty}, S.~D., \& {Dullemond}, C.~P. 2009,
  \aap, 495, 881, \dodoi{10.1051/0004-6361/200810846}

\bibitem[{{Walsh} {et~al.}(2014){Walsh}, {Millar}, {Nomura}, {Herbst}, {Widicus
  Weaver}, {Aikawa}, {Laas}, \& {Vasyunin}}]{Walsh14}
{Walsh}, C., {Millar}, T.~J., {Nomura}, H., {et~al.} 2014, \aap, 563, A33,
  \dodoi{10.1051/0004-6361/201322446}

\bibitem[{{Weaver} {et~al.}(2018){Weaver}, {Isella}, \& {Boehler}}]{Weaver18}
{Weaver}, E., {Isella}, A., \& {Boehler}, Y. 2018, \apj, 853, 113,
  \dodoi{10.3847/1538-4357/aaa481}

\bibitem[{{Williams} \& {Best}(2014)}]{williams14}
{Williams}, J.~P., \& {Best}, W. M.~J. 2014, \apj, 788, 59,
  \dodoi{10.1088/0004-637X/788/1/59}

\bibitem[{{Wilson}(1999)}]{wilson99}
{Wilson}, T.~L. 1999, Reports on Progress in Physics, 62, 143,
  \dodoi{10.1088/0034-4885/62/2/002}

\bibitem[{{Xu} {et~al.}(2017){Xu}, {Bai}, \& {{\"O}berg}}]{xu17}
{Xu}, R., {Bai}, X.-N., \& {{\"O}berg}, K. 2017, \apj, 835, 162,
  \dodoi{10.3847/1538-4357/835/2/162}

\bibitem[{{Yu} {et~al.}(2016){Yu}, {Willacy}, {Dodson-Robinson}, {Turner}, \&
  {Evans}}]{yu16}
{Yu}, M., {Willacy}, K., {Dodson-Robinson}, S.~E., {Turner}, N.~J., \& {Evans},
  II, N.~J. 2016, \apj, 822, 53, \dodoi{10.3847/0004-637X/822/1/53}

\bibitem[{{Zhang} {et~al.}(2016){Zhang}, {Bergin}, {Blake}, {Cleeves},
  {Hogerheijde}, {Salinas}, \& {Schwarz}}]{Zhang16}
{Zhang}, K., {Bergin}, E.~A., {Blake}, G.~A., {et~al.} 2016, \apjl, 818, L16,
  \dodoi{10.3847/2041-8205/818/1/L16}

\bibitem[{{Zhang} {et~al.}(2017){Zhang}, {Bergin}, {Blake}, {Cleeves}, \&
  {Schwarz}}]{zhang17}
{Zhang}, K., {Bergin}, E.~A., {Blake}, G.~A., {Cleeves}, L.~I., \& {Schwarz},
  K.~R. 2017, Nature Astronomy, 1, 0130, \dodoi{10.1038/s41550-017-0130}

\bibitem[{{Zhang} {et~al.}(2019){Zhang}, {Bergin}, {Schwarz}, {Krijt}, \&
  {Ciesla}}]{Zhang19}
{Zhang}, K., {Bergin}, E.~A., {Schwarz}, K., {Krijt}, S., \& {Ciesla}, F. 2019,
  \apj, 883, 98, \dodoi{10.3847/1538-4357/ab38b9}

\bibitem[{{Zhang} {et~al.}(2015){Zhang}, {Blake}, \& {Bergin}}]{zhang15b}
{Zhang}, K., {Blake}, G.~A., \& {Bergin}, E.~A. 2015, \apjl, 806, L7,
  \dodoi{10.1088/2041-8205/806/1/L7}

\bibitem[{{Zhang} {et~al.}(2020{\natexlab{a}}){Zhang}, {Bosman}, \&
  {Bergin}}]{Zhang20_hd163}
{Zhang}, K., {Bosman}, A.~D., \& {Bergin}, E.~A. 2020{\natexlab{a}}, \apjl,
  891, L16, \dodoi{10.3847/2041-8213/ab77ca}

\bibitem[{{Zhang} {et~al.}(2020{\natexlab{b}}){Zhang}, {Schwarz}, \&
  {Bergin}}]{Zhang20_evolution}
{Zhang}, K., {Schwarz}, K.~R., \& {Bergin}, E.~A. 2020{\natexlab{b}}, \apjl,
  891, L17, \dodoi{10.3847/2041-8213/ab7823}

\bibitem[{{Zhang} {et~al.}(2018){Zhang}, {Zhu}, {Huang}, {Guzm{\'a}n},
  {Andrews}, {Birnstiel}, {Dullemond}, {Carpenter}, {Isella}, {P{\'e}rez},
  {Benisty}, {Wilner}, {Baruteau}, {Bai}, \& {Ricci}}]{Zhang_S18}
{Zhang}, S., {Zhu}, Z., {Huang}, J., {et~al.} 2018, \apj, 869, L47,
  \dodoi{10.3847/2041-8213/aaf744}

\end{thebibliography}
\bibliographystyle{aasjournal}

\end{document}